\documentclass[manuscript]{aastex}
\usepackage{epsfig}
\usepackage{graphicx}
\usepackage{lscape}
\usepackage{float}
\usepackage{geometry}
\usepackage{epsfig}
\usepackage[usenames]{color}
\usepackage[stable]{footmisc}
\usepackage{natbib}
\usepackage{hyperref}

\def\l{$\lambda$}
\def\mbh{$M_{\rm BH}$\/}
\def\nh{$n_{\mathrm{H}}$\/}

\def\ne{$n_{\rm e}$\/}
\def\nc{$N_{\rm c}$\/}
\def\rfe{$R_{\rm FeII}$}

\def\feiiq{\rm Fe{\sc ii}$\lambda$4570\/}

\def\ltsima{$\; \buildrel < \over \sim \;$}
\def\simlt{\lower.5ex\hbox{\ltsima}}  
\def\gtsima{$\; \buildrel > \over \sim \;$}

\def\simgt{\lower.5ex\hbox{\gtsima}}

\def\lya{{ Ly}$\alpha$}
\def\civ{{\sc{Civ}}$\lambda$1549\/}

\def\cmq{cm$^{-2}$\/}
\def\cm3{cm$^{-3}$\/}
\def\hb{{\sc{H}}$\beta$\/}

\def\mgii{{Mg\sc{ii}}$\lambda$2800\/}

\def\niv{{\sc{Niv]}}$\lambda$1486\/}
\def\ciii{{\sc{Ciii]}}$\lambda$1909\/}

\def\o4363{{\sc{[Oiii]}}$\lambda$4363\/}

\def\siiii{ Si{\sc iii]}$\lambda$1892\/}
\def\aliii{Al{\sc iii}$\lambda$1860\/}
\def\heiiuv{He{\sc{ii}}$\lambda$1640}
\def\nv{{N\sc{v}}$\lambda$1240}

\def\feii{{Fe\sc{ii}}\/}
\def\siii{{Si\sc{ii}}$\lambda$1814\/}
\def\feiii{{Fe\sc{iii}}\/}
\def\fe{{\sc{Fe}}\/}

\def\fe76087{{\sc [Fe vii]}$\lambda$6087\/}

\def\kms{km~s$^{-1}$}

\def\ergss{ergs s$^{-1}$\/}
\def\hii{H{\sc ii}\/}

\def\heii{{{\sc H}e{\sc ii}}$\lambda$4686\/}

\def\rb{$r_{\rm BLR}$\/}

\def\siiv{Si{\sc iv}$\lambda$1397\/}
\def\oiv{O{\sc iv]}$\lambda$1402\/}

\def\siiiuv{Si{\sc ii}$\lambda$1533\/}
\def\feiiiuv{Fe{\sc iii}(UV34)}
\def\REF{\par\noindent\hangindent 20pt}

\slugcomment{To appear in ...}

\shorttitle{Broad Line Region of High $z$ Quasars}
\shortauthors{Negrete et al.}

\begin{document}
\title{Physical Conditions in the Broad Line Region of  $z\sim3$ Quasars:\\ 
A Photoionization Method to Derive \rb\ and \mbh\ \altaffilmark{\clubsuit}}
\author{C. Alenka Negrete\altaffilmark{1} and Deborah Dultzin\altaffilmark{1}}
\affil{Instituto de Astonom\'ia, Universidad Nacional Aut\'onoma de M\'exico, Mexico}
\author{Paola Marziani\altaffilmark{2}}
\affil{INAF, Astronomical Observatory of Padova, Italy}
\and
\author{Jack Sulentic\altaffilmark{3}}
\affil{Instituto de Astrof\'isica de Andaluc\'ia, Spain}
\altaffiltext{0}{Based on observations made with ESO Telescopes at  Paranal Observatory under programme ID 078.B-0109(A) }
\altaffiltext{1}{anegrete@astroscu.unam.mx, deborah@astroscu.unam.mx}
\altaffiltext{2}{paola.marziani@oapd.inaf.it}
\altaffiltext{3}{sulentic@iaa.es}

\begin{abstract}
We present high S/N UV spectra for eight quasars at $z\sim3$\ obtained with VLT/FORS. The spectra  enable us  to analyze in detail the strongest emission features in the rest-frame range 1400-2000 \AA\ of each source (\ciii, \siiii, \aliii, \siii, \civ\ and \siiv). Previous work indicates that a component of these lines is emitted in a region with well-defined properties i.e., a high density and low ionization emitting region). Flux ratios \aliii/\siiii, \civ/\aliii, \siiv/\siiii, \civ/\siiv\ and \siii/\siiii\ for this region permit us to strongly constrain  electron density, ionization parameter and metallicity through the use of  diagnostic maps built from  {\sc CLOUDY} simulations. Reliable estimates of  the product density times ionization parameter allow us to derive the radius of the  broad line region \rb\ from the definition of the ionization parameter. The \rb\ estimate and the assumption of virialized motions in the line emitting gas  yields an estimate for black hole mass. We compare our results with estimates obtained from the \rb\ -- luminosity correlation customarily employed to estimate black hole masses of high redshift quasars. 
\end{abstract}


\section{Introduction}

\subsection{Interpreting Quasar Spectra}

Measuring relevant physical parameters from the observed broad-line spectra of quasars 
is still an open challenge. Identification and intensity measurements of the strongest 
emission lines has made possible a rough inference of the typical conditions of the 
emitting gas since the earliest days of quasar spectroscopy. The very first quasars of 
intermediate redshift discovered in the 1960s showed a fairly high ionization spectrum, 
with prominent lines of \civ, and \heiiuv\ in addition to strong Balmer lines of the 
low-redshift quasars. Photoionization by the central continuum source was considered 
the preeminent heating mechanism of the emitting gas. Significant \ciii\ emission 
suggested electron densities (\ne) in the range $10^9 - 10^{10}$ \cm3. The observed 
intensity ratio \ciii/\civ\ indicated ionization parameter (defined by Eq. \ref{eq:u} later 
in this paper) values of the order of $10^{-1}$. This photoionization scenario was 
successful in explaining at least some quasar optical and UV spectra (see the review by 
Davidson and Netzer 1979 for a synopsis). More recent work emphasized the existence of  several problems with this simple scenario. Low ionization lines (LILs), and especially 
\feii\ are too strong to be explained by a photoionized region of moderate density and 
column density (see for example Dumont and Mathez 1981; Joly 1987; Collin-Souffrin et al.  1988; Dumont and Collin-Souffrin 1990). These authors stressed that the LILs required a  denser, low-temperature environment. 

We unfortunately lack a simple well-defined diagnostic measure of physical 
conditions in the broad line region. One strategy for estimating electron 
density in astrophysical sources involves using two emission lines of the same ion,
and with similar energies above the ground level, but with different radiative 
transition probabilities $A_{ki}$. In practice one often chooses two lines from the 
same term where one forbidden/semi-forbidden transition is associated with 
another that is semi-formidden/permitted in order to ensure very different 
values of $A_{ki}$ (for example, [Si{\sc iii}]\l1882 and \siiii).  
This technique is not straightforwardly applicable to 
the broad lines of quasars, precisely because the lines are broad, suitable 
candidates are too closely spaced in wavelength, and density is at least an order 
of magnitudes higher than the critical density for the forbidden transitions 
used in spectra of planetary nebul\ae\ and \hii\ regions. In addition the 
S/N and the resolution of quasar spectra are usually not very high. 

Feibelman and Aller (1987) used the \ciii/\siiii\ ratio to study the rather 
high-density environment  typical of symbiotic stars (\ne $ \sim 10^{7-10}$ \cm3).
In this case we have two semi-forbidden (intercombination) resonance lines 
with significantly different transition probabilities (see Table \ref{tab:lines}). 
The lines are emitted by two iso-electronic species with somewhat different 
ionization potentials (11 eV for C$ ^+$ vs. 8 eV for Si$ ^+$). The ionic 
fraction is dependent on the relative abundance of silicon-to-carbon (to be assumed) 
as well as on the ionization structure within the emitting region (to be computed). 
Using lines of different ions introduces additional potentially serious sources 
of uncertainty. It is perhaps not surprising that most workers believe 
that \ne\ in the BLR cannot be reliably estimated using quasar spectra. 
Strong  \ciii\  emission would imply that \ne\ cannot be very high 
(\ne $ \sim 10^{11-13}$\cm3). Very high density was invoked to explain the 
rich low ionization spectrum (especially \feii) seen in the spectra of most quasars. 
Several lines in the UV spectrum of I Zw 1 point towards high density at least 
for the LIL emitting zone: prominent \feii, relatively strong \aliii, and detection of 
C{\sc iii}\l1176 (Baldwin et al. 1996; Laor et al. 1997b). The region where these lines 
are produced cannot emit much \ciii. But is \ciii\  really so strong 
in most quasars?  BLR conditions are certainly complex and a single emitting 
region is  not sufficient to explain both LILs and high ionization lines (HILs).

\subsection{Quasar Systematics}

Quasar spectra are not all alike. There are significant differences in line 
intensity ratios and broad line profiles from object to object (Bachev et al. 
2004; Marziani et al. 2010). More importantly, these differences can be 
organized in a systematic way as has been realized since the early 1990s 
(Boroson and Green 1992). Since then several authors stressed the importance 
of the so-called eigenvector 1 (E1) of quasars (e.g. Gaskell et al. 1999). 
Sulentic et al. (2000, 2007) expanded the E1 trends into a 4-dimensional 
space involving optical, UV and X-ray measures. They also defined spectral 
types along a sequence occupied by AGN in an optical plane involving \feii\  
and FWHM H$\beta$ parameters. Objects at extreme ends of the E1 sequence 
are very different at almost all wavelengths and median spectra computed 
in spectral bins within this plane emphasize systematic changes in broad 
line properties (Sulentic et al. 2002, 2007). The differences have motivated 
the suggestion of a possible dichotomy between Narrow Line Seyfert 1 (NLSy1s) 
like sources and broader line objects that include almost all radio-loud 
quasars. The most effective divider of the two quasar types appears to be  
at FWHM of the H$\beta$\ broad component H$\beta _\mathrm{BC}  \approx$ 4000 \kms\ for low-to-moderate luminosity  sources (Marziani et al. 2009). This corresponds to to Eddington ratio $L/L_\mathrm{Edd} \sim 0.2 \pm$0.1 (Marziani et al. 2003b).

The distinction between NLSy1-like objects (hereafter Population A sources with FWHM(\hb) $\leq$ 4000 \kms) and the rest of quasars (Population B with FWHM(\hb) $\ga$ 4000 \kms) is of special relevance here. 
Pop. A sources show relatively low equivalent width lines with e.g. the \l1900 
blend $\sim$30 \AA. A close analysis of the \ciii\ blend in the prototypical 
NLSy1 I Zw 1 shows strong \siiii, and \aliii\ blended with rather weak \ciii\ 
along with prominent \feiiiuv (\l1895.5, \l1914.0, and \l1926.3) and \feii\  blends (Laor et al. 1997b; Vestergaard and Wilkes 2001). The \ciii\ line is apparently so weak that \feiii \l1914 may be the most prominent feature at $\lambda \approx$ 1910\AA\ (cf. Hartig \& Baldwin 1986). This interpretation is confirmed by detailed deblending of a source (SDSS J120144.36+011611.6) that 
can be considered a high luminosity analog of I Zw 1 (\S \ref{sec:sdss}). 

{\bf Median composite UV spectra  of low-$z$\ quasars show that \aliii\ and \siiii\ are more prominent in 4DE1 spectral types A2 and A3. Bins A1, A2, A3 are defined in terms of increasing FeII$\lambda$4570 (see Fig. 1 of Sulentic et al. 2002). I Zw 1, although belonging to the extreme type A3  (Bachev et al. 2004)  is not unique as a NLSy1, since the fraction of Pop. A sources in bin A3 and  A4 is $\approx$ 20 \%\ of all Pop. A sources in the sample of Zamfir et al. (2010). A significant  number of extreme NLSy1 are not classified as quasars in SDSS and must be collected from the galaxy catalog (Hu et al. 2008). }

\subsection{Emission Line Diagnostics and BLR Properties}

Emission lines and line ratios are used in diagnostic maps to estimate
temperatures and electronic densities in galactic and extragalactic 
photoionized regions. Examples include HII-regions and galaxies with 
HII-region nuclear spectra where electron densities are less than
\ne $\approx 10^4$ \cm3. This method has been successfully applied
to the narrow line region (NLR) in AGN. Application to the broad 
line quasars has yielded results that are difficult to interpret.
One recent exception is the  work by Maksuoka et al. (2008) who
succeeded in analyzing the partly ionized region thought to emit most 
of the LILs in quasar spectra.  They suggest that O{\sc i}  8446 
and the Ca{\sc ii} triplet are emitted by  dense, low 
ionization gas probably located in the periphery of the BLR.
If the electron density and ionization conditions are known it is possible to 
derive, with additional assumptions, the distance of the BLR emitting region 
from the central continuum source (as stressed earlier also by Baldwin et al. 1996).
 
The physical conditions of photoionized gas can be described by 
electron density \ne, hydrogen column density \nc, metallicity  (normalized to solar), 
shape of the ionizing continuum, and the ionization parameter $U$. 
The latter represents the dimensionless ratio of the number of ionizing 
photons and the electron density \ne\ or, equivalently, the total number density of hydrogen \nh, ionized and neutral.\footnote{In a fully ionized medium \ne $\approx 1.2$ \nh. We prefer to adopt the definition based on \nh\ because it is the one employed in the CLOUDY computations.} Both $U$\ and \nh\ are related through the equation
\begin{equation}
\label{eq:u}
U = \frac {\int_{\nu_0}^{+\infty}  \frac{L_\nu} {h\nu} d\nu} {4\pi n_\mathrm{H} c r^2}
\end{equation}
where $L_{\nu}$\ is the specific luminosity per unit frequency, $h$\ is 
the Planck constant, $\nu_{0}$\ the Rydberg frequency, $c$ the speed of light, and $r$\ can be interpreted 
as the distance between the central source of ionizing radiation and the 
line emitting region. Note that $Un_\mathrm{H}c$ is the ionizing photon flux
\begin{equation}
\label{eq:phi}
\Phi(H) =  \frac{Q(H)} {4\pi r^2}.
\end{equation}
If we know {\em the product of } \nh\ and $U$, we 
can estimate the radius $r$\ of the BLR from Eq. \ref{eq:u}.
The dependence of $U$\ on \rb\ was used by Padovani \& Rafanelli (1988) 
to derive central black hole masses assuming a plausible  average 
value of the product \nh $\cdot U$. The typical value of \ne\ was 
derived at that time from semiforbidden line \ciii\ which implied that  the 
density could not be much higher than  \ne $\approx 10^{9.5}$ 
cm$^{-3}$ (Osterbrock \& Ferland 2006).  Padovani (1988) derived an 
average value $<U \cdot$ \ne $ >  \approx 10^{9.8}$\ from several sources where 
\rb\ had been determined from reverberation mapping, and for which the number of ionizing photons could be measured from multiwavelength observations. The average value was then used to compute black hole masses for a much larger sample of Seyfert 1 galaxies and  low-$z$\ quasars (Padovani \& Rafanelli 1988; Padovani, Burg \& Edelson 1989).  Wandel, Peterson \& Malkan (1999) compared the results of the photoionization method with the ones
obtained through reverberation mapping, found a very good correlation for the masses computed with the two methods, and concluded that ``both methods measure the mass of the central black hole.''

\subsection{Outline of the paper}

The importance of the product  $U \cdot$\nh\ goes beyond knowledge of the physical 
conditions in the BLR if it can lead to  estimates of  BLR radius and black hole mass. 
This paper identifies suitable emission line ratios that overcome 
some of the major problems in the analysis of emission lines of quasars. 
It also defines a photoionization method that can be applied to even the highest 
redshift quasars making use of high S/N  near-IR spectroscopic data.
In \S \ref{sec:observations} we present the spectra of 8 pilot sources obtained with the VLT/FORS;  
in \S \ref{sec:reduction} we discuss data reduction; 
In \S \ref{sec:data_analysis} we describe our method of fitting broad emission line profiles; 
in \S \ref{sec:component_analysis} we give a phenomenological interpretation of the profile fits;  
in \S \ref{sec:physical_conditions} we describe  a method for deriving BLR physical conditions and give the results of our fits; 
in \S \ref{sec:extreme} we discuss two sources not belonging to our sample that show extreme behavior and that are helpful to understand more common quasar spectra; 
in \S \ref{sec:results} we give the results for the photoionization method;
in \S \ref{rblr} we derive the radius of the BLR (its distance from the ionizing source) and the mass of the black hole for each quasar in our sample; 
in \S \ref{sec:discussion} we discuss our results; finally, 
in \S \ref{sec:conclusions} we summarize our results and the prospect of a more extended application of our method. 
All the computations were made considering $H_0$ =70 km s$^{-1}$ Mpc$^{-1}$  and a relative energy density $\Omega_\Lambda=0.70$ and $\Omega_\mathrm{M}=0.3$.

\section{Observations}
\label{sec:observations}

\begin{figure}
\epsscale{1.1}
\plotone{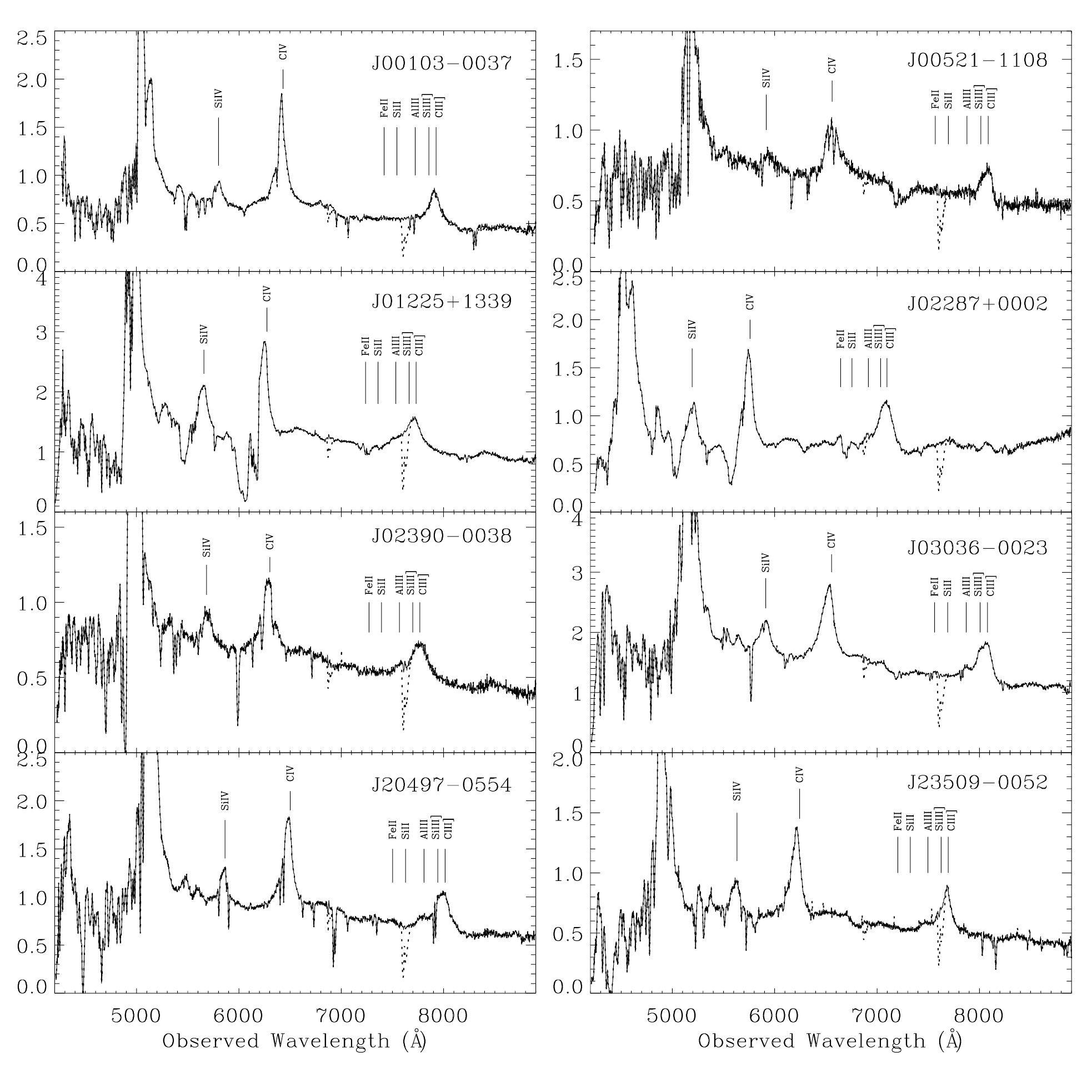}
\caption{Sample of 8 VLT spectra. Abscissa is obseved wavelength in \AA, ordinate is specific flux in units 10$^{-16}$ ergs s$^{-1}$ cm$^{-2}$ \AA$^{-1}$ corrected for Milky Way Galactic extinction. The superimposed dotted line is before atmospheric bands subtraction. We show the positions of the lines of our interest \ciii, \siiii, \aliii, \siii, Fe{\sc ii}$\lambda$1787,  \civ\ and \siiv. J01225+1339 and J02287 are BAL quasars.\label{fig:sample_ab}}
\end{figure}

Data were obtained between Nov. 2006 and Jan. 2007 using the VLT2/FORS1 
telescope operated in service mode. FORS1 is the visual and near UV focal reducer 
and low dispersion spectrograph of the Very Large Telescope (VLT) operated by 
European Southern Observatory (ESO) (Appenzeller et al. 1998).
Our VLT sample consists of 8 quasars with $z\sim3$. 
In Figure \ref{fig:sample_ab} we show the spectra uncorrected for redshift. Tab. \ref{tab:obs} 
provides a log of observations that is organized as follows. 

Column 1: object name, 
Col. 2: apparent B magnitude, 
Col. 3 redshift, 
Col. 4: line(s) used for redshift estimation: a)  O{\sc i}\l1304.8, b) \ciii; 
Col. 5: absolute B magnitude, 
Col.  6 flux at 6 cm taken from FIRST (Far InfraRed and Submillimetre Telescope), 
Col. 7: date (refers to time at start of exposure), 
Col. 8: Digital Integration Time, 
Col.  9: number of exposures with integration time equal to DIT, 
Col. 10: airmass at the beginning of each exposure, 
Col. 11 $S/N$ in the continuum around 1700\AA.

The observation of one of our 8 quasars, J00521-1108,  yielded only a low S/N spectrum which we
retain  because  observed features in the blend at $\sim 1900$ \AA\ are clear 
enough to fit the individual lines. Two sources, J01225+1339 and J02287+0002, 
are BAL quasars. We will keep them separate because \civ\ is  severely affected  
by absorption.

\section{Data Reduction} 
\label{sec:reduction}

Data were reduced using  standard {\sc iraf} tasks. All spectra were wavelength and flux 
calibrated in the observed frame and then corrected for Galactic extinction. Flux correction was applied using meteorological data provided by ESO. The observed flux was multiplied by the inverse of the light lost computed from the ratio seeing over slit width in arcsec. Correction to rest frame requires estimating the redshift $z$\ which is not a trivial task as outlined below. Rest frame correction also involved scaling the
specific flux in flux per unit wavelength interval by a factor $(1+z)^{3}$. Measurements were carried out on the rest-frame spectra. It is necessity to describe below two important aspects of the data reduction.

\subsection{A \& B Atmospheric Bands Correction}
\label{ab}

The A or B atmospheric band falls on top of the 1900\AA\ blend in many of the spectra.
This is an important region for this study especially because it involves  \siiii, 
\aliii, and  \siii. In order to remove these absorption features we created an 
A+ B band template from standard star spectra used as specific flux calibrators. 
We scaled this template to find a best fit. Fig. \ref{fig:sample_ab} shows the A and B absorption 
correction where we make a line identification to illustrate which lines are affected.  
In cases where the A or B bands overlap a weak line like \siii\ the 
effect is considerable and measures of \siii\ should not be considered at all or with extreme care.  This happens for sources J00103-0037, J03036-0023, and J20497-0554.
In  cases where one of the bands overlaps a stronger line like  \siiii\ or \aliii, 
the correction was good enough to permit accurate measures.

\subsection{Redshift Estimate}

Normally one uses strong narrow emission lines to set the rest frame in a quasar.
In our case no strong narrow lines are available so we consider the peaks of 
\lya, \civ\ and \ciii. The \lya\ peak is affected by absorption and  \civ\ 
is a HIL feature often showing blueshifts and/or asymmetries (Gaskell 1982, 
Espey et al. 1989, Corbin 1990, Tytler \& fan 1992; Marziani et al. 1996; Richards et al. 2002, Baskin \& Laor 2005;  Sulentic et al. 2007). This is especially true in Pop. A sources.
\ciii\ is  blended  with \siiii\ and \feiii\ that is especially prominent in this 
region and could well affect the peak. Pop. B sources show a rather weak Fe spectrum 
making the \ciii\ peak a more reliable $z$\ estimator. 

Our best option is to use the low ionization line O{\sc i}\l1304 whenever  it is strong. 
However it is blended with low ionization Si{\sc ii}\l\l1304,1309 
($^2 P^0_{3/2,1/2} - ^2S_{1/2}$). Both O{\sc i}\l1304 and Si{\sc ii}\l\l1304,1309 
are broad lines and in Pop. B sources might show large redshifts or even
significant blueshifts. Simulations in the (\nh, $U$) region of interest 
show O{\sc i}\l1304 $\approx$ 2 Si{\sc ii}\l 1304,1309 and this is confirmed 
in the spectrum I Zw 1 where O{\sc i}\l1304 and Si{\sc ii}\l\l1304,1309 are
resolved. The two components of the Si{\sc ii}\l\l1304,1309 doublet are set 
to the same intensity (i.e., we assume an optically thick case). 
We model the blend  O{\sc i}\l1304 + Si{\sc ii}\l\l1304,1309 with 5 Gaussians; the three components 
of the O{\sc i}  feature are produced by Bowen florescence mechanism, and 
should show ratios consistent with their transition probabilities. 
Generating a model spectrum in {\sc iraf} (lines broadened to 4000 \kms) yields a rest 
frame peak wavelength of 1304.8 $\pm$\ 0.2 \AA\ (in vacuum) which we use
as a reference for our VLT spectra assuming that there is no hint of 
systematic BC shifts as is the case for all Pop. A sources (the majority in our sample) and many Pop. B sources (Marziani et al. 2003a).

\begin{figure}
\epsscale{1.1}
\plotone{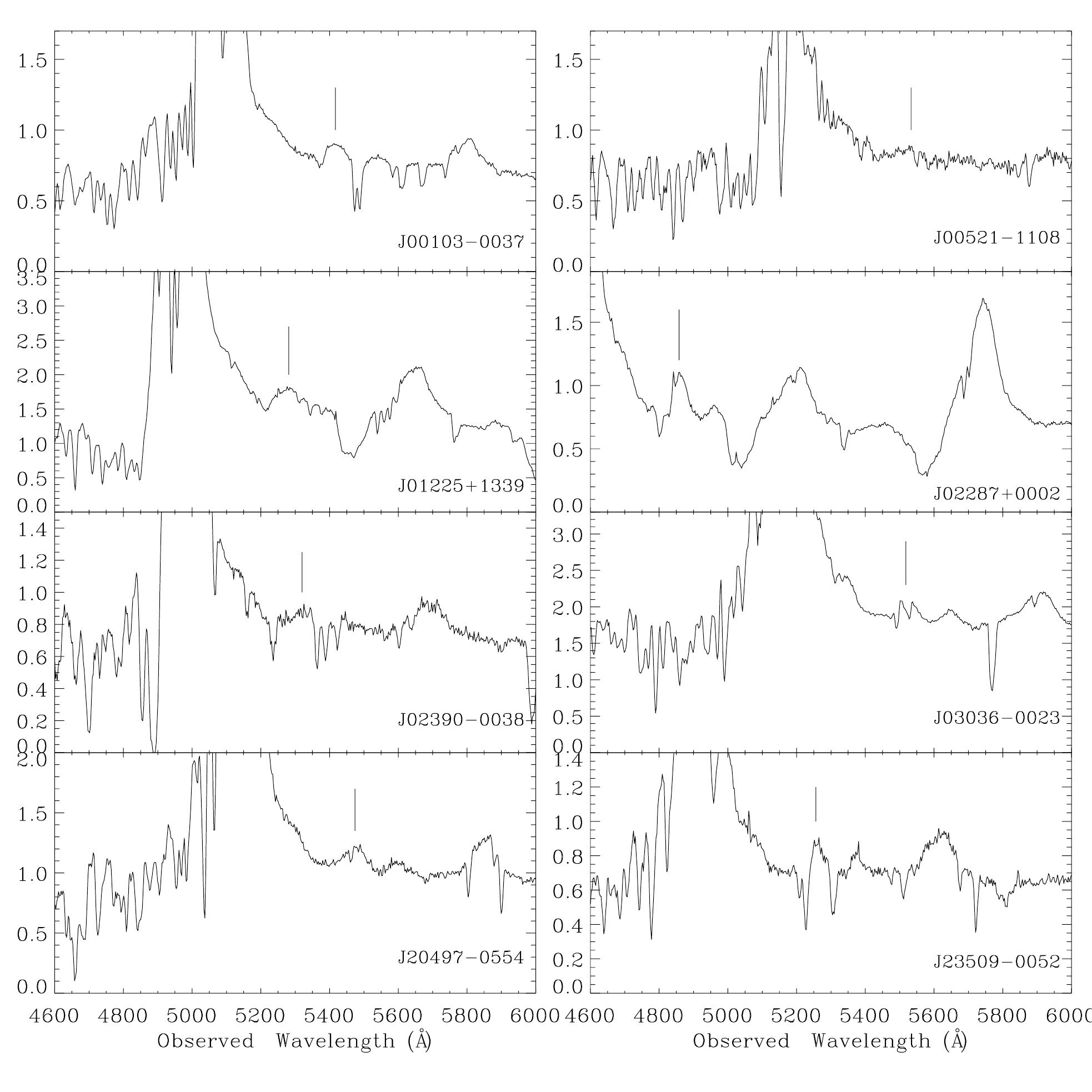}
\caption{O{\sc i}$\lambda$1304.8 used to place the restframe showed by the mark. Abscissa is rest frame wavelength in \AA, ordinate is specific flux in units 10$^{-16}$ ergs s$^{-1}$ cm$^{-2}$ \AA$^{-1}$ corrected for Milky Way Galactic extinction. In J00521-1108 and J02390-0038 the peak of O{\sc i}$\lambda$1304.8 is not observed clearly. In J00103-0037, J02287+0002 and J20497-0554, the redshift results, using both O{\sc i}$\lambda$1304.8 and \ciii, are ambiguous. In J01225+1339, J03036-0023 and J23509-0052, the redshifts obtained using O{\sc i}$\lambda$1304.8 or  \ciii\  are consistent. \label{fig:oi}}
\end{figure}

\begin{figure}
\epsscale{1.1}
\plotone{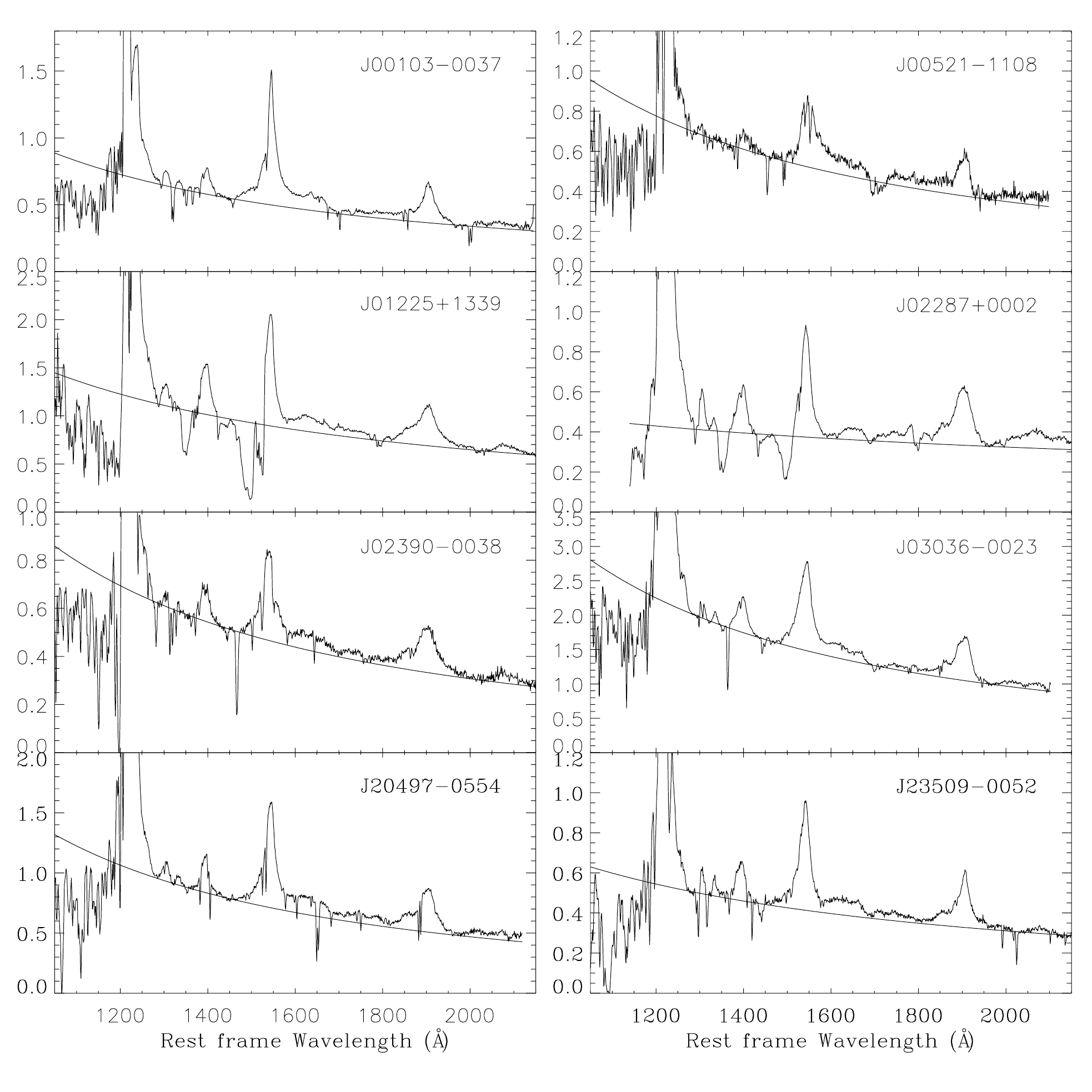}
\caption{Sample of 8 VLT spectra in rest frame wavelength.  Abscissa is rest frame in \AA, ordinate is specific flux in the rest frame in units 10$^{-13}$ ergs s$^{-1}$ cm$^{-2}$ \AA$^{-1}$.\label{fig:sample_z}}
\end{figure}

Examination of Fig. \ref{fig:oi} reveals that the peak of O{\sc i}\l1304 in source J00521-1108 is not observed clearly. We use \ciii\  to set the rest frame in this case. There are other sources J00103-0037, J02287+0002, J02390-0038 and J20497-0554 where the redshift estimation using both O{\sc i}\l1304 and \ciii\ are not in good agreement. 
The largest  disagreement  was found for J02287+0002. Redshifts obtained for 
the three remaining quasars, J01225+1339, J03036-0023 and J23509-0052, 
were obtained from O{\sc i}\l1304. Fig. \ref{fig:sample_z} shows the 
deredshifted VLT-FORS spectra for our sample of 8 quasars.

\section{Data Analysis}
\label{sec:data_analysis}

\subsection{Methodological Considerations on Multicomponent Fits}

The {\sc specfit}  {\sc IRAF} task (Kriss 1994)  allows us to fit the continuum, 
emission and absorption line components, \feii\  and \feiii\  blends, etc. 
We fit two spectral ranges: (1) 1450--1680 \AA\ for analysis of \civ\ and (2)
1750--2050 \AA\ for analysis of the 1900 \AA\ blend. Significant \feii\  
and \feiii\  emission are expected close to and underlying the 1900 \AA\ blend.  
Study of the 1900 \AA\ blend is especially difficult  in quasars because the
lines are broad and the blending severe. We therefore need to take advantage 
of several previous results.

\subsection{\feii\ and \feiii\ Emission}
\label{sec:fe2_fe3}

Our approach is completely empirical and employs an  \feii\  + \feiii\  template taken from templates successfully used in previous works.  Our \feii\  template is based on a {\sc CLOUDY} simulation and  is not very far from the preferred model of Bruhweiler \& Verner (2008).

\feiii\ lines are common and strong in the vicinity of \ciii\ as is evidenced by their
presence in average LBQS (Francis et al. 1991) and SDSS (Vanden Berk et al. 2001)
spectra. They appear to be strong when  \aliii\ is also strong (Hartig and Baldwin 1986). Vestergaard \& Wilkes (2001) produced an \feiii\  template based on the 
UV spectrum of I\,Zw\,1.  Since then,  Sigut et al. (2004) have modeled the \feiii\  
BLR spectrum.  See also Verner et al. (2003) for a plot of emission around 1900 \AA. 
\lya\ pumping enhances Fe{\sc iii} (UV 34)\l 1914.0  (Johansson et al. 2000) and this line can be a major contributor to the blend right on the red side of \ciii\  (see Fig. 2 of Vestergaard \& Wilkes). We reproduced the option B of the empirical \feiii\  template of Vestergaard \& Wilkes (2001), taking advantage of the line identifications from Ekberg (1993). When detected we can  use \feii\  UV 191  to set a rough \feii\ level while the feature at 2080 \AA\ 
is helpful for a more precise estimation of the intensity of \feiii. The continuum was 
fitted using the regions around 1450\AA\ (1750 and 1960\AA) that are relatively 
free of \feii\ emission (Vanden Berk et al. 2001). We used the same power-law to 
describe the continuum at both the \civ\ and 1900\AA\ regions. 

\subsection{Line Components}

We base our {\sc specfit} analysis on several previous observational results. The most  important ones are as follows:
\begin{itemize}

\item Sulentic et al. (2002) gridded the broad component of FWHM \hb$_{BC}$ versus \rfe=W(\feiiq blend)/W(\hb$_\mathrm{BC}$) parameter plane into bins of fixed $\Delta$ FWHM = 4000 \kms\ and $\Delta$ \rfe = 0.5. Quasar spectra  in different bins are different in many measures. 
As mentioned earlier, the largest differences are found between NLSy1-like objects, Pop. A, 
and  broader  sources of Pop. B  with FWHM(\hb) $\ga$ 4000 \kms. The gridding of Sulentic et al. (2002)  is valid for low $z$\  ($<$0.7) quasars. At higher $z$\ an adjustment must be made  since no sources with FWHM H$\beta_{BC} < 3500$ \kms\ exist above luminosity $\sim 10^{48}$ \ergss\ (Marziani  et al. 2009).   

\item Median spectra were computed for spectral bins from the atlas of Marziani 
et al. (2003a) who found that \hb\ can be described by a Lorentz function 
in Pop. A sources and by the sum of 2 Gaussians in Pop. B sources 
(unshifted + broader redshifted components) (Zamfir et al. 2010; Marziani 
et al. 2010).

\item A careful \feii\  subtraction reveals a blue-shifted \hb\ component in some
bin A3 sources (i.e. the stongest \feii\  emitters; Zamfir et al. 2010).

\item \civ\ (HIL) and H$\beta$ (LIL) profiles show significant differences in Pop. A. 
Large  \civ\ blueshifts ($\la -1000$ \kms) are observed in Pop. A {\em only} 
(Sulentic et al. 2007). HIL and LIL profiles are more similar in Pop. B sources.

\item We do not have \hb\ observations for our high-$z$\ objects since there are  no near IR spectra. We  use the results of Marziani et al. (2003b, 2010): they show that the BC of \siiii, \aliii\ and \civ\ lines is similar to the one of \hb, including the FWHM and profile shape, either Gaussian or Lorentzian. The similarity helps us to define whether an object is Population A or B in this paper.

\end{itemize}

These observational results point toward three different components in broad line 
profiles (see Marziani et al. 2010) which can be described as follows:

\begin{enumerate}

\item A broad component (BC) showing a roughly symmetric profile with 
FWHM in the range 1000-5000 \kms. It is consistent with the 
component identified by Matsuoka et al. (2008).  This broad component dominates 
LILs in Balmer lines of Pop. A sources while it becomes less prominent 
in Pop. B.   The profile is best modeled by a Lorentzian function in Pop. A
sources  while Pop. B profiles are better described by a Gaussian 
(Marziani et al. 2003b).

\item A very broad component (VBC), as seen in LILs and HILs of most pop B sources but is absent from Pop. A profiles.  The VBC can be modeled as a Gaussian (FWHM $\sim$ 10000 \kms) often with a significant shift to the red. It can be called a defining property of Pop. B sources (Marziani et al. 2010). 
{\bf This component is clearly identified  in the \civ\ line of Pop. B objects, and is also appreciable on the red side of  \ciii\ of Pop. B  objects J00103-0037 and J02390-0038.}

\item A blueshifted broad component (BBC), defined as the residual emission in the \civ\ line after subtracting a scaled BC Lorentzian profile (Marziani et al. 2010).  This blueshifted component is often prominent in \civ\ and \lya\ of Pop. A sources. It is much less intense in radio-loud Pop. B sources (Marziani et al. 1996; Punsly 2010; Richards et al. 2010). We model this profile as a blueshifted Gaussian. The  Gaussian approximation is probably inappropriate especially if the BCC is strong: this component is believed to be produced in a partially-obscured radial flow, not in a virialized emitting system.
{\bf In the present work we do not even try to fit the blueshifted in the doubly ionized lines. It is quite obvious from the fits that a possible contribution of this component would be negligible.}
\end{enumerate}

{\bf Baldwin et al. (1996) 
presented a similar analysis. Their Fig. 2 organizes spectra in a sequence that is roughly corresponding to E1, going from \aliii-strong sources to objects whose spectra show prominent \ciii\ along with weak \aliii\ (Bachev et al. 2004). 
Two of the three line components they isolated correspond to the ones we consider in this paper: a blue-shifted feature, and a more symmetric, unshifted and relatively narrow component that we call LIL-BC. Less obvious is the correspondence of a third feature, although it appears to be the redshifted part of what we call the VBC. 
}

{\bf Several improvements have been introduced since the paper of Baldwin et al. 1996. These improvements are expected to make our analysis easier.  First, the definition of a template of \feiii\ emission (Vestergaard \& Wilkes 2001),
along with the possibility to model \feii\ in the 1400--2000 \AA\ spectral region with {\sc cloudy} (Verner et al. 1999, 2004).
The analysis of spectra along the E1 sequence allows one to see trends that make the interpretation of the emission line blends easier (Marziani et al. 2010). 
}

\subsection{Expected emission from the various components} 
\label{Expected}

{\bf We looked for evidence of three possible components as described above: BC, VBC and BBC only for the most intense HILs: \civ\ and \siiv. We did not include the contributions of the BBC for the doubly-ionized lines. In the case of Pop. A sources, we indeed assume that the BC contains the vast majority of the light. In the case of Pop. B sources we consider the contribution of the VBC.  The \aliii\ doublet shows no evidence of either a BBC or VBC. There is no obvious BBC for \siiii\ in the blend. There is evidence of a VBC of \ciii\ extending on the red side of the blend, and this is taken into account. Moreover, we expect that the \aliii\ doublet is emitted exclusively in the BC, the region where \feii\ is also emitted. This is empirically confirmed  by the aspect of the 1900 blend in many sources, where we do not see any evidence of BBC nor VBC. We remark that the \aliii\ doublet is relatively unblended, and that a BBC  feature as strong as in \civ\ would not easily escape visual detection.The same is also true for \siiii.  Several fits that included BBC in \ciii\ yielded 0 intensity, implying a large \civ/\ciii\ (Marziani et al. 2010).   The \siiv\ + \oiv\ blend closely resembles the shape of \civ, suggesting that  BBC is relevant, especially for \siiv\ (\oiv\ is expected to give  a minority contribution to line emission at the high density derived for the BC; any \oiv\ contribution to the BBC is not relevant to our method). BBC is very weak or undetectable in the vast majority of the \hb\ profiles analyzed in Marziani et al. 2010 (but see Zamfir et al. 2010 for several cases of \hb\ BBC), while prominent in \lya; the \lya/\hb\ ratio in this component is high.  In summary, BBC is visually strong in \lya, \civ,  and \siiv. \heiiuv\ in BBC is needed for a self-consistent fit of the \civ+\heiiuv\ blend. The VBC of \siiii\ is poorly constrained, but in the fits where \ciii\ VBC is visible, we always find \siiii\ $<$ \ciii,  consistent with the high ionization level expected in the VBC region. Some \siiv\ VBC emission is assumed in the Pop. B fits.  No VBC emission is expected in \aliii\ and \feii.  These constraints help also to make the fits less ambiguous.
}

 {\bf The absence of a VBC makes the decomposition of  Pop. A spectra easier. As said earlier, the \feii\ intensity scale of the template (but \feii\ is in general weak) is anchored to the UV 191 intensity; similarly the \feiii\ intensity is set by a feature external to the 1900 \AA\ blend (2080 \AA). There is no evidence of BBC of \aliii\ and \siiii; the \aliii\ doublet profile is mostly unblended and defines the LIL-BC.   The additional complication here is the \feiii\ 1914 line whose intensity is affected by \lya\ pumping.  Since the shift and FWHM are assumed the same for all lines (and templates) in the 1900 blend, the only  free parameters in addition to shift and FWHM are the intensities of  6 components, including the two  from the templates, and \siii\ and \aliii\ that are not heavily blended.  The {\sc specfit} analysis is especially helpful to measure in a non-subjective way, taking all constraints into account, the  two parameters that are most affected by the blend: the intensity of \siiii\ and \ciii\ (we repeat that any \feiii\ $\lambda$1914 contribution in excess to the one of the adopted template is included in the estimated \ciii\ intensity). 
 }

{\bf In the case of Pop B, the presence of a VBC does not really complicate the fit as a matter of fact. The \ciii\ line undeniably shows a VBC protruding on the red side of the 1900 blend.  In any case, considering that we can expect the  VBC to be assimilable to a shifted Gaussian with FWHM $\sim 10000$\kms, the unblended part of the \ciii\ VBC provides a strong constraint.  The \siiii\ VBC  is certainly the most difficult feature to ascertain, as it buried under \siiii\ and \ciii\ emission. We rely on the {\sc specfit} results. that indicate negligible \siiii\ VBC. }

\subsection{Errors} 
\label{errors}
We identify five sources of error from the conditions for data reduction and methodological considerations described above:

\begin{enumerate}

\item A \& B atmospheric bands correction (already described in \S \ref{ab}). 

\item Line profile shape, Gaussian or Lorentzian (Pop. A or B).  The distinction between Pop. A and B is based on line width with the boundary at  FWHM H$\beta \approx$  \kms\ in low luminosity quasars and around $\approx$ 5000 \kms\  at higher luminosity such as the eight sources presented here.  Most of our quasars are unambiguously Pop. A or B because of line width and because Pop. B sources show an H$\beta$\ VBC and pop A sources a prominent \civ\ BBC. In these cases only one profile shape (Gaussian or Lorentzian) was fitted.  

\item Rest-frame determination using O{\sc i}\l1304 or \ciii.  
In some cases the redshift estimates derived from the two lines do not agree, most likely because of absorptions present in O{\sc i}\l1304.8 and because this is not a very intense line. The principal impact  of uncertainty in the rest frame placement is estimation of the peak wavelength of \civ. If the line peak  differs from \l1549\AA, the BC intensity is diminished and we infer a greater contribution from the blue (BBC) component. Similarly, for the blend 1900\AA, the rest  frame shift may increase or decrease our estimate for the strength of  \ciii\  with consequent decrease or increase of the \siiii\ contribution. This additional source of uncertainty affects  J00103-0037, J02287+0002 and J20497-0554.

\item \feii\  intensity (continuum placement). Broad \feii\  emission can produce a pseudo-continuum affecting our estimates of emission line intensities.  \siii\ is especially affected in our spectra because it is weak.  \aliii\ is similarly affected when it is weak. The effect is less noticeable for \civ\ since expected \feii\  emission  underlying the \civ\ line  is weak also for strong \feii\ emitters.

\item Broad absorption lines (BALs) in quasars principally affect the blue side of  \civ. We also find an absorption feature between FeII\l1787 and \siii\ (eg. Fig.  \ref{fig:fits_bal}). In  sources J01225+1339 and J02287+0002\ Êwe can only set upper limits for line intensities (since we fit unabsorbed components).

\end{enumerate}

There are other sources of error such as small BC shifts and FWHM variations.  {\bf We assume that the BC of  the \siiii, \aliii, \siii\ and \civ\ line has the same FWHM and wavelength shift, although we allow for variations in their relative flux strengths and adopted profile type (Lorentzian for Pop.\,A and Gaussian for Pop.\,B). For  \feiiiuv\ and \feii\ we slightly  relax this constraint.  The \feiiiuv\ and \feii\ emission is not very strong and the FWHM of individual features is poorly constrained by {\sc specfit}. In the case of \ciii\ we need to consider the possibility that the profile is narrower because there might be a contribution from different regions: indeed, the {\sc specfit} routine usually converges toward a narrower profile if the \ciii\ width is not constrained. The effect depends on the strength of the \feiii\ $\lambda$ 1914 feature. }

The goal of this study is to estimate diagnostic line ratios. A posteriori, we can say that the estimated line ratios are rather insensitive to the emission component profile shape: assuming a Gaussian or Lorentzian profile yields the same ratio for the strongest lines (i.e.,  \civ, \siiii, \aliii) upon which our analysis is based (with an uncertainty of $\sim$10\%). The same conclusion applies to the redshift uncertainty. Only in the case of  J02287+0002 the redshift difference produces a significant effect due to a $\Delta z \approx 0.16$. 	However, we adopt the fit based on O{\sc i}\l1304: the alternative fit produces line ratios that are of not obvious interpretation and at variance with respect to the other sources.  The most serious sources of error remain effects of A/B band overlap for \siii\ and the presence of a BAL for \civ. 

\section{Results of Line Component Analysis on Individual Objects}
\label{sec:component_analysis}

In Figures from \ref{fig:fitsA} to \ref{fig:fits_bal} we show our best fits for the VLT sample taking into account the considerations described in \S \ref{sec:data_analysis}. The fits of \siiv\ are shown in Fig. \ref{fig:fitss4} and the intensity values and equivalent widths are in Tables \ref{tab:fluxes} and \ref{tab:ew}. We present here a phenomenological description of the fits.  {\bf The line profiles, intensities and line ratios usually follow the trends of Pop. A or B sources, although in some  objects there are features that are ambiguous (for example the line shape). In these cases however, we have assigned Pop. A or B type on the basis of the FWHM (see Cols.  5 and 6 of Table \ref{tab:r_m}). }

\subsection{Pop. A Objects}

\begin{figure}
\epsscale{0.6}
\plotone{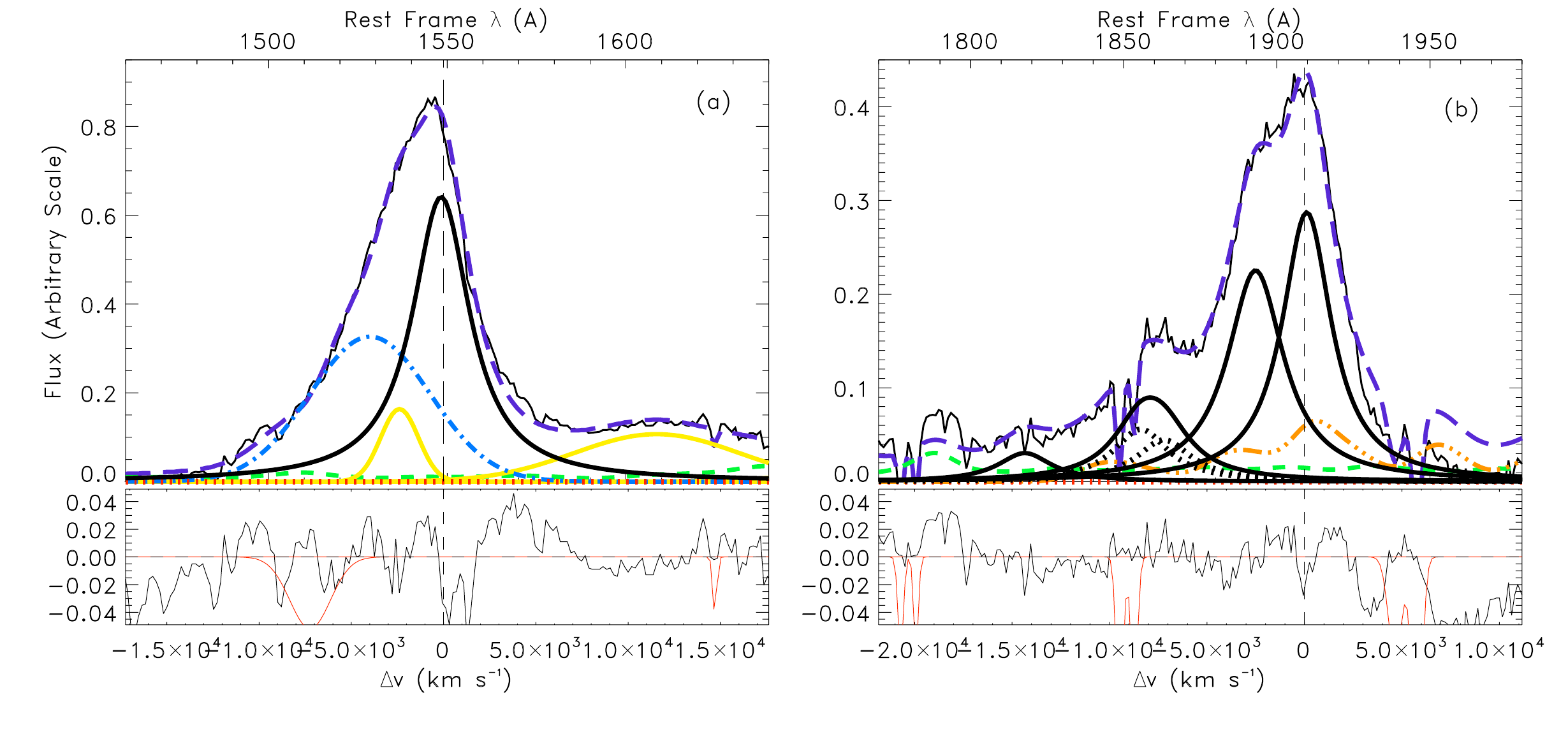}
\plotone{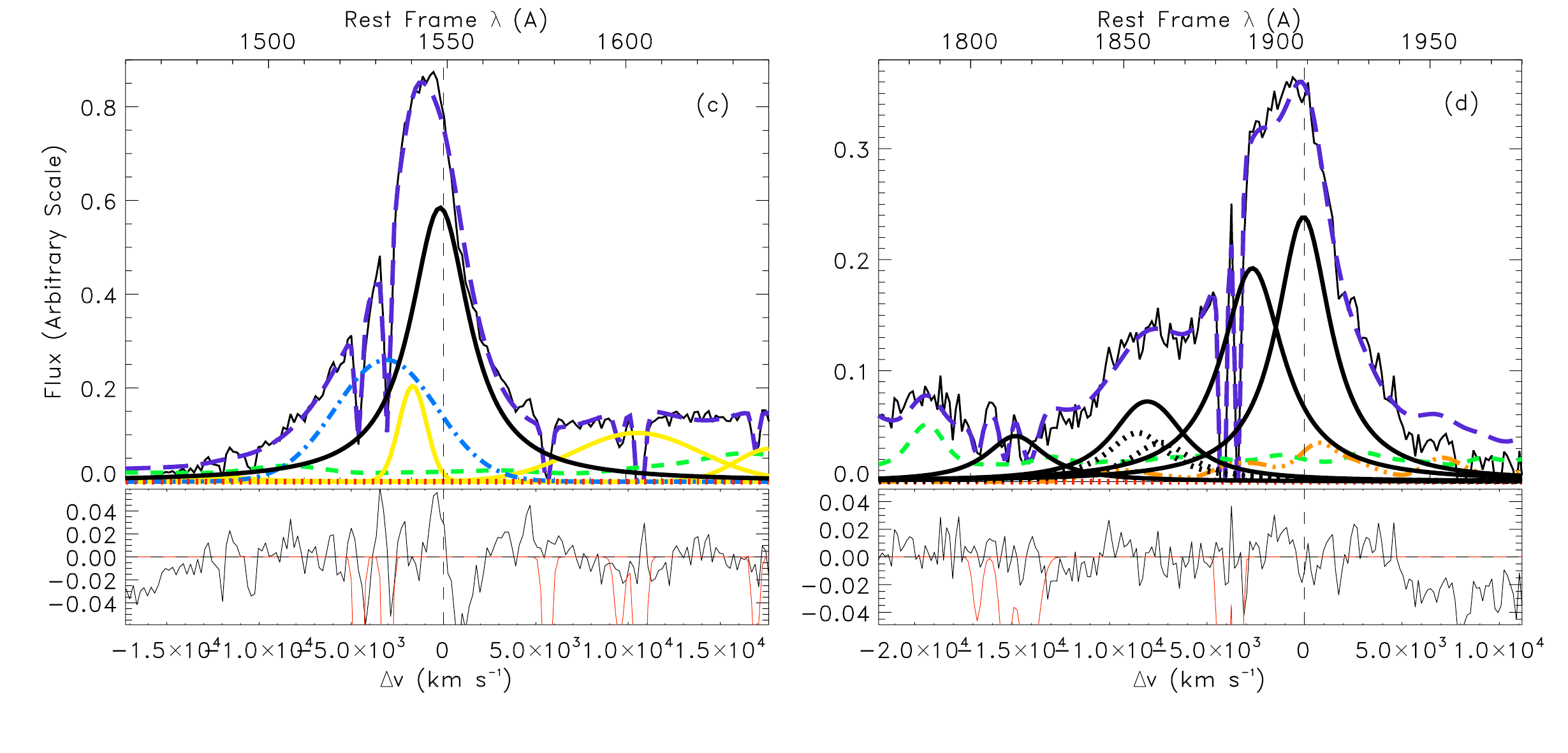}
\plotone{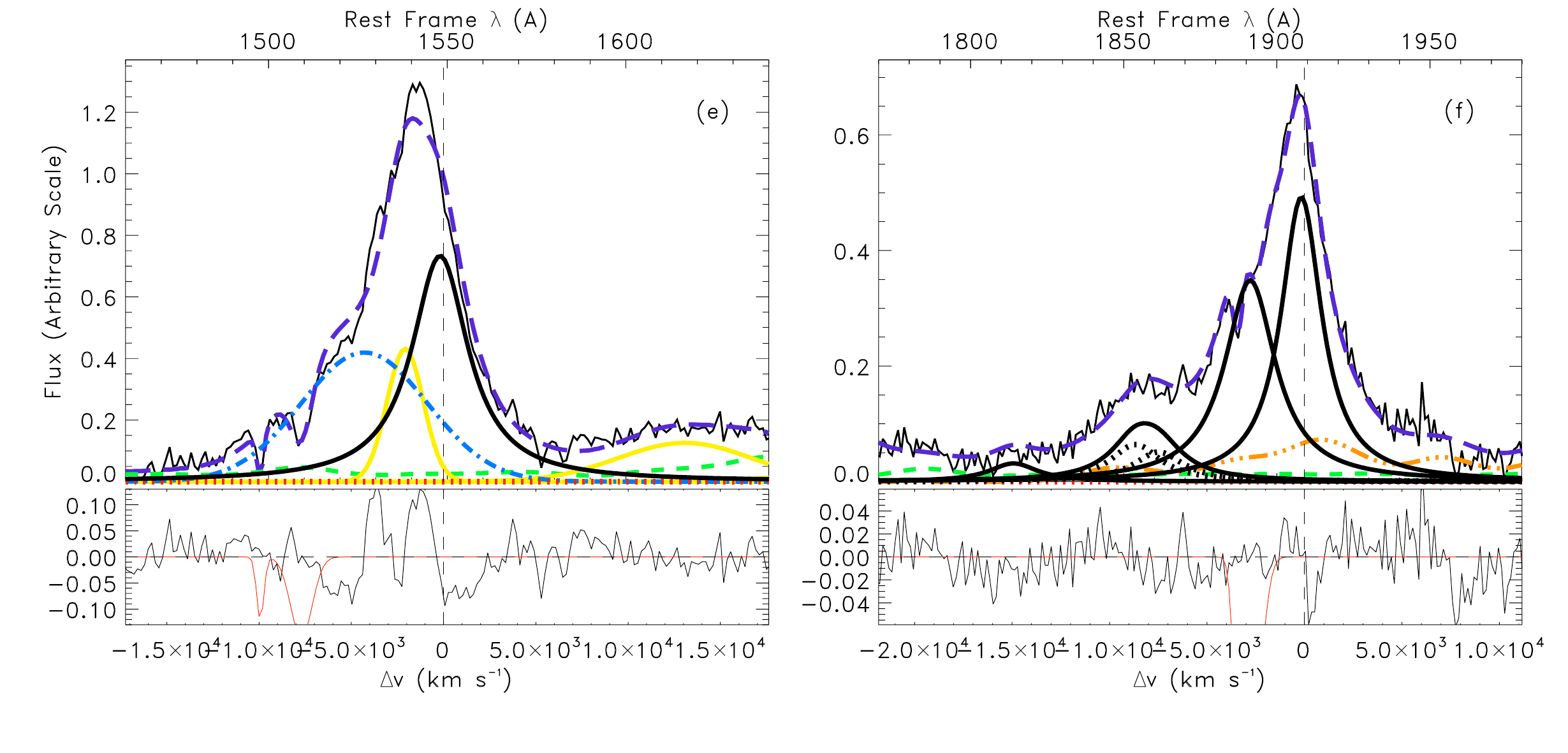}
\caption{Fits for Pop. A objects: J03036-0023 (a, b),  J20497-0554 (c, d), J23509-0052 (e, f). Upper panels show the fits and the lower pannels under the fits show the residuals and also the fitted absorptions lines. Upper abscissa is rest frame wavelength in \AA, lower abscissa is in velocity units, ordinate is specific flux in arbitrary units. Vertical dashed line is the restframe for \civ\ and \ciii. Long dashed line is the fit, solid dark lines are the broad components: \civ\ in left panels and \ciii, \siiii, \aliii, \siii\ in right panels. Dotted dark lines under \aliii\ show the doublet. Short dashed line is \feii. \feiii\ is shown in dash-triple-dot line in the right panels. Dash-dot line in the left panels is the blue-shifted component of \civ\ while dotted line is the very broad component, present also in \ciii\ for Pop. B objects. In the left panels we show with faint lines the contribution of {N\sc{iv}}$\lambda$1486, {Si\sc{ii}}$\lambda$1533 and {He\sc{ii}}$\lambda$1640 core and blue-shifted components. For colors see online figures. \label{fig:fitsA}}
\end{figure}

\begin{itemize}

\item  J03036-0023 -- We estimate for this source a FWHM(BC)$\sim$3700 km s$^{-1}$\ and we use a Lorentzian function to fit the broad lines.  The peak of \civ\ is blueshifted and requires a strong BBC (Fig. \ref{fig:fitsA}(a)). 
The bump on the red side of \civ\ can be accounted for by \heii\ BC and BBC. 
There is no evidence for a red shifted component in \ciii\ (Fig. \ref{fig:fitsA}(b)). \aliii\ is prominent. Unfortunately the blue wing of \siii\ and the red wing of {Fe\sc{ii}}\l1787 are affected by A band absorption. 

\item J20497-0554 --  This source shows FWHM(BC)$\sim$3800 \kms. 
As for J03036-0023, the \civ\ line can be accounted for by an unshifted BC (assumed Lorentzian) of a considerably contribution of a BBC (Fig. \ref{fig:fitsA}(c)). We see a prominent \aliii\ line and FeII\l1787 (Fig. \ref{fig:fitsA}(d)). \siiii\ is affected by several narrow absorption lines; however, it is obviously strong. The lack of a red wing on  \ciii\  suggests that no VBC is present.

\item  J23509-0052 -- This source has  FWHM(BC)$\sim$3600  \kms. \civ\ shows a slight blue asymmetry with a BBC (Fig. \ref{fig:fitsA}(e)) required to model it. The contribution of \feii\  is small and {Fe\sc{ii}}\l1787 is weak (Fig. \ref{fig:fitsA}(f)). \ciii\  is very strong.  \aliii\  is affected by A band absorption;   the profile  we fit is probably an upper limit. This object could well belong to spectral type A1 that includes Pop. A sources with the lowest \rfe ($\simlt 0.5$).
\end{itemize}

\subsection{Pop. B Objects}.

\begin{figure}
\epsscale{0.9}
\plotone{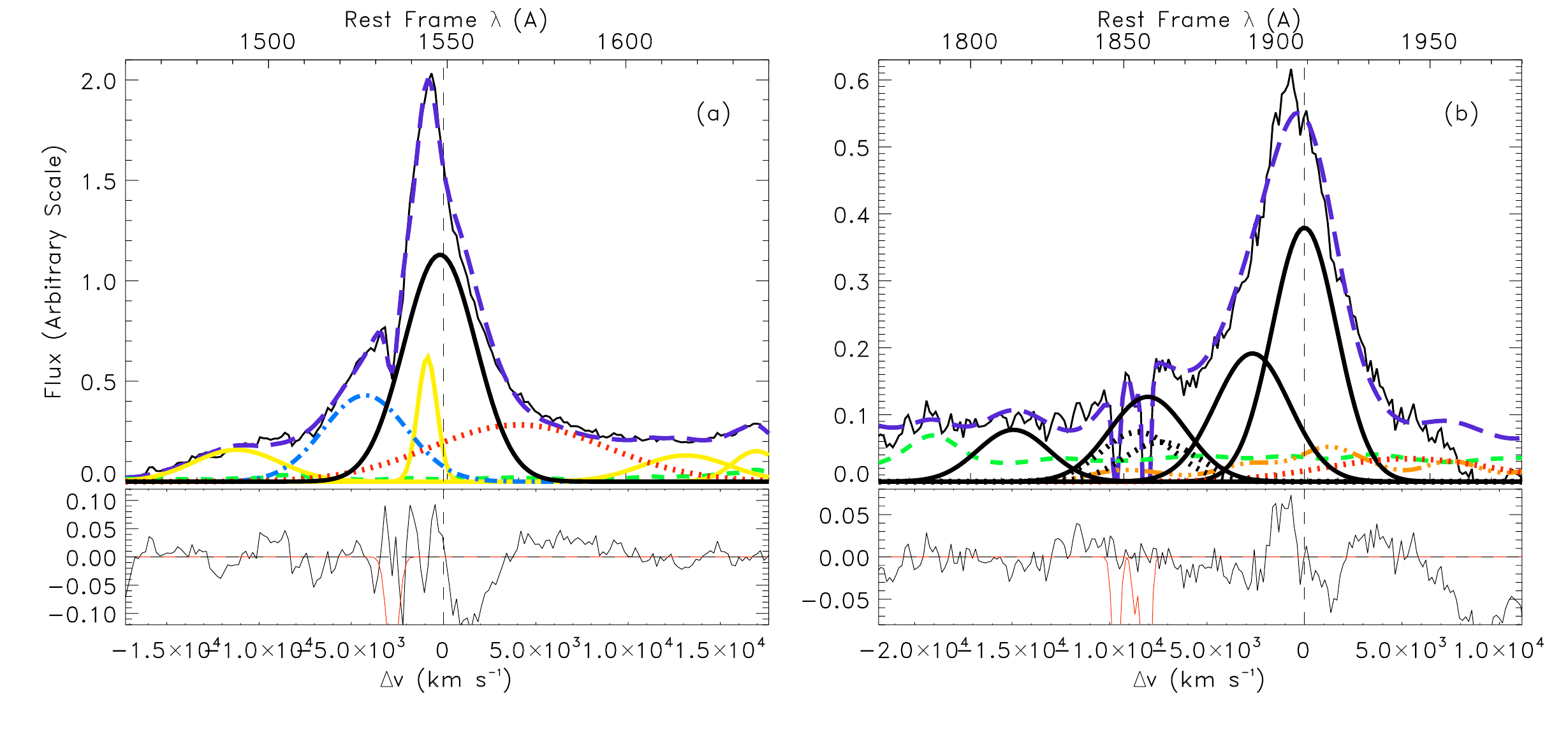}
\plotone{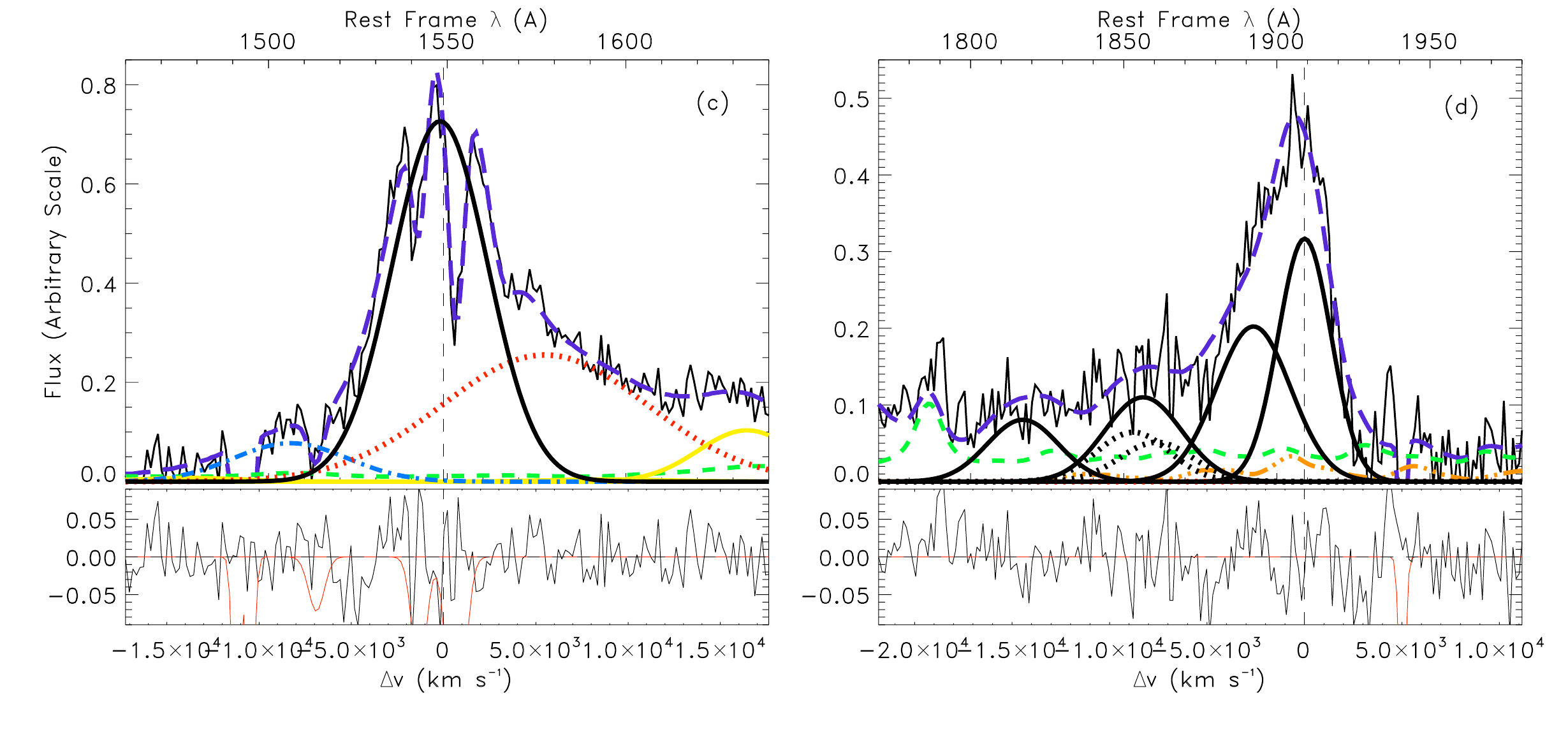}
\plotone{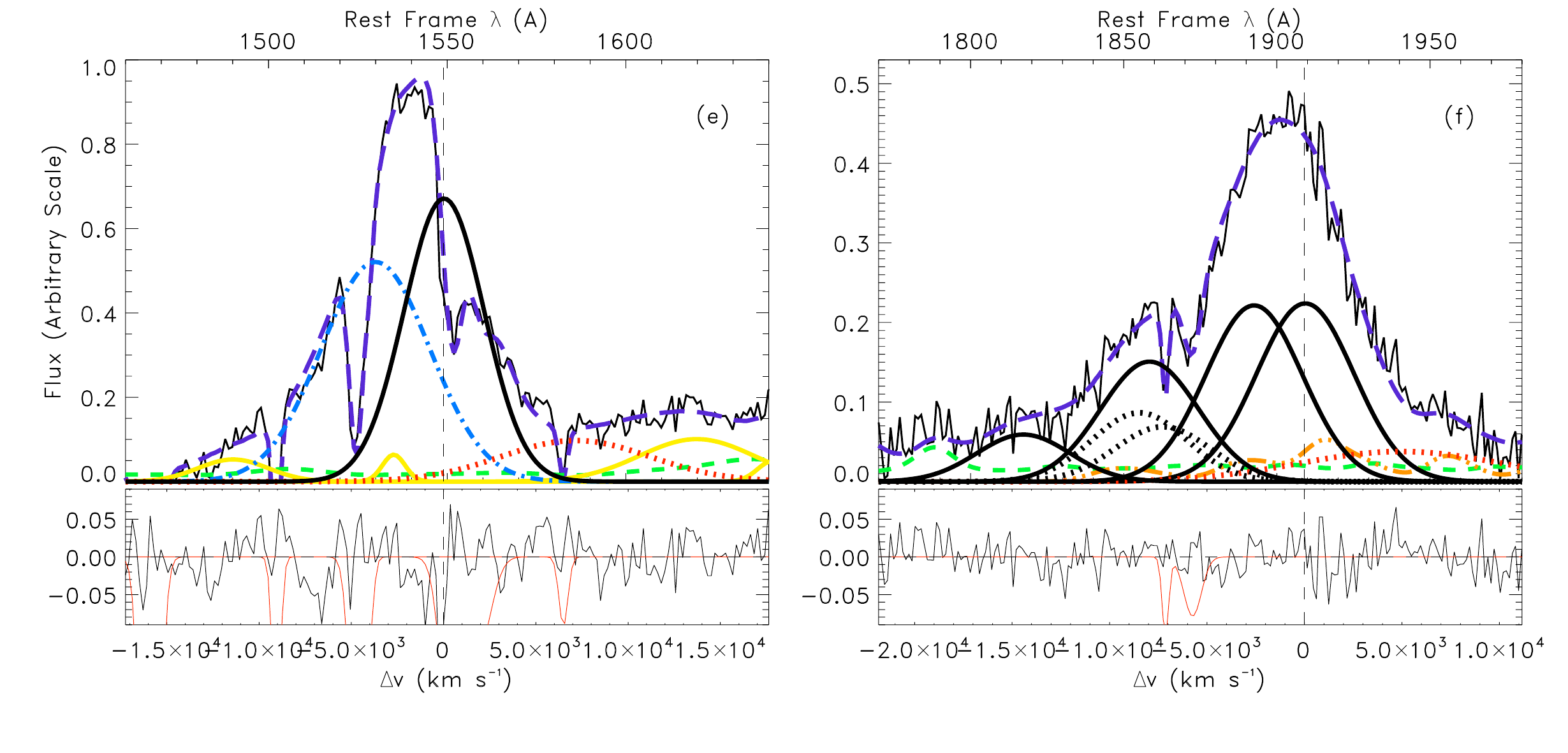}
\caption{Fits for Pop. B objects: up J00103-0037, middle J00521-1108, low J02390-0038. Units and meaning of symbols are the same of Fig. \ref{fig:fitsA}. \label{fig:fitsB}}
\end{figure}

\begin{itemize}

\item J00103-0037 --  This source has a FWHM(BC)$\sim$4500 kms$^{-1}$. The red side of \civ\ is blended with \heii.  Fitting a BC with no shift plus a BBC to \civ\ leaves a very large residual on the red side. A very broad redshifted component (VBC) is needed to model the spectrum (Fig.\ref{fig:fitsB} (a)).  The faint narrow line under \civ\ can be explained as the narrow component (NC) of \civ (see Sulentic et al. 2007). The presence of a similar NC in \ciii\ could possible explain the large residual seen $\sim$1900.  We specifically note the prominent  \ciii\  emission and weak (but detected) \aliii\ (Fig.  \ref{fig:fitsB}(b)). The \feii\ ``bump'' at 1787 \AA (UV 191) is appreciable. Fainter \feii\ emission is relatively unimportant because \feii\ creates a pseudo-continuum. \siii\ is compromised by A-band absorption. The blend at 1900\AA\ includes a  \ciii\  VBC and the fit indicates \civ/  \ciii\  (VBC) $\approx$ 7 which is reasonable.

\item J00521-1108 -- This source shows the noisiest spectrum in the sample.  We fit a FWHM(BC) $\sim$ 5300 \kms\ with \civ\ requiring a large VBC to account for the red wing. (Fig.  \ref{fig:fitsB}(c)). Absorption features seriously affect the \civ\ profile. The profile of \ciii\ is strongly asymmetric due to some sort of absorption on the red side. 
\aliii\ is weak consistent with pop. B (Fig. \ref{fig:fitsB}(d)). 

\item  J02390-0038 -- This objects has a somewhat atypical Pop. B spectrum due to a very strong BBC in \civ\ (Figure \ref{fig:fitsB}(e)). This source has a FWHM(BC)$\sim$ 5400 \kms. Consistent with pop. B we find  {Fe\sc{ii}}\l1787 to be weaker than \siii.  \ciii\  is flat topped and has a very similar intensity as \siiii. Both \civ\ and \ciii\ it show    red wings indicating VBC emission. The A band absorption lies between \siiii\ and \aliii\ (Fig. \ref{fig:fitsB}(f)) producing a possible overestimation of \aliii.  

\end{itemize}

\subsection{BAL QSOs}

\begin{figure}
\epsscale{0.9}
\plotone{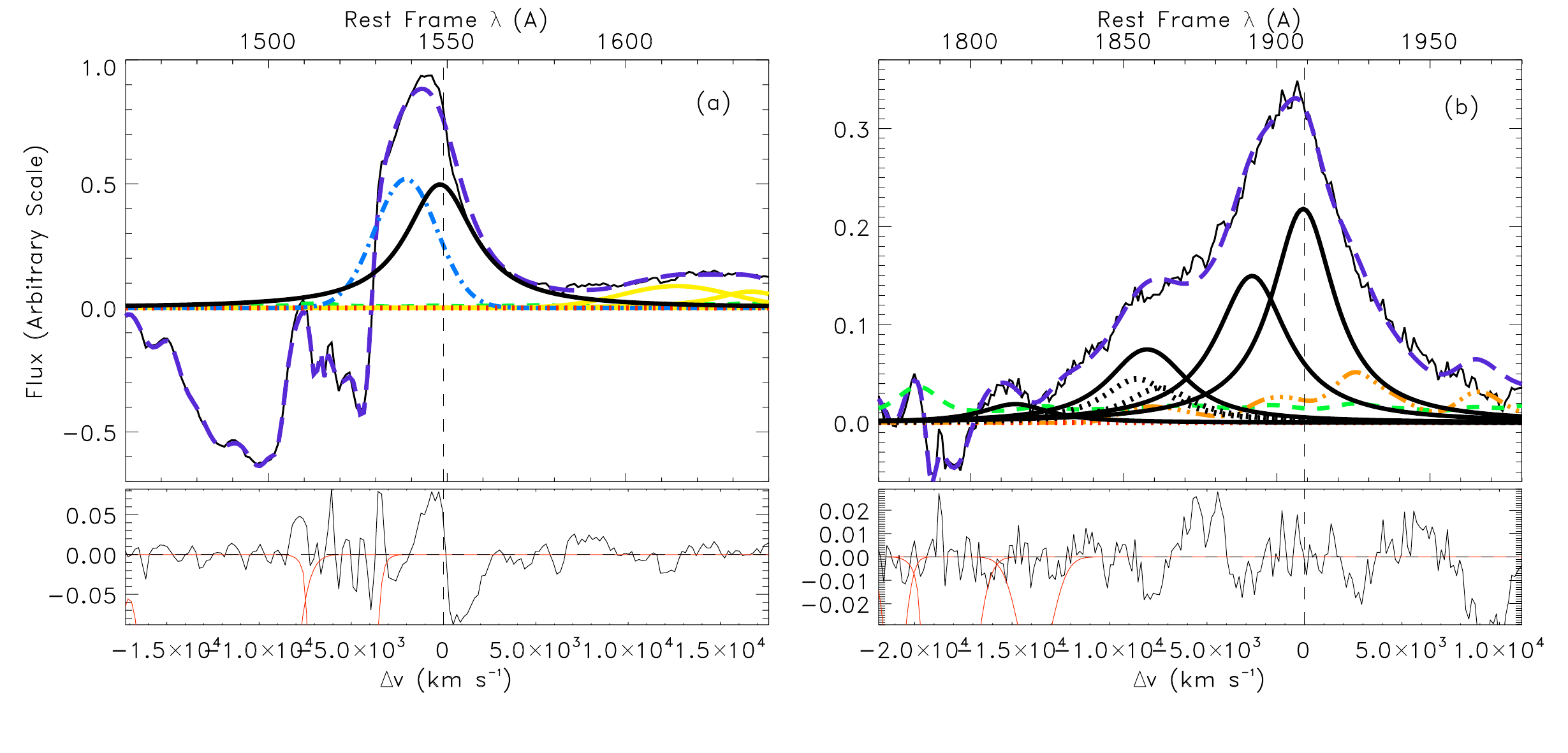}
\plotone{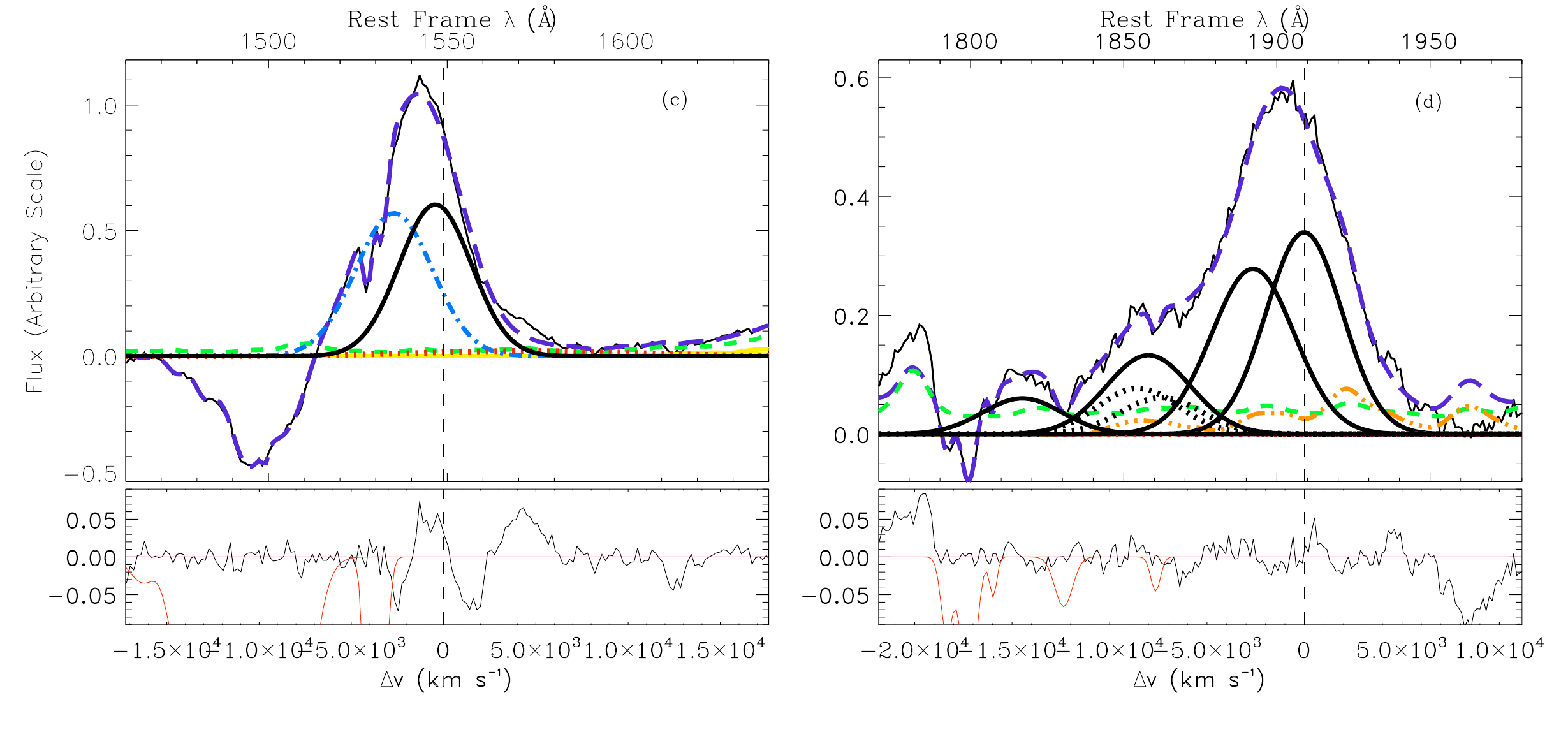}
\plotone{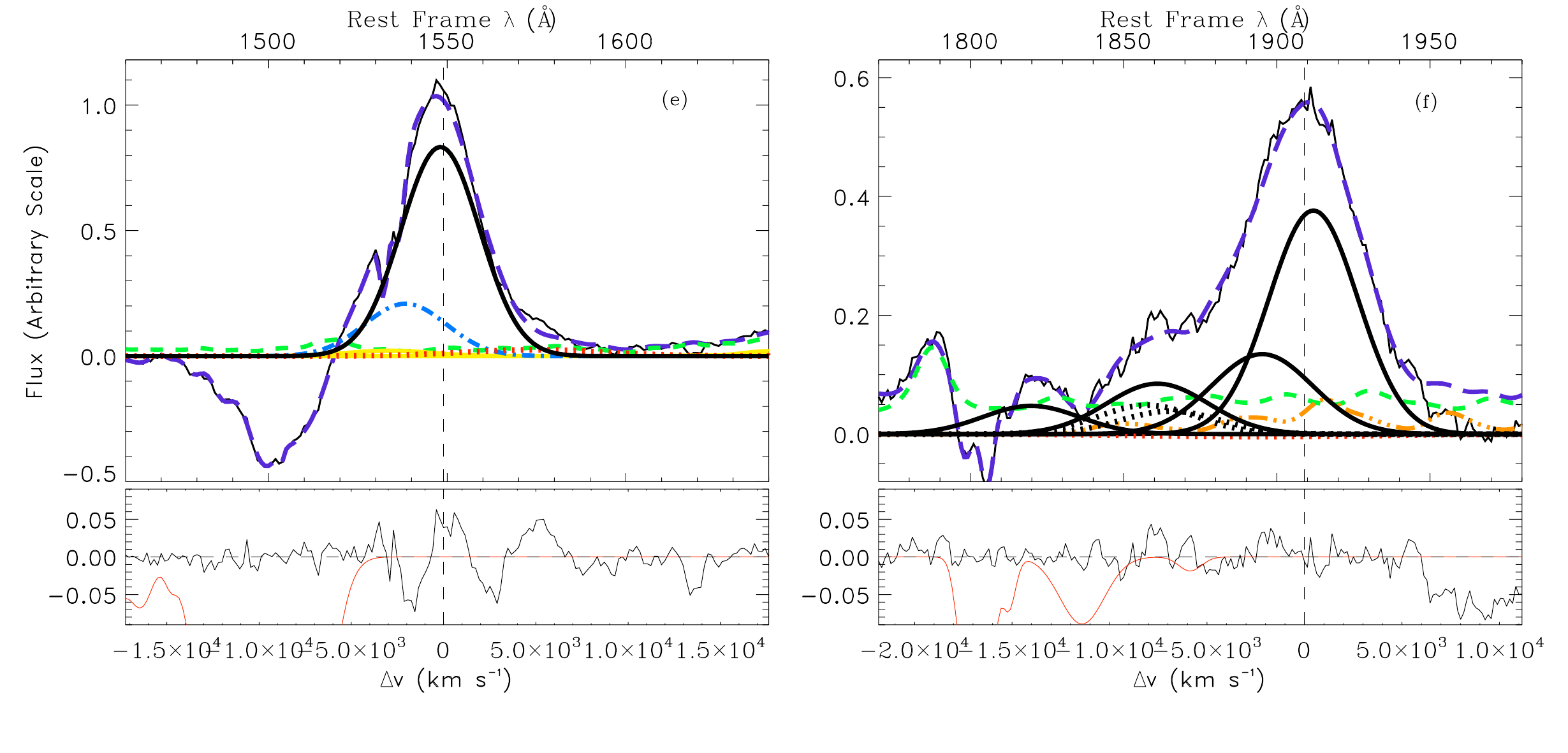}
\caption{Fits for BAL quasars: up J01225+1339, middle J02287+0002 using $z_{\mathrm O{\sc i}\lambda1304}$ rest frame, low J02287+0002 using $z_{\mathrm C{\sc iii}\lambda1909}$ rest frame. Note in J02287+0002 the line displacement  with the consequently line intensity changes, specially in \ciii, \siiii\ and \civ\ broad and blue-shifted components. Units and symbols are the same as in Fig. \ref{fig:fitsA}.   \label{fig:fits_bal}}
\end{figure}

\begin{figure}
\epsscale{0.9}
\plotone{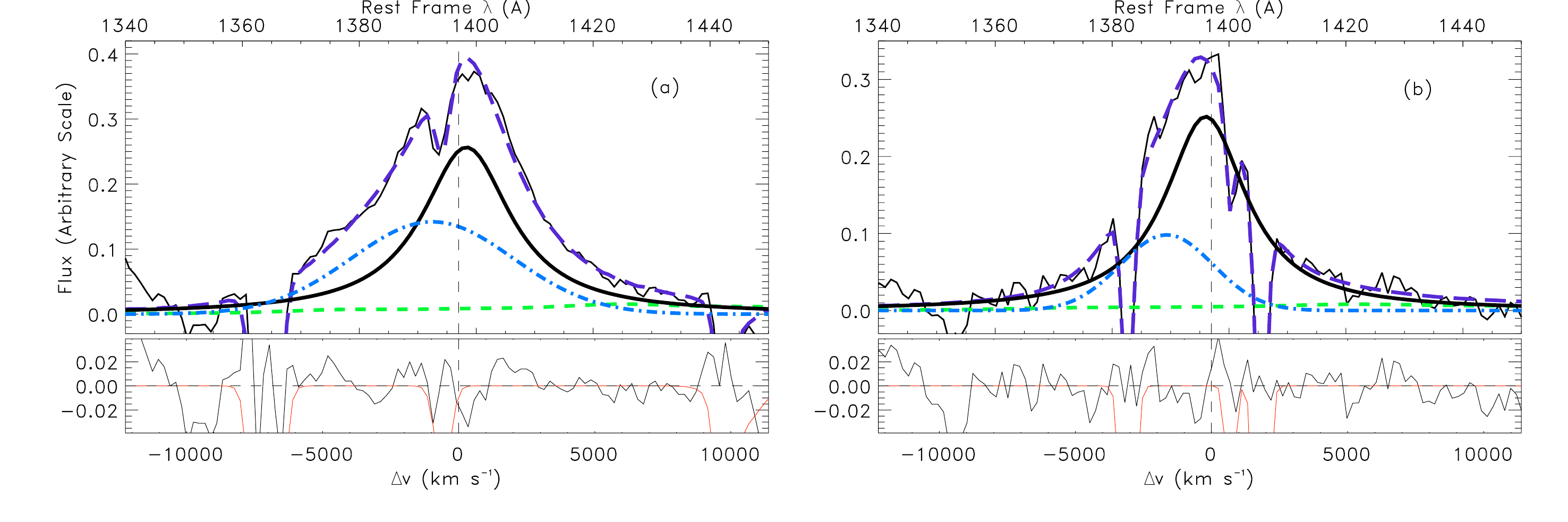}
\plotone{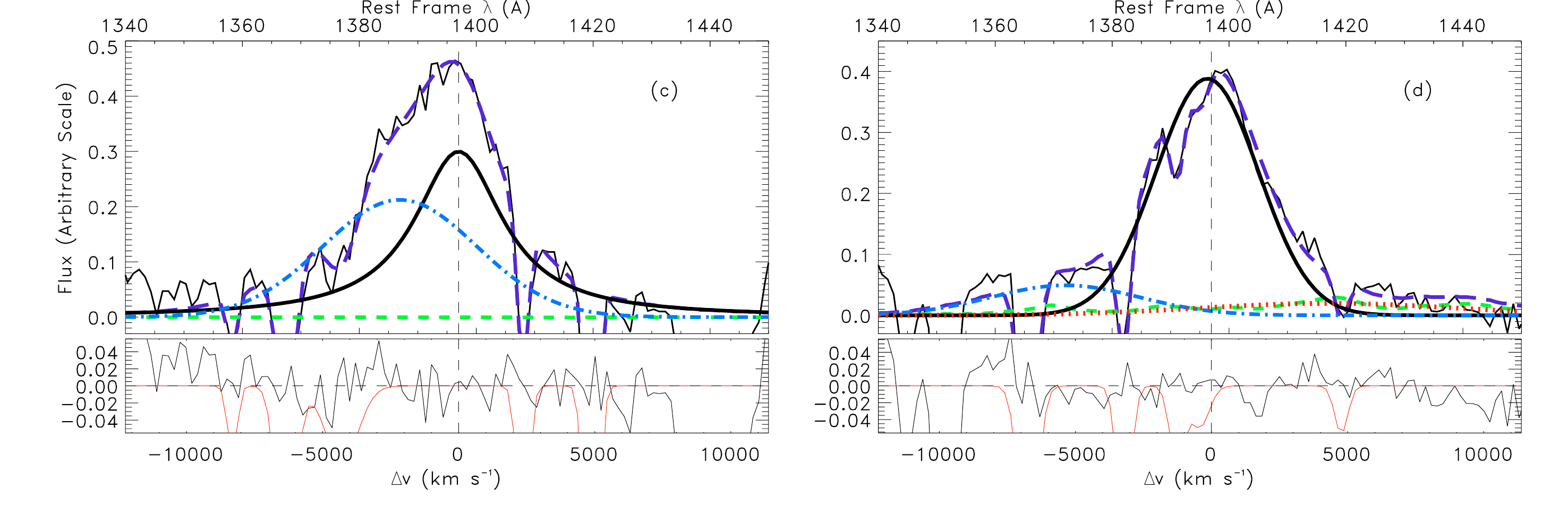}
\plotone{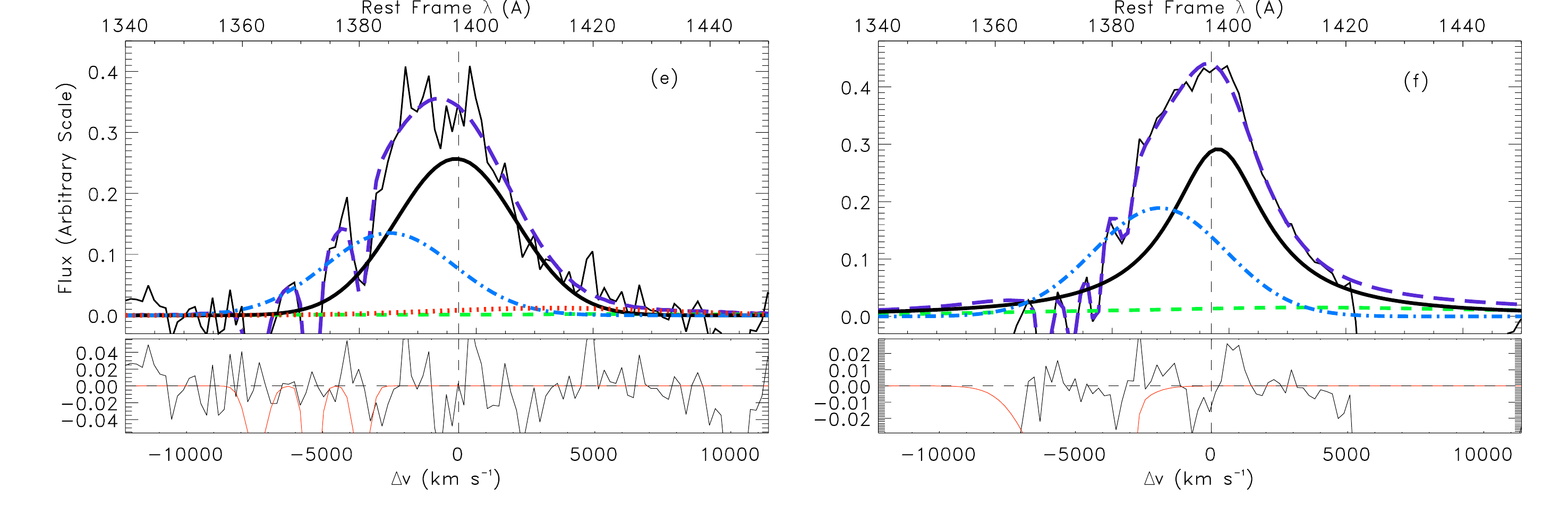}
\plotone{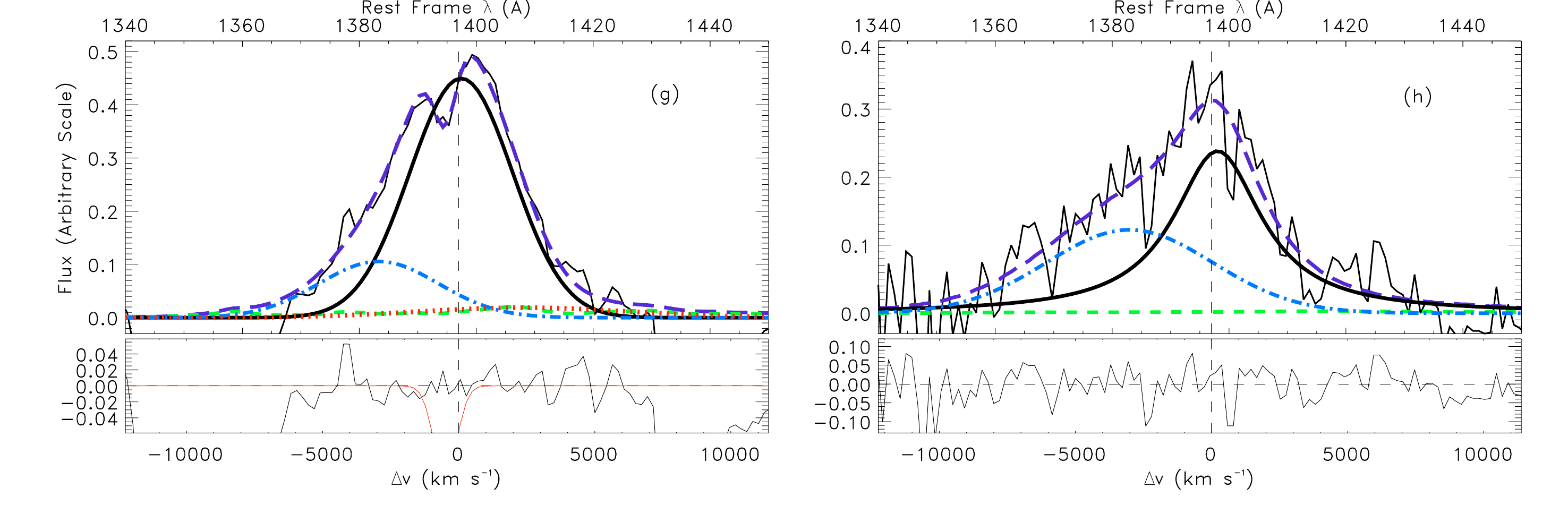}
\caption{Fits for \siiv: (a) J03036-0023, (b) J20497-0554, (c) 23509-0052, (d) J00103-0037, (e) J02390-0038, (f) J01225+1339, (g) J02287+0002 and (h) J12014+0116. We do not measure \siiv\ for J00521--1108 and 3C\,390.3 because they have low S/N. We asume that the profile of \siiv\ can be fitted with the same three possible components (core, blue-shifted and red-shifted) as in the case of \civ. Units and symbols are the same as in Fig. \ref{fig:fitsA}. \label{fig:fitss4}}
\end{figure}

\begin{itemize}

\item J01225+1339 --  \civ\ is highly affected by two broad absorption lines (Fig.  \ref{fig:fits_bal}(a)) with blueshifts of 5200 and 10800 \kms\ at peak absorption with equivalent widths/FWHM  -12\AA\ / 3900 \kms\ and -25\AA/5200 \kms, respectively.  The blueshift of the \civ\ peak  leads us to suspect a large blueshifted BBC emission component. The 1900 \AA\ blend shows absorptions coincident with  {Fe\sc{ii}}\l1787  and the blue side of \siii\ which is however unambiguously detected (Fig. \ref{fig:fits_bal}(b)).  \aliii\ is prominent which implies that this a Pop. A source.  The FWHM(BC) $\sim$ 4400 \kms\ is consistent with a high-luminosity Pop. A source.  {\bf It is also possible  that this BAL QSO is an outlier like Mark 231 at low-$z$\ (Sulentic et al. 2006), in other words an extreme Pop. A object. }
The  \ciii\  is well fitted with a Lorentzian profile. Broad A band atmospheric absorption lies over \siiii.

\item J02287+0002 -- This object has a very complex spectrum. On one side it has a FWHM(BC) $\approx$ 4700 \kms\ {\bf  which is typical of low-luminosity\ Pop. B and the lines profiles are better fitted with Gaussians. On the other side, however, it shows features that are typical of extreme Pop. A sources:  prominent {Fe\sc{ii}}\l1787, strong  \aliii, no  \ciii\  VBC  (Fig. \ref{fig:fits_bal}(d)).  Considering that the FWHM limit  between Pop. A and B is increasing with luminosity, the FWHM(BC) is within the limit of Pop. A. The \ciii\ line is not very flat topped but the similar intensities of \ciii\ and \siiii\ remind the case of  J02390-0038.  Also because it has a strong blue-shifted component in \civ\ atypical to Pop. B objects  (Fig. \ref{fig:fits_bal}(c)).}

The estimated rest frame of this quasar differs by $\sim$ 1400 \kms\ using  O{\sc i}\l1304 and \ciii.  This is the  largest discrepancy in our sample. In order to evaluate the effect of the $z$\ discrepancy we  performed two fits  using both rest frames. Figs. \ref{fig:fits_bal}(a) and (b) use the  \ciii\  restframe.  In  the 1900 \AA\ blend, we found a contribution of \siiii\ similar to \ciii.   If we use the O{\sc i}\l1304 inferred rest frame we show in Figs. \ref{fig:fits_bal}(e) and (f) that  \ciii\  becomes stronger with a resultant decrease of \siiii. A similar effect occurs for \civ\ broad and blue-shifted components.

The \civ\  BAL shows a blueshift of 9100 \kms at deepest absorption, a EW of --14 \AA\ and a FWHM of 4600 \kms.  

\end{itemize}

{\bf Summing up, we are able to assign an A/B identification to all sources in our sample. The two BAL QSOs appear as objects of extreme Pop. A. We remind that the identification of \ciii\ in the BAL QSOs and  in sources with strong \aliii\ is debatable (Hartig and Baldwin 1986): strong \feiii\ 1914 could take the place of most \ciii\ emission. }

\section{Estimation of Physical Conditions in the Emitting regions} 
\label{sec:physical_conditions}

\subsection {{\sc CLOUDY} Simulations}

We computed a multidimensional grid of {\sc CLOUDY} (Ferland et al. 1998) simulations, (see also Korista et al. 1997) to derive $U$\ and \nh\ from our spectral measurements. Simulations span the density range $7.00 \leq \log$ \nh$ \leq 14.00$, and $-4.50 \leq \log U \leq 00.00$, in intervals of  0.25. Each simulation was computed for a fixed ionization parameter and density assuming plane parallel geometry. The 2D grid of simulations was repeated twice assuming \nc$ = 10^{23}$\ and $10^{24}$ cm$^{-2}$. Several cases were computed also for \nc =  $10^{25} $cm$^{-2}$.  Metallicity was assumed to be either solar or five times solar. Two alternative input continua were used: 1) the standard AGN continuum of {\sc CLOUDY} which is equivalent to the continuum described by Mathews and Ferland (1987) and 2) the low-$z$\ quasar continuum of Laor et al. (1997a). Computed line ratios are almost identical for fixed ($U$, \nh). However  the ionizing luminosity differs by more than a factor of 2 for a fixed specific continuum luminosity. The contour plots showing the distributions of \ciii/\siiii, \aliii/\siiii, \siii/\siiii,  {\bf \siiv/\siiii}, \civ/\aliii and \civ/\siiii\ (Fig. \ref{fig:contours}) are generated from 29 $\times$ 19 = 551 simulations and assume a  standard set of simulations using a  Mathews and Ferland (1987) continuum, $N_c = 10^{23}$ cm$^{-2}$, and solar metallicity. The {\sc CLOUDY} 08.00 computations included a model of the Fe$^+$\ ion with 371 levels. The UV \feii\  template, described in \S \ref{sec:fe2_fe3} is based on a suitable {\sc CLOUDY} simulation. Even if a relationship is very likely between the dense low ionization gas producing our diagnostic lines and \feii\ emission (supported observationally) our diagnostics do not use any \feii\  computation. The weak  line {\sc Oi}\l 1304 is used only for rest frame estimation and not for diagnostic considerations.

Apart from the hypothesis of plane-parallel geometry, no inferences are made about the actual distribution, location, and kinematics of the  line emitting gas. The assumption of constant density is crude. If gas is distributed in clouds then magnetic confinement appears to be unnecessary to avoid cloud dispersion from pressure imbalance or cloud shear associated with a hot confining medium. Magnetic confinement could make density uniform within the cloud (Rees 1987;  Bottorff and Ferland 2000).

\subsection{Intermediate Ionization Lines in the Blend at \l 1900} 
\label{intermediate}

The ratio \ciii/\siiii\ is density dependent because the transition probabilities of the two lines are so different: 114 s$^{-1}$ vs 12600 s$^{-1}$ (see Table \ref{tab:lines}). The forbidden lines at 1883\AA\, and 1907\AA\ have such low transition probabilities that they are collisionally quenched at much lower density and will not be considered. Line ratios like \ciii/\siiii\ are useful diagnostics in a rather narrow range of density which depends on the transition probabilities. Above the critical density, emission lines originating from forbidden or semi-forbidden transitions become collisionally quenched, and hence weaker than lines for which collisional effects are still negligible. \ciii\ is clearly unsuitable as a diagnostic for \ne $ \gg 10^{11}$ \cm3, as the \ciii/\siiii\ $\rightarrow$ 0.  Feldman et al. (1992) gives critical density of \siiii\ $n_e \sim 2 \cdot 10^{11} $ \cm3. \aliii\ is a permitted transition with large transition probability ($A \sim 5 \cdot 10^8$ s$^{-1}$) and has very-high critical density (i.e., its equivalent width goes to zero toward thermodynamic equilibrium, which occurs at very high density, when all line emission is collisionally quenched). Our 2D array of {\sc CLOUDY} simulations shows that the ratio \aliii/\siiii\ is well suited to sample the density range $10^{10} - 10^{12.5} $ \cm3.  Within this range the \siiii\ intensity decreases smoothly by a factor 10;  above the upper limit in density, the predicted  intensity of  \siiii\ decreases.  This corresponds to the densest, low ionization emitting regions likely associated to the production of \feii.  

Intermediate-ionization lines such as \ciii, \aliii, \siiii\ and \siiv avoid the issue of collisional ionization ({\bf invoked  for the production of low-ionization species by S. Collin and collaborators, as mentioned in the introduction}). 

The widely separated doublet \aliii\ is expected to be produced entirely in the fully ionized zone (Laor et al. 1997b). {\sc CLOUDY} simulations confirm this suggestion. Figure \ref{fig:ionic_emissivity} (top panel) shows the ionic fraction as a function of the geometrical depth in a cloud (or slab) of fixed column density (\nc $= 10^{23}$ cm$^{-2}$) and density (\nh $= 10^{12.5}$\cmq). As it can be seen, Al$^{++}$, Fe$^{++}$, and Si$^{++}$\ share a region of dominance deep in the cloud, close to the end of the Str\"omgren sphere. Beyond, in the partially-ionized zone (PIZ) there is, as by definition, a significant fraction of ionized hydrogen. The dominant ionic stages of Si and Fe become Fe$^+$ and Si$^+$. It is very appropriate to consider the \aliii/\siiii\ ratio since the two lines are apparently emitted in the very same zones within the gas slab. Available reverberation mapping results may support this interpretation but are rather difficult to extrapolate since they are limited to  an handful of low luminosity objects.  {\bf The main finding is that  \ciii\  responds to continuum changes  on timescales much longer than \civ\ and other HILs. This results comes from the analysis of total \ciii\  + \siiii\ in NGC 3783 (Onken \& Peterson 2002),  from the \ciii\ of NGC 4151 (Metzroth et al. 2006) and of NGC 5548. It is intriguing that  the \ciii\ cross-correlation delay in  NGC 5548 (by far the best studied object) is even larger than that of \hb\ (32 ld vs. 20 ld; Peterson et al. 2002; Clavel et al. 1991). For fixed density, lines of higher  ionization form at higher photon flux.  The C$^+$,   Al$^+$,  Si$^+$\  ionization potential are 24, 18, and 16 eV respectively. These  comparable ionization potentials  are well below the one of HILs, $X^{i -1} \ga$  50 eV. However, the much lower \ciii\  critical density  implies that the \ciii\ line should be  formed farther out than \siiii\ and \aliii\   if all these lines are produced  under similar ionization conditions. }

\subsection{The Si{\sc ii} Contribution to the UV Spectrum of Quasars } 

AGN have a rather rich singly-ionized silicon spectrum in the range 1000-2000 \AA, due to several resonant transitions from the ground term $3s^2  3p^1  {}^2P^0$ to terms associated to the  $3s^1 3p^2$ (\l1814 and \l1304), $3s^2 4s$ (\l1531), $3s^2 3d$ (\l1263) electronic configurations. A feature on the red side of Ly$\alpha$, at 1263 \AA\ is detected in several Pop. A sources like I Zw 1 (Marziani et al. 2010).

The \l1309 feature is partially blended with the O{\sc i}  triplet associated with a Bowen fluorescence mechanism from HI Ly$\beta$. The \l1309 line is well resolved in sources like I Zw 1 with the extension of the broad \l1309 feature suggesting significant Si{\sc ii}  emission around \l1309. The \l1531 feature is blended with  \civ. High S/N spectra of low redshift sources from HST (Laor et al. 1994) as well as for quasars at $z \approx$ 4 (Constantin et al. 2002) indicate that the feature is very weak in most sources.

The Si{\sc ii}  feature at \l1814 is detected in at least four of the five quasars studied by Laor et al. (1994). In the 6 objects used for the E1 sequence of Marziani et al. (2010), it is detected without reasonable doubt in I Zw 1 only. However this has to do more with the poor S/N of the 1900 \AA\ region than of anything else; good S/N HST spectra show an unambiguous detection in Akn 120 and Mark 509, for example. Early detection of the UV Si{\sc ii}  lines from IUE observations suggested collisional excitation and no relevant fluorescence effects (as observed in type Ia supernov\ae; Dumont and Mathez 1981). Fluorescence effects and recombination are revealed by optical lines which are emitted through several cascade routes to the ground state. However, a median spectrum of A2 quasars with S/N $\approx$ 200  (Zamfir et al. 2010) barely detects any expected emission feature  in the range 4000-6500 \AA. The prominence of the \l1814 feature may then be associated to the relatively low temperature believed to exist in the innermost part of the line emitting ``cloud'', $T_e \sim 5000 - 8000^\circ $ K. We may consider the equivalent width ($W$) ratio between the doublet lines at 1264 and at 1814 \AA\ both due to a $^{2}D_\frac{5}{2} \rightarrow ^2P_\frac{3}{2}$\ transition; in the simplest case we have:  ${W_{1263}}/{W_{1814}} \approx  \left( {N_\mathrm{u,1264}}/{N_\mathrm{u, 1814}} \right)\left({f_{1264}}/{f_{1816}} \right)\left({\lambda_{1264}}/{\lambda_{1816}}\right)^2
\approx 0.485 \frac {f_{1264}}{f_{1816}} e^{-\frac{\Delta E}{k T}} \approx 0.27$ if $T = 5000^\circ$,   where  $f$\ is the oscillator strength, $N_\mathrm{u}$\ are the  density of the upper electronic levels giving rise to the two lines, and  the energy difference between the transitions is  $\Delta E \approx$ 3 eV.

\subsection{Use of \siiv, \civ\ and \siii} 
 
The same caveats mentioned for the ratio \ciii/\siiii\ apply to the \aliii/\siiii\ ratio as well. The ground term $^1S_0$ has energy 16 eV and 24 eV for Si$^{+2}$ and C$^{+2}$ respectively. A dependence on the ionization parameter is expected, as already mentioned. However, given the similarity in the ionization structure of the photo-ionized slab, it is after all not surprising that the ratio \aliii/\siiii\ is almost insensitive to the ionization parameter over a wide range of density. We remind that the detection of strong \aliii\ {\em alone} already suggests that we are considering very high density emitting gas even if metallicity is super-solar. Simulations indicate that \aliii\ should increase smoothly with density and be weakest in canonical BLR if the density is \nh $ \sim 10^9$ \cm3\ (cf. Korista et al. 1997). 

The ratio \aliii/\siiii\ is therefore diagnosing high density gas, while the \ciii/\siiii\ ratio covers the domain of  \nh $ \sim 10^{10}$\cm3.   The ratio \aliii/\siiii\ alone is, generally speaking, insufficient to determine \nh.  A second diagnostic ratio is needed to constrain $U$\ and to unambiguously derive \nh. We consider   \siii/\siiii,  \civ/\siiii, \siiv/\siiii\  as three diagnostic ratios suitable for constraining $U$. 

The \siii\ doublet is conveniently placed, although weak or undetectable in most sources. The \siii/\siiii\ ratio is anti-correlated with $U$\ in a regular form, as our {\sc cloudy } simulations show, especially for $\log U \la -2$, and for density not much above the critical density of \siiii. The behavior of the  ratio \siii/\siiii\  resembles the distribution of \ciii/\civ\ in the plane (\nh, $U$) which shows a very regular dependence and a smooth anti-correlation with $U$\ until the collisional quenching of  \ciii\  sets on. However, the IP of Si$^0$ is just 8 eV. {\bf The majority of the \l1814 doublet is emitted in the partially ionized zone (PIZ) near the hydrogen ionization front.} Fig \ref{fig:ionic_emissivity} (bottom panel) shows the behavior of the volume emissivity $\epsilon$\  times the geometrical depth within an emitting  slab of gas as a function of the depth itself. Since the actual line emission is proportional to $\epsilon \cdot h$\ reported in Fig. \ref{fig:ionic_emissivity}, the emission of the \aliii, and \siiii\ lines is negligible in the PIZ, but the one of \hb\ certainly is not, and may be well dominating if \nc$  \gg 10^{23}$cm$^{-2}$. Significant emission in the PIZ is expected also for the \siii\ doublet. However, for this latter line, total emission should not depend strongly on column density, since the emissivity becomes very low at  \nc $  \ga 10^{23}$cm$^{-2}$.  
If higher \nc\ are considered (up to 10$^{25}$cm$^{-2}$), ratios including \siiii, \aliii, and \siii\ show a very weak dependence on column density, with changes a few percent at worst.

We attempted  to isolate a \civ\ and a \siiv\ component that corresponds to the \aliii\ and \siiii\ lines. This can be a small part of the total \civ\ and of \siiv\ emission, but there is no point to consider the whole \civ\ emission when \civ\   shows a large blueshift and is much broader than \hb\ (Fig. 5 of Sulentic et al. 2000) and \siiii\ and \aliii\ (Fig. 2 of Marziani et al. 2010).  Taking into account various sources of ambiguity (mainly uncertainty in the quasar rest frame, S/N, blending with HeII\l1640), the  \civ\ BC component {\bf we measure with our fits} is constrained within $\pm$ 50\% at worst.  Thus we find significant \civ\ emission from the low-ionization gas, {and the basic assumption is that its \civ\ profile is  the same as the \aliii, \siiii\ lines}.  If one considers the emissivity behavior, \civ\ and \siiv\ are  obviously favored within the fully ionized zone. 

{\bf The ratio of  \civ\  and \siiv\  over \ciii\  or over \siiii\ increases with ionization parameter in a way that is roughly independent on density until the collisional quenching of the semi-forbidden lines sets on (Fig. \ref{fig:contours}).  The previous considerations are helpful to understand why the \civ\ and \siiv\ ratios provide clear diagnostics of the ionization parameter and  seem to be in (at least qualitative) agreement with ratios employing  lower ionization lines  like  \siii/\siiii\ that is also mainly sensitive to $U$.  The striking fact that the \civ\ and \siiv\  emission confirms  low-ionization supports the hypothesis that the four lines are emitted by the same region.  The ratios involving Si only have the obvious advantage that the the determination of the physical parameters is not dependent on metallicity; the drawback of \siiv\ is that  this line is weaker and blended with \oiv. However, as already pointed out, the \siiv\ BC intensity should be slightly affected by \oiv. }

\section{Results on two Extreme, Elucidating Cases}
\label{sec:extreme}

We analyze two  extreme objects -- one extreme Pop. A and one extreme Pop. B --  with the same methodology we used previously. The aim of this analysis is to help the interpretation of the line components measured on the spectra of the 8 $z\approx$3 quasars. 

\subsection{SDSS Weak  \ciii\  source: SDSS J120144.36+011611.6}
\label{sec:sdss}

\begin{figure}
\epsscale{1.1}
\plotone{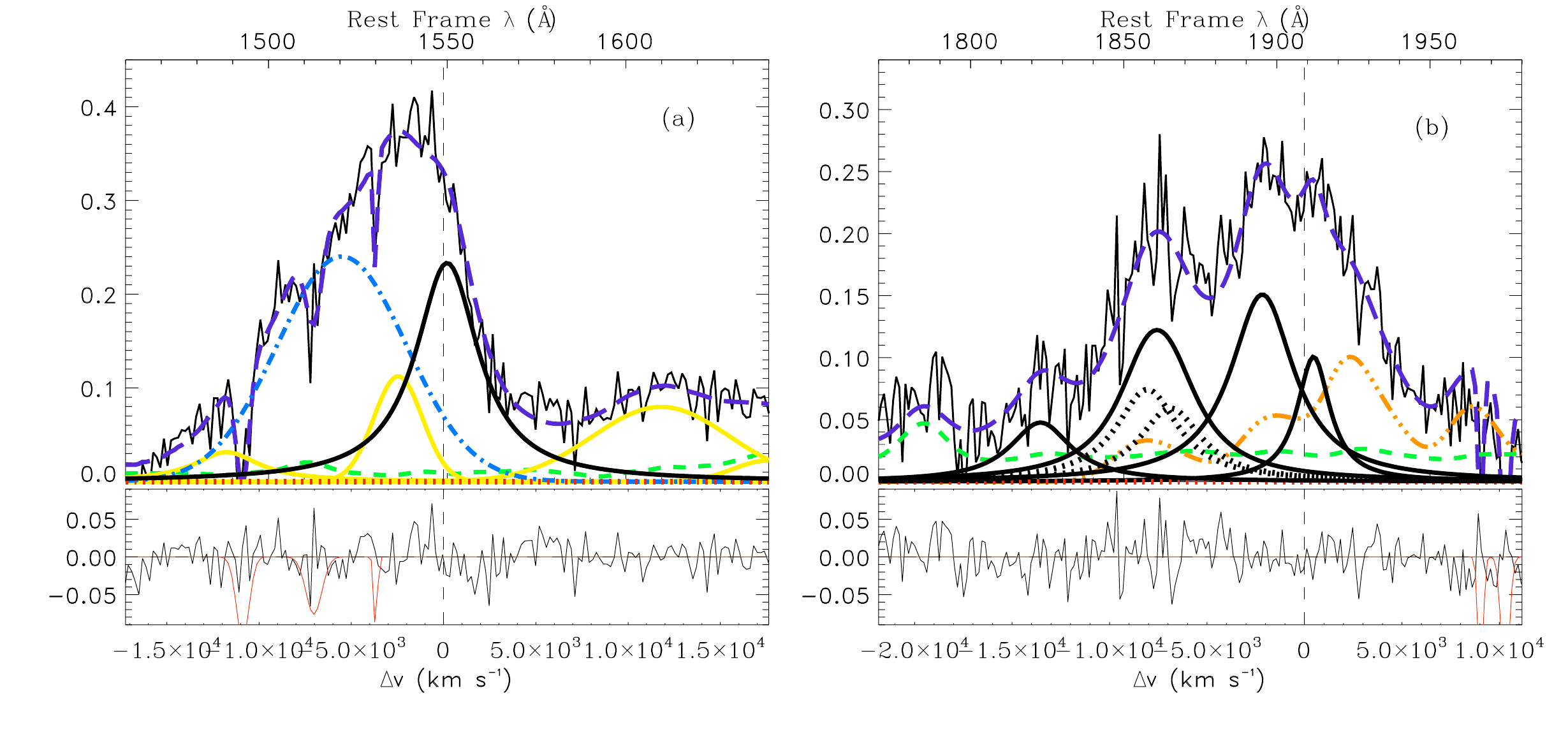}
\plotone{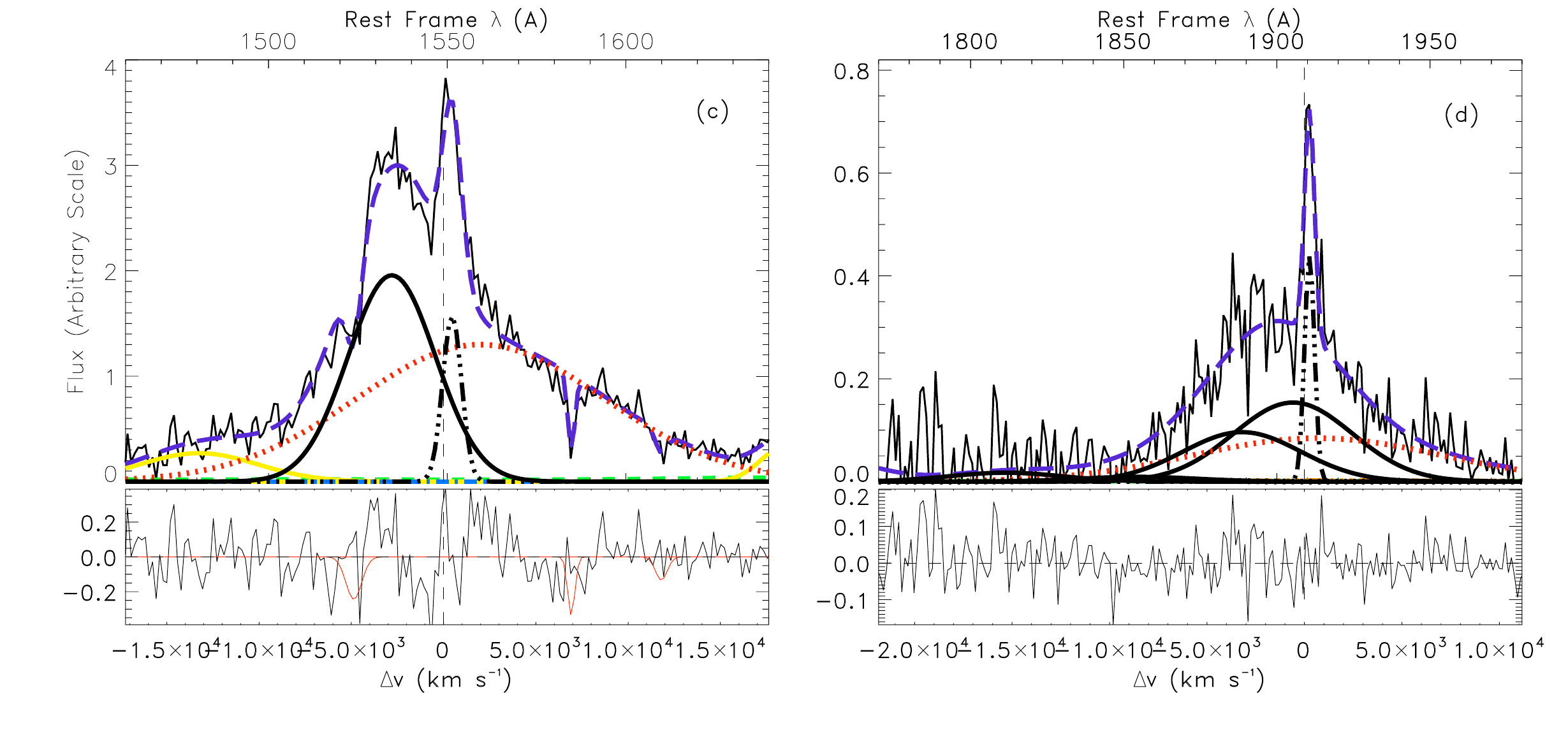}
\caption{Fits for the two extreme objects: up J120144.36+011611.6, low 3C390.3. Units and meaning of symbols are the same of Fig. \ref{fig:fitsA}. For 3C390.3 it was needed to fit a narrow unshifted component that we show in dash-dot-dot line. Note for 3C390.3 \siii\ is almost absent. \label{fig:extreme}}
\end{figure}

\begin{figure}
\epsscale{1.1}

\plotone{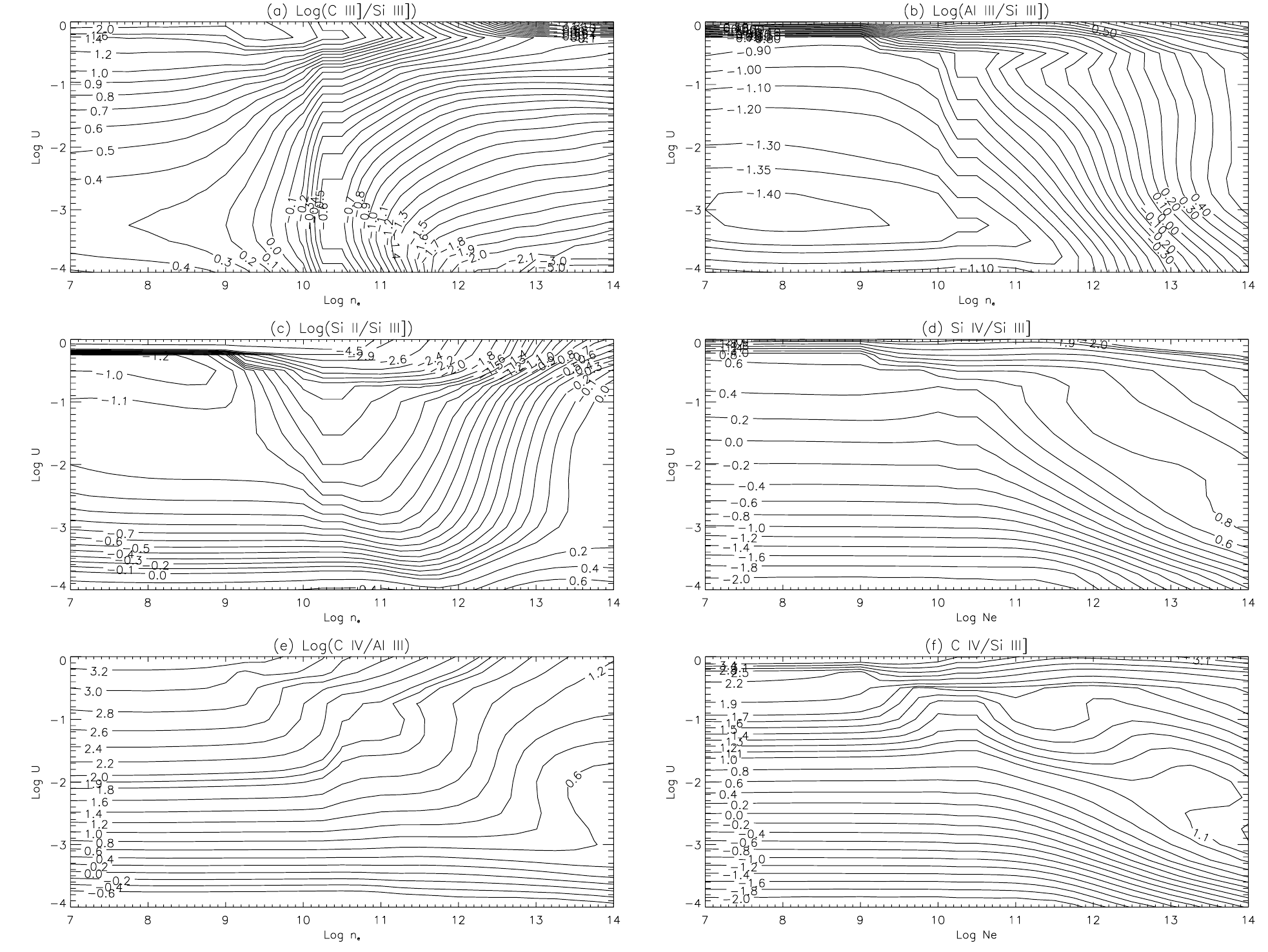}
\caption{Isocontours for the ratios (a)  $\log$(\ciii/\siiii), (b) $\log$(\aliii/\siiii), (c) $\log$(\siii/\siiii), (d) $\log$(\siiv/\siiii), (e) $\log$(\civ/\aliii), and (f) $\log$(\civ/\siiii)  derived from CLOUDY simulations. Abscissa is electron density in cm$^{-3}$, ordinate is the ionization parameter, both in logarithm scale.\label{fig:contours}}
\end{figure}

\begin{figure}
\epsscale{0.65}
\plotone{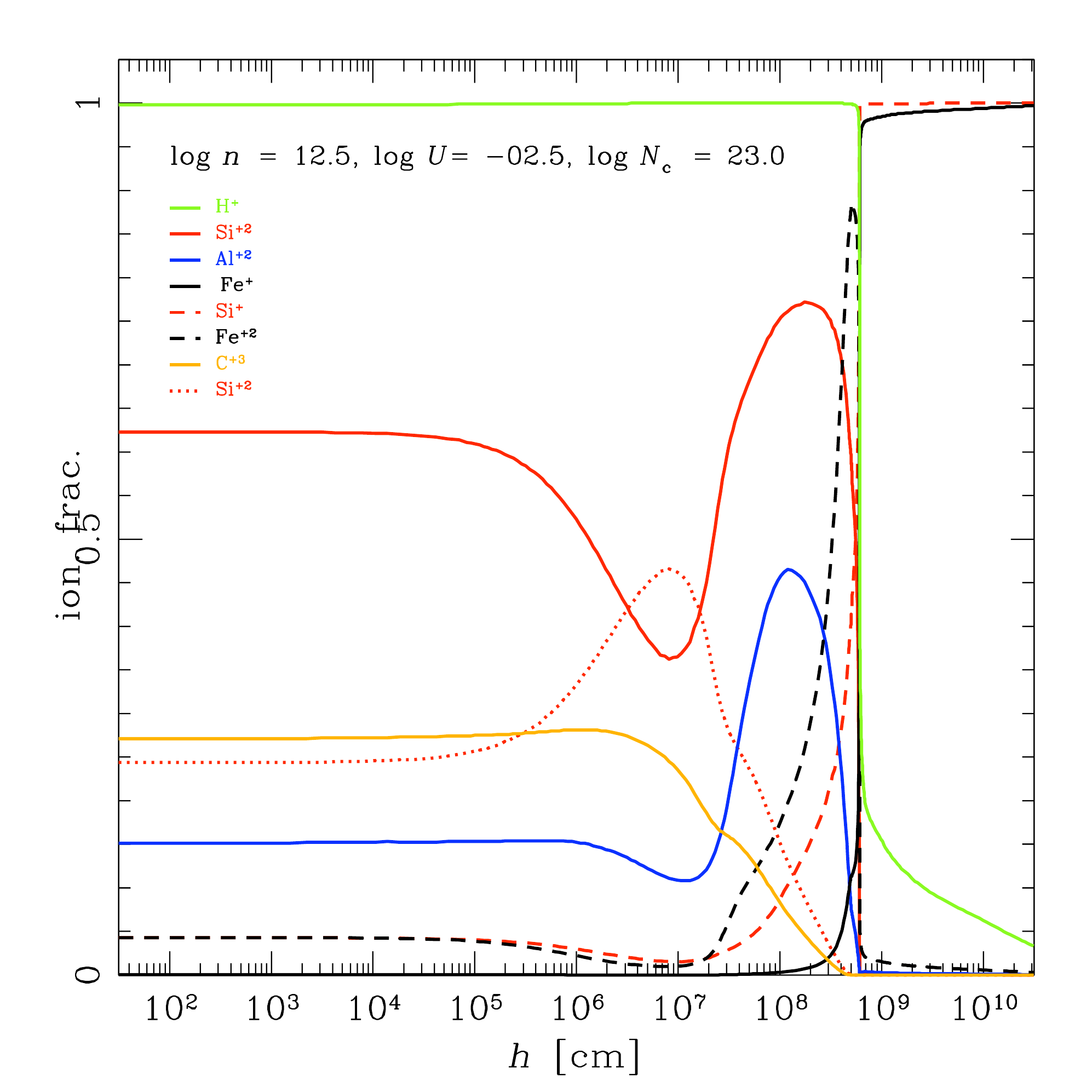}
\plotone{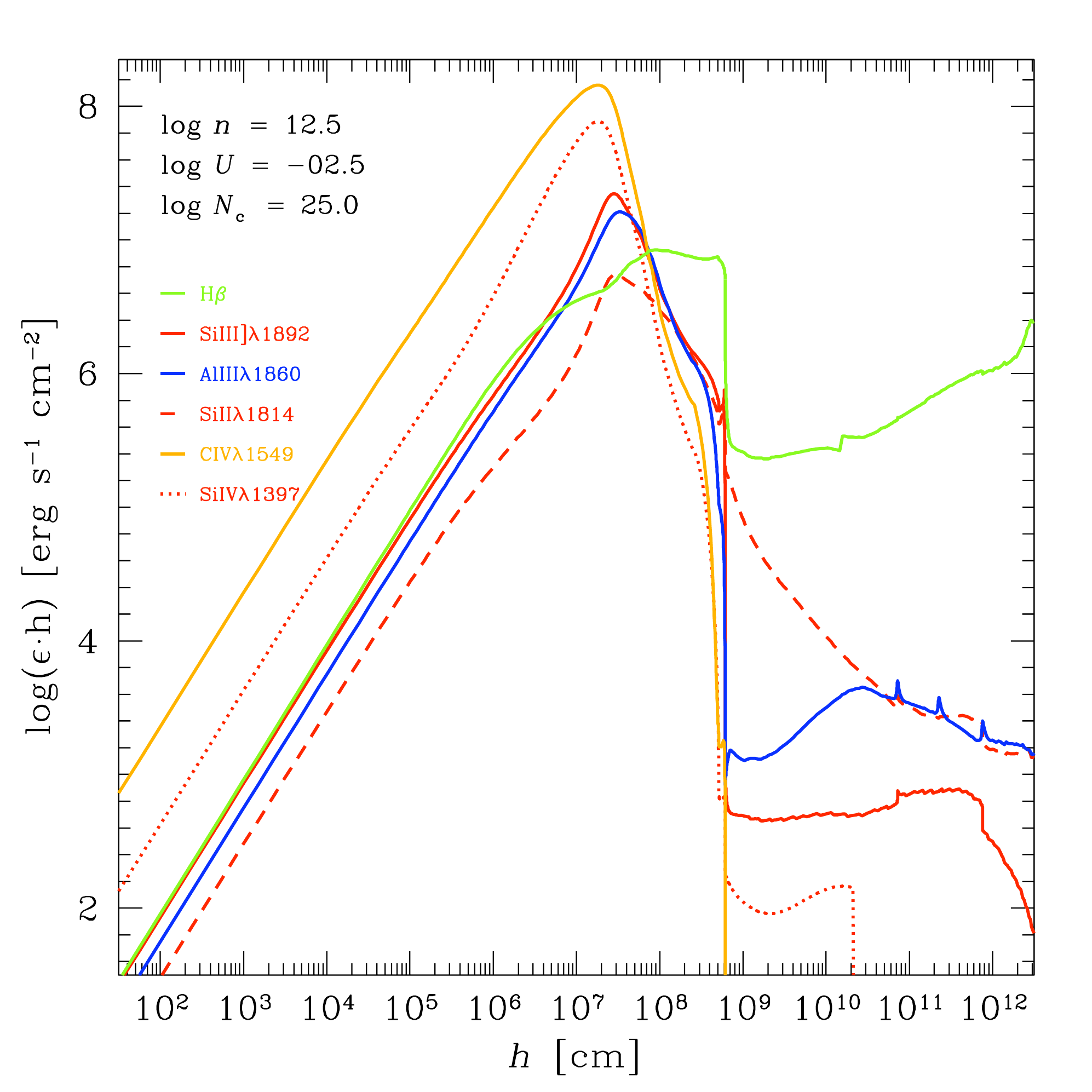}
\caption{Top: Ionic Fractions as a function of the geometric depth $h$\ in the gas slab. Bottom: Line emissivity per unit volume in units of \ergss\ \cm3\ multiplied by depth.  \label{fig:ionic_emissivity}}
 \end{figure}

Our project stems from the careful analysis of the I Zw 1 UV spectrum by Laor et al. (1997b), and the evidence of a well-defined trend in the \l1900\AA\ blend along the E1 sequence (Aoki \& Yoshida 1999, Wills et al. 1999, Bachev et al. 2004, Marziani et al., 2010). The NLSy1 I Zw 1 is known to be a sort  of extremum in the E1 sequence: it shows strong \feii\  and \feiii, prominent \aliii\ and \siiii\ emission. It is an example of the A3 spectral type, whose median 1900\AA\ blend in shown in Fig. 3 of Bachev et al. (2004). Definitely, these sources are present also at intermediate to high redshift.  We describe here the  analysis of one source, SDSS J120144.36+011611.6 (Fig. \ref{fig:extreme}), which seems to be a high-redshift, high-luminosity replica  of I Zw 1, with broader lines (the FWHM limit of NLSy1s and Pop. A sources is luminosity dependent; see Netzer \& Trahktenbrot 2007 and Marziani et al. 2009).  These sources are Pop. A, and we assume that the profile of the BC is Lorentzian. Hereafter unshifted Lorentzian part of the line is said to be BC. Pop. A sources are also free of any VBC, making the analysis of the blend simpler. 

Several considerations can be made from Fig. \ref{fig:extreme}. 

\paragraph{\aliii}The lines less ambiguous to measure are the \aliii\ doublet  because the two lines are less blended with other features, and they  are remarkably strong.  \siiii\ is more heavily blended with  \ciii\  and \feiii. However, the {\sc specfit} routine allows usually a plausible deconvolution of \siiii, making the \aliii/\siiii\ ratio very  reliable.  \feii\  and \feiii\  are obviously strong. We can use \feii\  UV 191 to set roughly the level of \feii, while the feature at 2080\AA\ is helpful to estimate intensity of \feiii. 

\paragraph{\feiii} Our {\sc specfit} analysis produces a very weak  \ciii\  component. A precise measurement of  \ciii\  is cumbersome, since its intensity depends on the actual \feiii\  emission. The strong feature at \l1914 could be dominating, and the template may seriously underestimate it (see Vestergaard and Wilkes 2001 for several alternatives in the empirical \feiii\ emission of I Zw 1). In any case the residual  \ciii\  emission is small, suggesting that the \l1900 blend, once believed to be mostly \ciii, is actually almost void of  \ciii\  emission in these A3-type sources.

\paragraph{\siii}  The \siii\ line is well visible in the spectrum of SDSS J120144.36+011611.6 and can be used as a substitute of \civ\ to measure the ionization parameter. The ratio \siii/\siiii\ is mainly sensitive to the ionization parameters, as it is the ratio \civ/\siiii. The ratio \siii/\siiii\ has the considerable advantage of being weekly dependent on metallicity.  If metallicity is known (see also \S \ref{metals}), the ratio \civ/\aliii\ should be in principle preferred since \aliii\ is emitted through a permitted resonance transition while \siii\ is emitted in the collisionally excited, partially-ionized zone (PIZ). 

\paragraph{\civ}  We fit \civ\ with a Lorentzian component that is unshifted + a \civ\ BBC. Note that the lack of shift in \aliii\ and \siiii\ imposes a strong, determinant condition on the strength of the Lorentzian-component in  \civ. It is important to stress that some  \civ\ BC emission is expected to be present according to our array of simulations. The ratio  \civ/\aliii\ is rather modest, as we can easily see even by eye.  We note in passing that the with of BCC is $\approx 10000$ \kms, with a peak blueshift indicatively of $-6000$\kms, significantly larger than the one measured on the \civ\ of I Zw 1 by Marziani et al. (2009). 

{\bf \paragraph{\siiv} We fit \siiv\ with a Lorentzian component that is unshifted + a  BBC (that may include \siiv\  and \oiv\ contribution). Results that are consistent with the ones of \civ. The ratio \siiv/\siiii\  has the advantage that is almost independent on metallicity. In principle, the crossing point between the \siii/\siiii\ and \siii/\siiii\ can set a metallicity independent point in the (\nh, $U$) plane. If the accuracy of the \siii\ intensity is good then this point can be used to retrieve information on metallicity (\S \ref{metals}). }

\begin{figure}
\epsscale{1.1}
\plotone{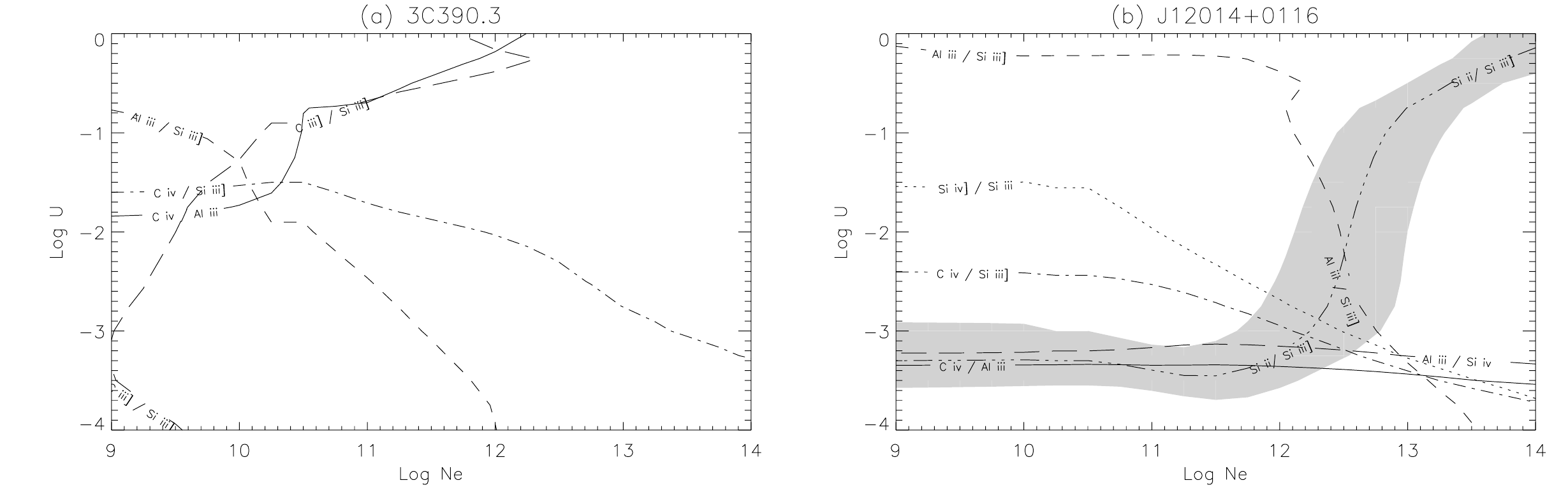}
\caption{Contour plots for the extreme objects (a) 3C390.9 and (b) J12014+0116.  Abscissa is electron density in cm$^{-3}$, ordinate is the ionization parameter, both in logarithm scale. Solid line is for $\log$(\civ/\aliii), dot line is for $\log$(\siiv/\siiii), dash line is for $\log$(\aliii/\siiii), long dash line is for $\log$(\aliii/\siiv), dash dot line is for $\log$(\civ/\siiii) and dash-triple-dot line is for $\log$(\siii/\siiii). The point where the isocontours cross determines the values of Log$n_e$ and LogU. The shaded area is the error bands for \siii. In 3C390.3 we add a long dash line for $\log$(\ciii/\siiii). In J12014+0116 the shaded area is the error bands for \siii. See text for further details. \label{fig:neu_extreme}}
\end{figure}

We consider the measured line ratios in the plane $U$\ vs. \nh\ (initally assuming metallicity equal to solar), and see where they cross. We find very high density and low ionization (Fig. \ref{fig:neu_extreme}). 

\subsubsection{Along the E1 Sequence: A More Complex Scenario}

Looking at the fit solution (Fig. \ref{fig:extreme}(b)), it seems that our spectrum has almost no \ciii. In many ways this is not surprising. The physical solutions in the ($U$, \nh) plane of Fig. \ref{fig:neu_extreme}, points toward very low ionization ($U \sim 10^ {-3} - 10^{-2}$) and high density (\nh $ \ga 10^{12}$\cm3). At such high values of \nh\ we expect that  \ciii\  is collisionally quenched, and no significant emission. The ratio \ciii/\civ\ is expected to be just $\sim$ 10$^{-2}$ in the dense LIL-BLR. Therefore we can say that sources like SDSS and I Zw 1 are extreme because all of the \l1900\AA\, blend is emitted by very low ionization, dense gas. 

However, as soon as we move away from spectral types A3+ along the E1 sequence, we see that the prominence of \aliii\ diminishes greatly. The emission disappears altogether at the other end of the E1 sequence, where several lobe-dominated radio-loud sources are found. For most quasars  we see that  \ciii\  is rather strong and unmistakably present. At a first glance this complicates the interpretation of the spectrum. However, one has to consider that  \ciii\  can be emitted only by gas of much lower \nh\ than that of the region where the bulk of \siiii\ and \aliii\ is emitted. Our simulations show that    \aliii\ intensity grows smoothly as a function of density in the ionization parameter range of interest $-3 \la \log U  \la -1$. This said, and considered the smooth trend we see from A3 to B1$^+$, the most reasonable conclusion is that a dense region emits significantly whenever \aliii\ emission is detected, although the relative prominence of the dense BC changes along the E1 sequence: it accounts for the entire LIL emission {\em only} in the most extreme Pop. A sources. The sequence of Fig. 3 of Bachev et al. (2004) seems to be mainly a sequence of prominence of \aliii+ \siiii\ vs \ciii. 

So, a very important conclusion is that a very dense, low-ionization region exists in the wide majority of quasars. It is associated with \feii\  prominence, as such gas is expected to emit a strong low-ionization spectrum. The most extreme \feii\  emitters are also the most extreme \aliii\ emitters; in some cases where no \aliii\ emission is measured, we also fail to detect any \feii\  emission (Marziani et al. 2010). Where is this region located? Why there are such distinctive line profiles as Lorentzian? The second issue goes beyond the present paper; for the moment our aim is  to measure  line components of \aliii, \siii, \siiii\ and  \civ\ that come from this region and to estimate its distance from the central black hole (\S \ref{rblr}).

\subsection{The Other E1 Extremum: 3C 390.3} 
\label{sec:3c390}

3C 390.3 is a lobe-dominated (LD) RL source, with very broad emission lines and no optical \feii\  within  detection limits (see Marziani et al. 2003, for the criterion used). It also shows a large peak shift in its \hb\ profile, and prominent narrow lines ([O{\sc iii}]\l5007, \hb), but also narrow components of \civ\ and \ciii\ which are all well separated from the broad \hb\ profile. 

To deconvolve the blends around \ciii\ and \civ, we assume  the same emission line profiles as for \hb. The \hb\ broad profile can be described as the sum of a BC and a VBC. An unusual property of the BC is its large peak blueshift (it is unusual because of the shift amplitude: even if median spectra of Pop. B sources show a small BC redshift, there is a pretty large scatter with both red- and blueshifted peaks observed in individual sources; Zamfir et al. 2010). Applying in a self consistent way the BC and VBC to the \ciii\ and \civ\ blends, leads to very interesting results as shown in Figure \ref{fig:extreme}: 

\begin{itemize}
\item  [$\bullet$] \aliii\ is very weak or even absent within the S/N limits; 
\item  [$\bullet$] the \civ\ profile is very similar to the one of \hb: BC+VBC accounts for more than 90\% emission, with a possible, very weak contribution of the blueshifted component which is usually dominating in Pop. A sources; 
\item  [$\bullet$] the red wing observed in the 1900\AA\ blend is accounted for only if there is a strong \ciii\ VBC, \feii\  and \feiii\ emission being below the detectability limit in this source; 
\item  [$\bullet$]  there is no evidence of a VBC in \aliii;  
\item  [$\bullet$] the ratio  \civ/\ciii\  BC is $\approx$ 10, a far cry from the ratios observed in extreme Pop. A sources like I Zw 1 and  SDSS J120144.36+011611.6; 
\item  [$\bullet$] even more interesting we find a  \civ/\ciii\  VBC $\approx$10, as for the BC. 
\end{itemize}

We conclude that this object is fundamentally different from Pop. A sources. The low ionization, high density BC seems to be absent. The similarity in the  \civ/\ciii\  ratios suggest that we are observing a gas in conditions similar to the one associated to the VBC. We predict that \feii\  will remain undetected or found to be  weak even with S/N $\rightarrow \infty$.  

In the plane ($U$, \nh) of Figure \ref{fig:neu_extreme}, the line ratios converge to a point at $\log U \approx$ -1.5, and log \nh $\approx$ 10.1 (the \siii\ line cannot be measured accurately since S/N is poor). In this case, the \ciii/\civ\   ratio also converges toward a ($U$, \nh) value consistent with the one indicated by the \aliii/\siii. In most other cases, this does not happen because the \aliii\ doublet is too strong to be accounted for by gas of \nh $ \sim 10^{10}$\cm3\ even at super-solar metallicity. Emission occurs at pretty high ionization, in conditions that once upon a time were thought to be standard in quasars (Davidson and Netzer 1979). This Pop. A and B difference at the extrema was already pointed out, in a semi-quantitative way, by Sulentic et al. (2000) and Marziani et al. (2001). 

It is however important not to generalize the case of 3C 390.3 to the remaining Pop. B sources. In most of them, \aliii\ is detected, and there is evidence of strong \ciii. Actually  \ciii\  emission seems to be appreciable in all of our VLT quasars. This means that we are in a composite situation, where low-ionization, high-density emission is present, along with significant VBC and other emission. The two extreme cases help us however to understand these more complex cases. 

\subsection{The Contribution of Lower Density Gas}
\label{sec:ciii_contr}

{\bf Once the true intensity of the BC components is retrieved, the presence of significant \ciii\ emission complicates  the analysis. As pointed out, the photoionization solution for the BC suggests very high density, and in this region no \ciii\ emission is expected. Whenever \ciii\ is observed, we need to reverse the question: how much does any \ciii\ emitting gas contributes to the lines used for diagnostic ratios? Negligible contribution is expected to \aliii. However, this is not true for \civ\ and \siiii. Especially among Pop. A2 and A3 objects, it is not so obvious that the profile of \ciii\ and \siiii\ is the same. It could be well that the \ciii\ profile is narrower than the ones of \siiii\ and \aliii\ (as found for SDSS J1201+0116), justifying the idea of \ciii\ emission from a disjoint region (\S \ref{intermediate}). For the BAL QSOs in our sample and \aliii-strong sources most of what we ascribe to \ciii\ could be actually \feiii, as suggested by Hartig and Baldwin (1986). 

However, objects like the typical A1 sources show \siiii\ and \ciii\ with similar profiles, so in the following we will assume the worst condition, that is all \ciii\ emitting gas is contributing to the BC lines. 
}

{\bf To estimate the contribution of the \ciii\ emitting gas to \siiii\ and \civ, we can consider the trends observed along E1. {\bf \siiii\ is strong when \aliii\  is strong}, and the \siiii/ \ciii\ ratio is lower when \ciii\ is strong, as  visible in both Baldwin et al. (1996)
 and 
Bachev et al. (2004), as well as  in the present paper.  The observed \siiii/\ciii\ ratio in the A1 median bin seems to be as low as in Mark 335, $\approx 0.4$.   Pop. B sources that often show prominent \ciii\ also show a low \siiii/\ciii\ ratio, as also appreciable in the median spectra of  Bachev et al. (2004). These trends suggest that most \siiii\ is emitted where \aliii\ is also emitted. As a consequence, any correction due to gas in different physical conditions  (lower density) emitting \siiii\ is expected not to be dominant unless \ciii\ is extremely strong. The \ciii\ emitting gas should be at density $\log n \sim\ 9$, or lower. Higher density would imply increasing \siiii/\ciii\ to values that would exceed the observed ones even if the LIL-BC is not emitting any significant \siiii\ (see Fig. \ref{fig:linesblend}).
}

\subsubsection{Preliminary analysis of a low-$z$\ sample}

\begin{figure}[h]
 \includegraphics[scale=0.4, angle=0]{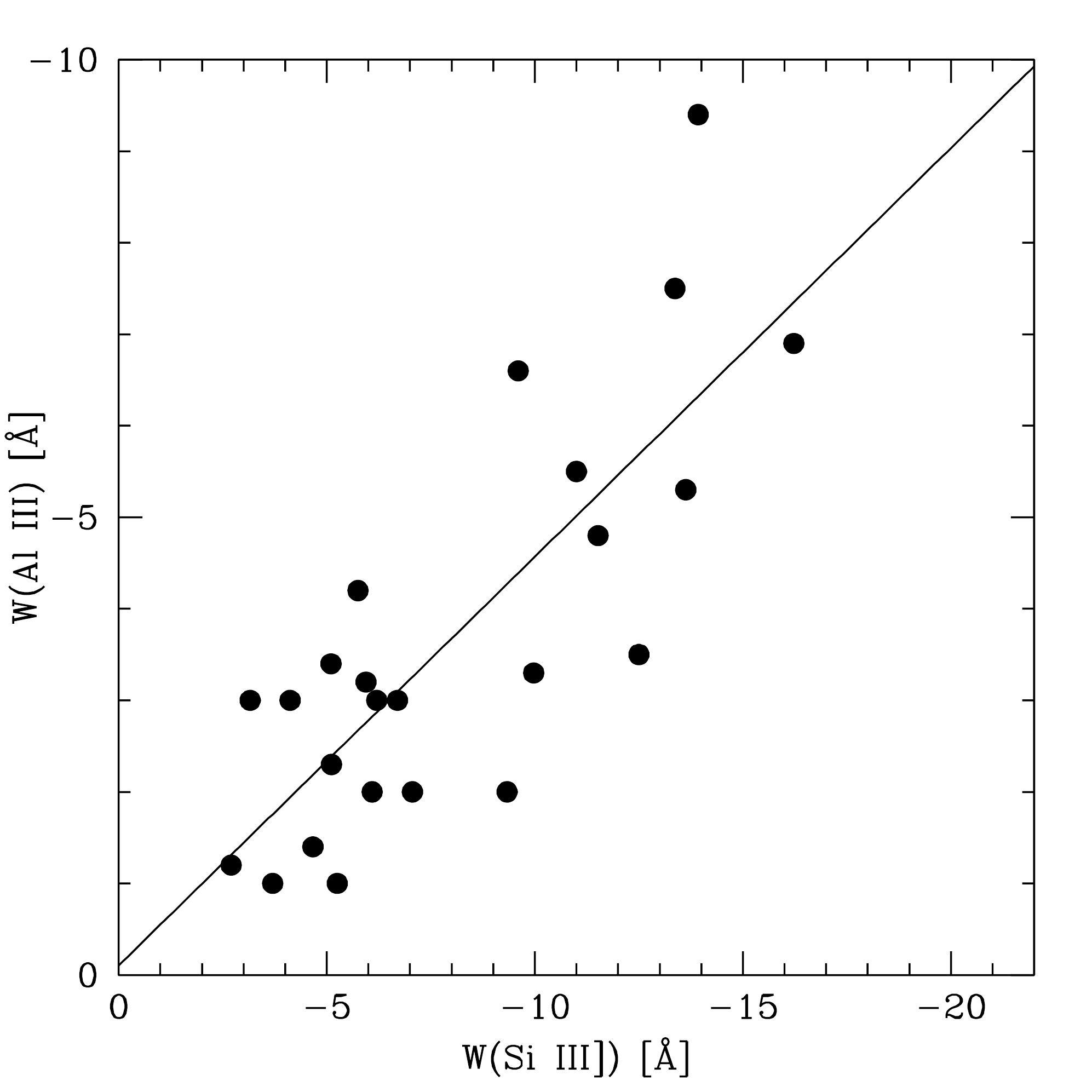}
\caption{Correlation between rest frame equivalent width of \aliii\ and \siiii. Line  in unweighted least square best fit. \label{fig:ratio}  }
\end{figure}

\begin{figure}[h]
 \includegraphics[scale=0.75, angle=0]{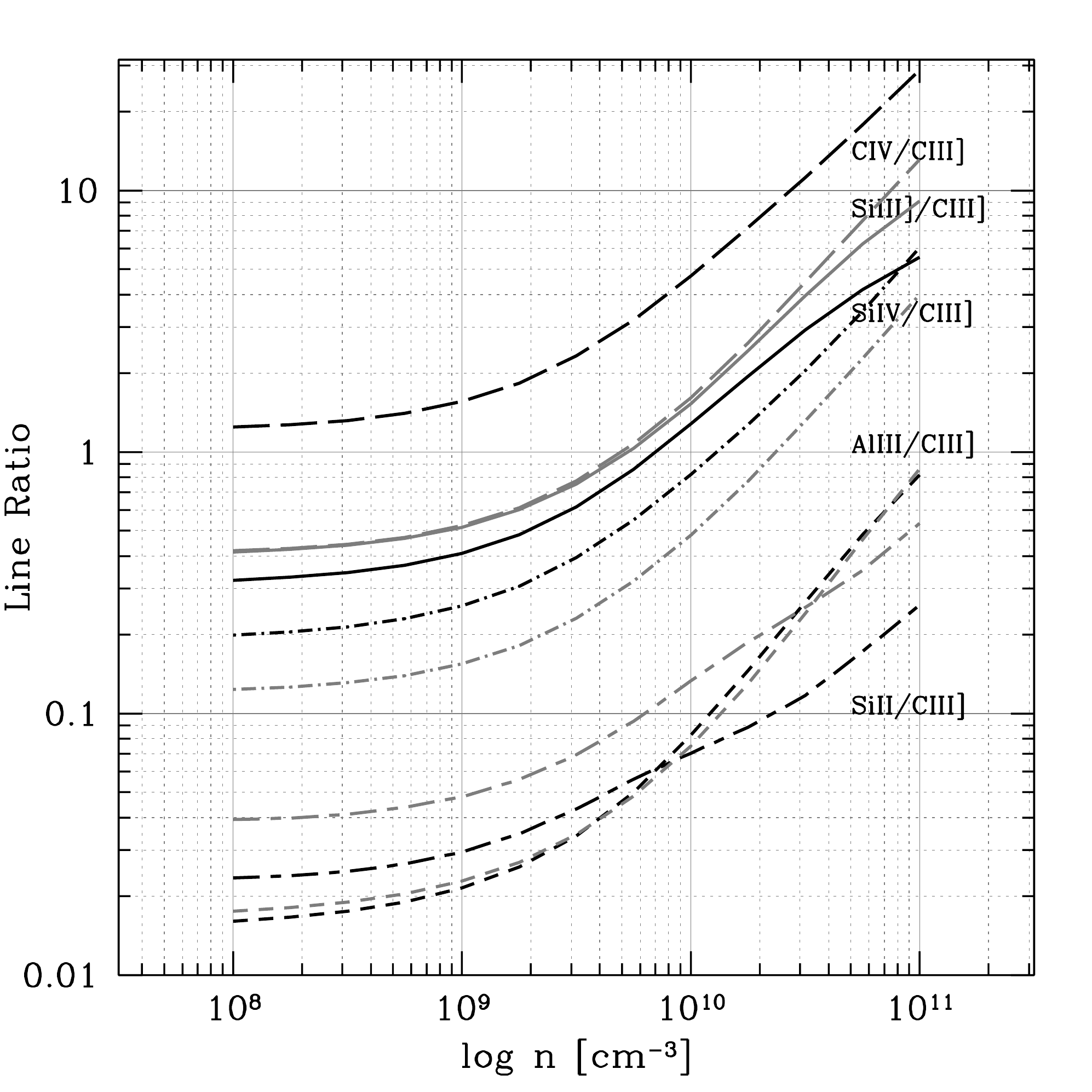}
\caption{Expected contribution from moderate density emitting gas as a function of density for ionization parameters $\log U = -2$ (black lines) and $\log U = 2.5$ (grey lines).\label{fig:linesblend}}
\end{figure}

{\bf To set these trends on a more quantitative basis we considered the set of pre-{\sc costar} recalibrated sources, for which \ciii\ and \civ\ data are publicly available (Kuraszkiewicz et al 2002, Evans \& Koratkar 2004).We performed measurements (interactively with the task {\sc splot} of {\sc iraf})  for about $30$\ sources with the highest S/N,  holding a \ciii\ blend that could be relatively easily deblended, and following the expectations described in \S \ref{Expected}. The rest frame equivalent width of \aliii\ and \siiii\ are found to be highly correlated (Fig. \ref{fig:ratio}). The correlation is due to \siiii\  being stronger when \aliii\ is strong, not necessarily because \ciii\ is strong: actually, the \siiii/\ciii\ ratio can be low when \ciii\ is strong, and the \aliii/\siiii\ ratio achieves maximum values when \ciii\ is faintest. This is consistent with the bulk of \aliii\ and \siiii\ originating in the same region. At the same time the presence of \ciii\ lowers the \aliii/\siiii, \civ/\siiii\ and \siiv/\siiii\  because of ``excess'' \siiii\ emission associated to the gas emitting \ciii. 
}

\subsubsection{The effect of low-density gas on the product ($U$\nh)}

\begin{figure}[h]
 \begin{center}
\includegraphics[scale=0.45, angle=0]{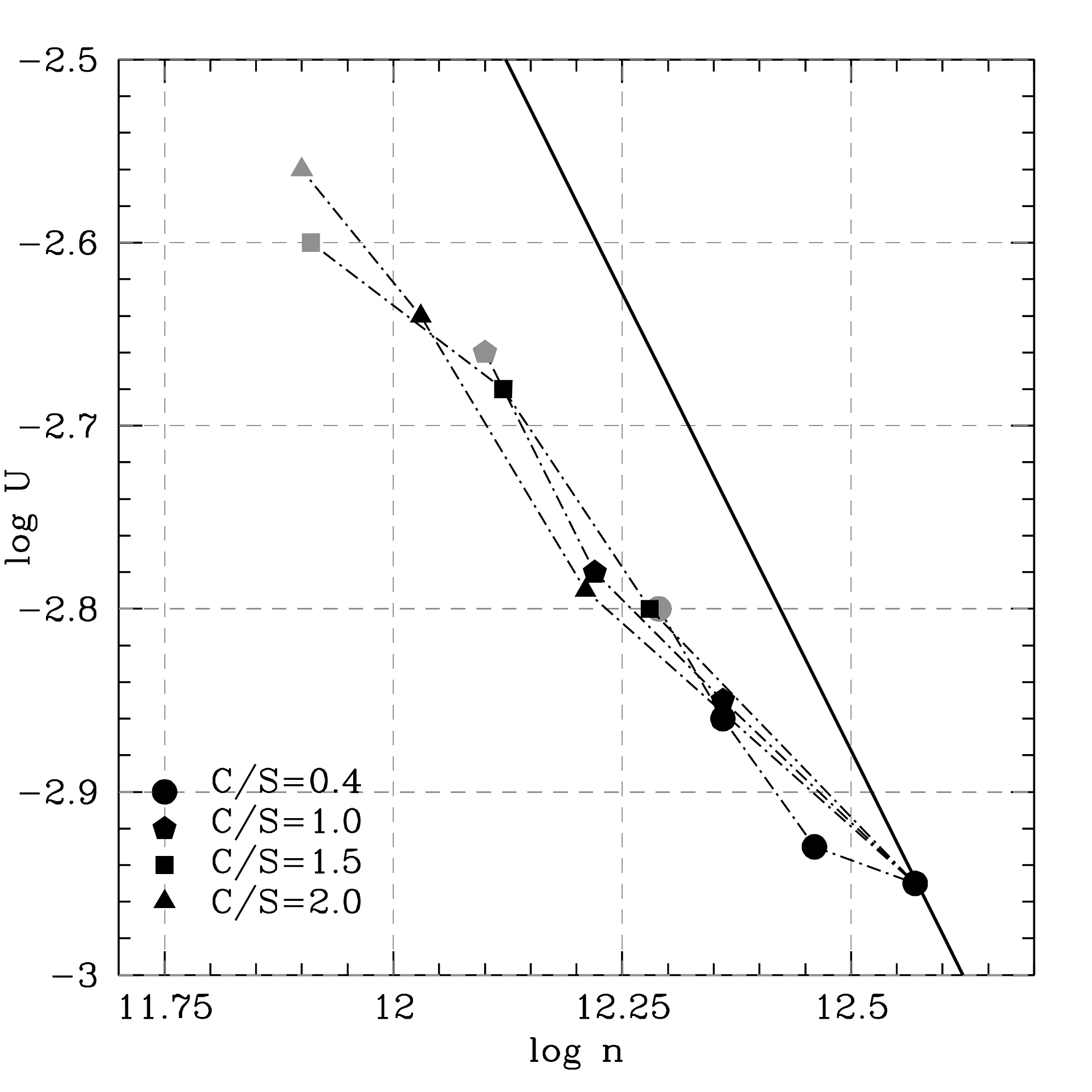}
\caption{Plane $\log U - \log n$\ in an expanded scale. The lower-left dot corresponds to a high density solution. Moderate density emitting gas is then added with increasing intensity of a \ciii\ component with the four \ciii/\siiii\ ratios reported in the figure.  Connected positions corresponds to increasing density of \ciii\ emitting gas, $\log n = 9,9.5,10$. Note that the case $\log n = 10$\ (grey symbols) is not appropriate for the observed data. \label{fig:product2}}
\end{center}
\end{figure}

\begin{figure}
\epsscale{1.1}
\plotone{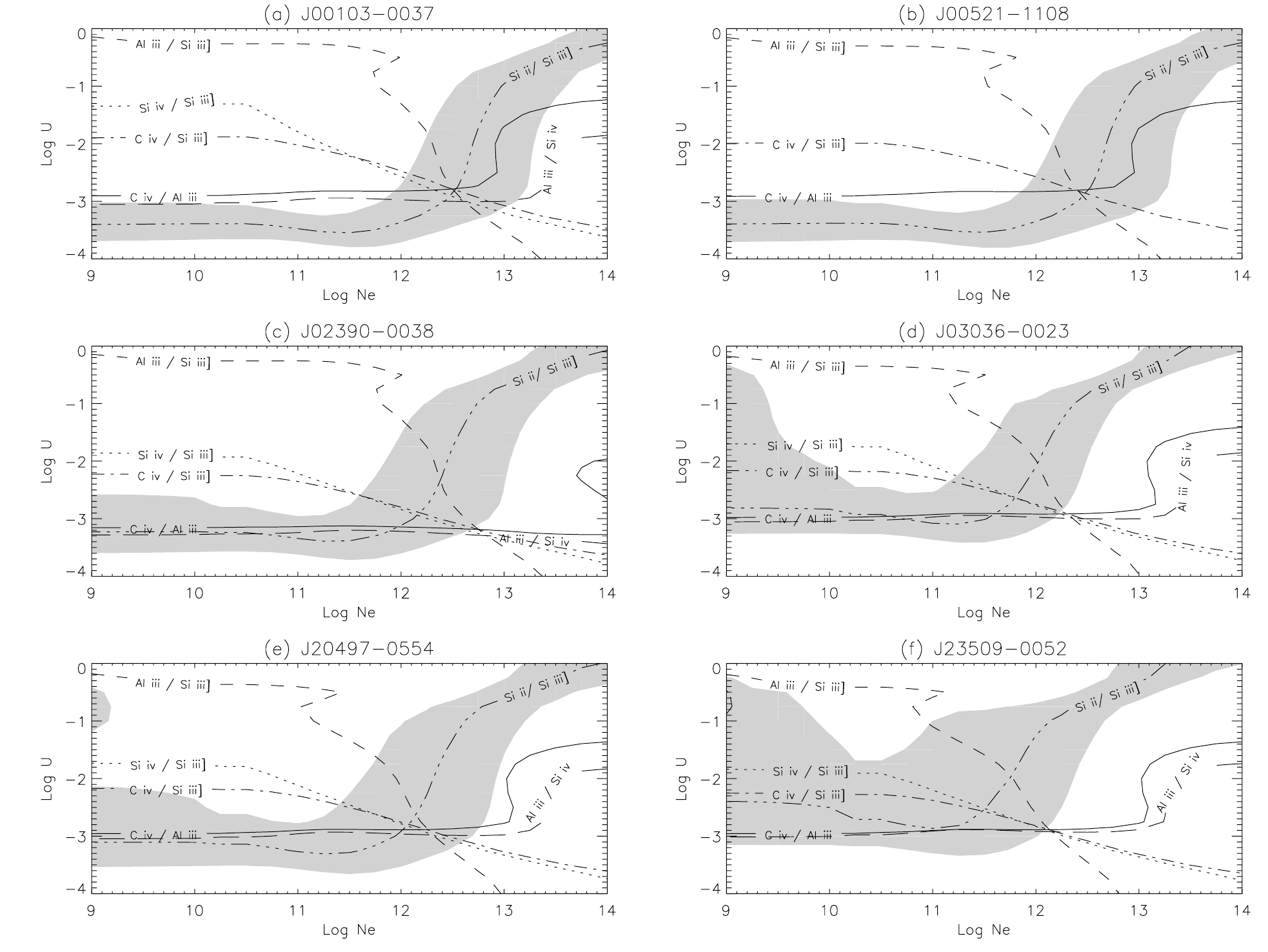}
\caption{Contour plots for (a) J00103-0037, (b) J00521-1108, (c) J02390-0038, (d) J03036-0023, (e) J20497-0554 and (f) 23509-0052.  Units and meaning of symbols are the same of Fig. \ref{fig:neu_extreme}. In the case of the objects shown in panels (c) and (f)  the \siii\ line is very weak and thus unreliable. For the objects in panels (a), (d) and (e) the line is also affected by the telluric absorption. For these objects we rely on the \siiv\ line. \label{fig:neu}}
\end{figure}

\begin{figure}
\epsscale{1.1}
\plotone{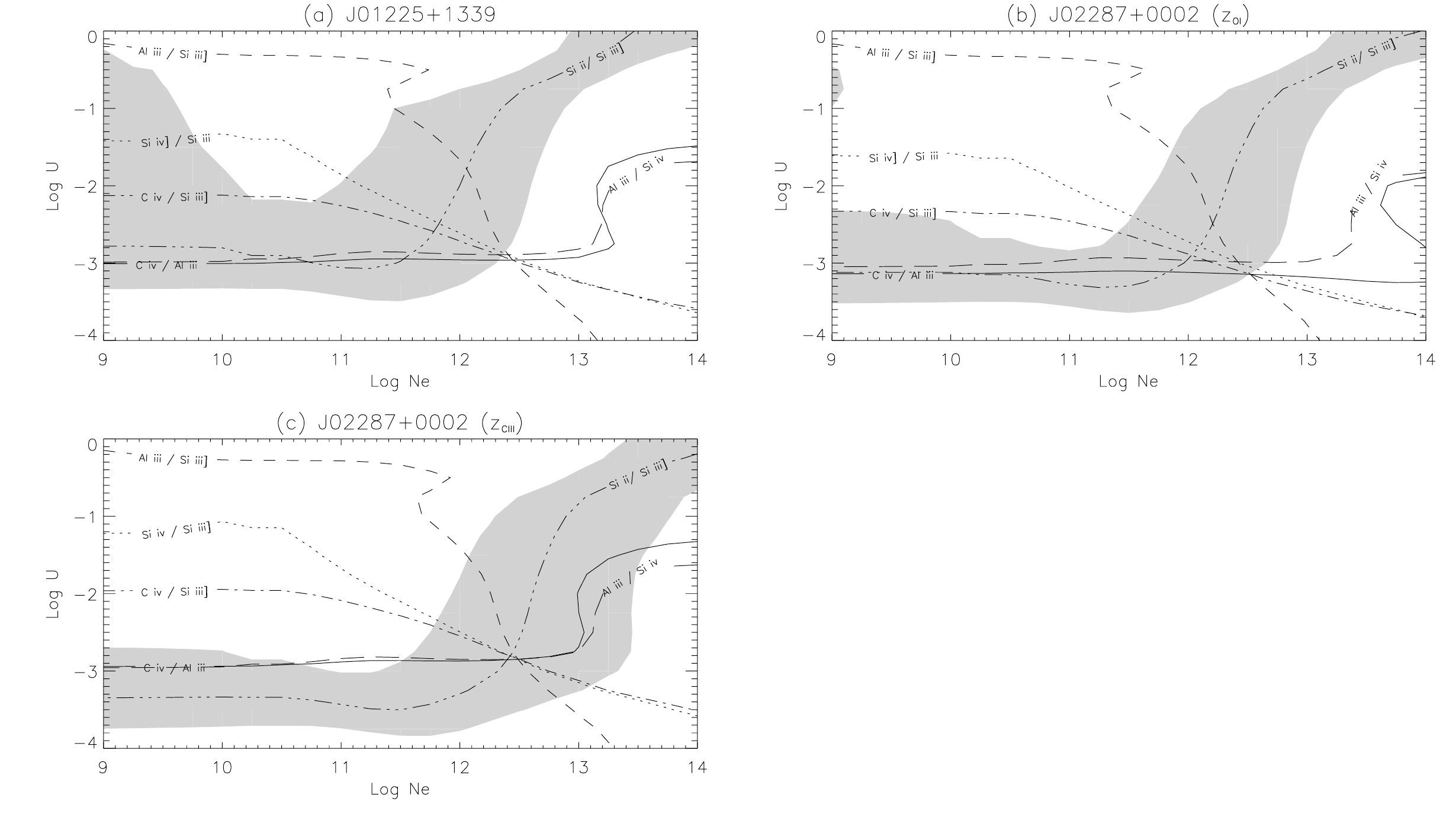}
\caption{Contour plots for BAL quasars (a) J01225+1339 and (b) J01225+1339 using $z_{OI\lambda1304}$ and (c) J01225+1339 using $z_{CIII]\lambda1909}$. Units and meaning of symbols are the same of Fig. \ref{fig:neu_extreme}. \label{fig:neu_bal}}
\end{figure}

{\bf We expect that  any correction will increase density (increase \aliii/\siiii\ lowering \siiii) and decrease ionization parameter (lowering \civ\ more than \siiii), but that their product will be less affected.  To show the amplitude of the effect we performed an experiment, adding to a pure, high-density solution, contribution from moderate density gas ($\log$ \nh $ \sim 9 - 10$). The Fig. \ref{fig:product2} shows the displacement in the density -- ionization parameter plane for several values of the \ciii/\siiii\ intensity ratio. The values refer to the \ciii\ addition over \siiii\ in the ideal case, corresponding to observed ratios of 0.34, 0.7, 1.0, 1.1 for an addition of a \ciii\ component whose intensity is 0.4, 1.0 ,1.5 ,2.0 the intensity of the \siiii\ component associate to the high density gas. As one can see from the figure, even if deviation for $U$\ and \nh\ taken separately are significant, deviations for the product are by far less important. The largest change for the product is found to be $\approx$ 0.15, if we exclude the gray dots corresponding to $\log$ \nh = 10\ (an unlikley case, since this would imply a correction \siiii/\ciii\ $>$ 1, inconsistent with what we observe when \ciii\ is strong). Following the expected line ratios of Fig.\ref{fig:linesblend} we apply a correction to the BC fluxes that is 0.4 and 1.5 \ciii\ for \siiii\ and \civ\ respectively  (Figures \ref{fig:neu}, \ref{fig:neu_bal} and \ref{fig:neu_extreme}), corresponding to $\log U = -2$ and $\log$\nh = 9. We remark that that if $U$\ is lower, the correction will have negligible effect, while assuming a larger $U$\ will lead to \civ\ flux in excess to the one observed.

To constrain the ionization parameter we can first consider that, since the \ciii\ gas comes from a (relatively) low density region, the contribution to \siii\ is small: for $\log n \sim\ 9$\ and $\log U \approx -2$, the contribution should be $\approx 0.03$ \ciii. Second, another powerful feature is the \siiv\ doublet (Baldwin et al. 1996):  the \siiv\ doublet is  less affected by the \ciii\ correction, the contribution from lower density gas being estimated $\approx$ 0.25 \ciii. The line ratio \siiv/\aliii\ is also sensitive to ionization and less affected by any lower density correction (provided that the relative abundance of S and Al stays the same, as it seems to be the case).

 In summary, corrected BC line  intensities are computed as follows: $I^\mathrm{c}$(\siii)$_\mathrm{BC}$ = $I$(\siii)$_\mathrm{BC}$ -- 0.03 $I$(\ciii)$_\mathrm{BC}$;  $I^\mathrm{c}$(\siiii)$_\mathrm{BC}$ = $I$(\siiii)$_\mathrm{BC}$ -- 0.4 $I$(\ciii)$_\mathrm{BC}$;   $I^\mathrm{c}$(\siiv)$_\mathrm{BC}$ = $I$(\siiv)$_\mathrm{BC}$ -- 0.26 $I$(\ciii)$_\mathrm{BC}$;  $I^\mathrm{c}$(\civ)$_\mathrm{BC}$ = $I$(\civ)$_\mathrm{BC}$ -- 1.5 $I$(\ciii)$_\mathrm{BC}$.
}

\section{Results on the  $z \approx$ 3 Quasars}
\label{sec:results}

To estimate $\log$\nh\ and $\log U$\ values, we use the {\sc cloudy} contour plots of the ratios \aliii/\siiii, \siii/\siiii,  \civ/\siiii,  \siiv/\siiii\ showed in Fig.  \ref{fig:contours}\footnote{Note that there are regions  where the ratio values are actually undefined:  close to the high $U$\ limit ($\log U \ga -0.3$),  ratios \aliii/\siiii\ and \siii/\siiii (with \nh $\la 10^9$\cm3) should not be considered.}. The data points of our objects are in regions were the ratios are well-defined. The ratios \civ/\siiii, \siiv/\siiii, and \siii/\siiii\ are mainly sensitive to the ionization parameter $U$, while \aliii/\siiii\ and \ciii/\siiii\ are mainly sensitive to the electron density. We know that \ciii\ is collisionally quenched at $\log$ \ne $\ga$ 10 and in the contour plot for \ciii/\siiii\ we see a step around this value. 

We measure the BC intensity of \siiii, \aliii, \siii, \siiv\ and \civ; with them we compute the diagnostic ratios (for  3C390.3 we use \ciii\ in place  of \siii; however there is no object similar to 3C390.3 among the $z \approx$ 3 quasars).  We present the fluxes of the line components in Tab. \ref{tab:fluxes} and the equivalent width in Tab. \ref{tab:ew}. Table \ref{tab:weak} shows the weak lines around \civ. For \civ\ line we show the core, blue shifted and the very broad components. Errors are at a $2 \sigma$ confidence level, and include the sources of uncertainty described in \S \ref{errors}.  Errors are then quadratically propagated according to standard practice to compute intensity ratios and their logarithm.

From Table \ref{tab:fluxes} we can derive the diagnostic ratios. As we see in Fig. \ref{fig:sample_ab}, \siii\ is absorbed by telluric B band in J00103-0037, J03036-0023, J20497-0554 (most affected). We will not consider \siii\ to compute \nh\ and $U$\ on those cases. However,  if we take at face value the \siii\ measure on J00103-0037 and J20497-0554, it will converge close to the point set by the remaining two ratios.  We  display  on a graph  a line representing the behavior of  each ratio under the assumption of solar metallicity; the ideal point where the lines representing different diagnostic ratios cross  determines the values of $\log$\nh\ and $\log U$. Figs. \ref{fig:neu} and \ref{fig:neu_bal} shows the contour plots were we can see that the diagnostic ratios converge to rather well defined values. The cross point is very precise for the objects J00103-0037, J00521-1108, J02287+0002 (using $z_{CIII}$), and J20497-0554 ; for the remaining objects J01225+1339, J02287+0002 (using $z_{OI}$), J02390-0038, J03036-0023, and J23509-0052 the cross point is slightly different. We must not forget the errors involving the fits, such as the changing of the shape profile that makes large the peak intensity if is Lorentzian or it could be less intense if it is Gaussian;  the \feii\  pseudo-continuum contribution that affects principally to \siii\ or the \feiii\  that in some cases affects  \ciii.  

{\bf In principle, the crossing point of the ratios \siiv/\siiii\ and \siii/\siiii\ is independent on metallicity. Therefore, any significant disagreement between this crossing point and the ratios based on \civ\  may indicate  chemical composition different from the assumed solar one (\S \ref{metals}). The difficulty here is  the large uncertainty of the \siii\ line. In all contour plots we show the $\pm 2 \sigma$ interval as a shaded band. So, it is proper  to consider deviations from metallicity only in the case of 4 sources where the crossing point excluding \siii\ is outside the uncertainty band.} 

For the objects J00103-0037, J03036-0023 and J20497-0554 we exclude the \siii\ line from the diagnostic ratios because it is affected by absorption. For the remaining objects, we take the average of the crossing contour plots of $\log$(\aliii/\siiii) crossing with  $\log$(\civ/\siiii) and $\log$(\aliii/\siiii) crossing with $\log$(\siii/\siiii). For 3C390.3 we use $\log$(\ciii/\siiii) instead of $\log$(\siii/\siiii). Table \ref{tab:neu} summarizes the $\log n_\mathrm{H}$ and $\log U$ values including their uncertainty. Since $U$\ and \nh\ are not independent quantities (their correlation coefficient is found to be 0.55),  we adopt the appropriate formula for the errors on the product $U$\nh\ (following Bevington 1969).  
We present average values  of the crossing points  for extreme objects of Fig. \ref{fig:neu_extreme} and \ref{fig:z5}(d). In the SDSS J12014+0116 case we give full weight to the \siii\ measurement. The crossing points disagree somewhat for SDSS J12014+0116.  This could be due to an underestimate of {\em both} \siii\ and \civ\ in the fits.  On the one hand, \siii\ is clearly seen and strong but is contaminated by \feii\ blend; on the other hand, \civ\ is strongly affected by the blue-shifted component and by the assumption that it is of Gaussian shape. An increase by 30\%\ in the measurement of the intensity of the two line would lead to a better agreement but this is somewhat an ad-hoc speculation. Rather, the significant difference between the crossing point of the \siii/\siiii\ and \siiv/\siiii\ ratios and the other ones point toward strong metal enrichment. We will show in \S \ref{metals}\ that this is probably the case.

At any rate, the convergence is toward a value of $\log U \approx -3$,  lower than for the other $z \approx $3 quasars.  This is reflected in the \civ\ EW of this source, also significantly lower.  {\bf It is intriguing to note that the correction because of lower density drives the other sources  toward values of $U$\ and \nh\ that are closer to the ones of SDSS J12014+0116.  }

Table  \ref{tab:r_m} reports the values of the \rb\ and the \mbh\ of our 8 objects  {and the extreme objects in the last two rows. }  Column 1 identifies the quasar name; Col. 2 gives the quasar proper distance in {\bf mega parsecs [Mpc]}; Cols. 3 and 4 are the continuum specific flux value at 1350\AA\, and 1700\AA\, respectively, Col. 5 reports the FWHM in km s$^{-1}$\ for the broad components, Col. 6 is the Population designation.  
{\bf Cols. (7) to (10) report the logarithm of the size of the BLR in cm obtained from: a) $1Z_\odot$, b) $1Z_\odot$ line ratios corrected because of low density emission, c) $5Z_\odot$, and d) $5Z_\odot$ line ratios corrected because of low density emission.}  Cols. (11) -(14) list   the logarithm of the black hole mass in solar masses in the same order as for \rb.  Finally Col. (15) is \mbh\ computed from Vestergaard and Peterson (2006) formula (Equation \ref{eq:vestergaard}). We will explain in \S \ref{sec:discussion} how these quantities are computed.

\subsection{Effects of Metallicity }
\label{metals}

The strength of \nv\ relative to \civ\ and \heiiuv\ suggests supersolar chemical abundances (Hamann and Ferland 1993; Hamann \& Ferland 1999).  Chemical abundances may be well 5 to 10 times solar (Dhanda et al. 2007), with Z $\approx$ 5$Z_{\odot}$ reputed  typical of high $z$\ quasars (Ferland et al. 1996). The E1 sequence seems to be mainly a sequence of ionization in the sense of a steady decrease in prominence of the low-ionization BC toward Population B (Marziani et al. 2001, 2010). However, this is not to neglect that metal-enrichment also plays a role, especially for the most extreme Pop. A sources i.e., those in bin A3\ and higher (Sulentic et al. 2001). 

The lines employed in the present study come from carbon, silicon and aluminium; all these element can be significantly depleted from gas if dust grains are formed (e. g., Mathis 1990). However, the emitting regions where our lines are produced are thought too hot to contain  significant amount of dust (a definition of broad line region is right the central engine region below the dust sublimation region: e.g., Elitzur 2009). In addition Si and Al are expected  to be produced under similar circumstances in the late stage of evolution of massive stars (Clayton 1983, Ch. 7). We note also that the \civ/\siiii\ and \civ/\aliii\ usually give results that are in perfect agreement in the plane (\nh,$U$). These findings support our assumption that, if metallicity variations are present, the relative  abundance Al to Si remains constant. 
{\bf We considered two cases for enhanced metallicity: (1) constant  solar abundance ratio Al:Si:C with $Z=5 Z_{\odot}$\ (5Z) ; (2) an overabundance of Si with respect to carbon by a factor 3, again with $Z= 5 Z_{\odot}$\ (5ZSA). This condition comes from the yields listed by Woosley and Weaver (1995) from type II Supernov\ae. The Si overabundance is also supported by the chemical composition of the gas returned to the interstellar medium by an evolved population with a top-loaded initial mass function simulated using {\sc Starburst 99}  (Leitherer et al. 1999).  The abundance of Al should scale roughly with the one of Si. While some cases with Al scaling with C are possible from the Woosely and Weaver (1995) yields, they are rarer than cases in which Al scales with Si. This latter case is appropriate for the most massive progenitors. Also, the assumption of Al scaling with C with [Si/C] = 0.477 would yield to implausible high density and lack of convergence to a well-defined solution for $\log$\nh$\simlt 14$. We therefore assume in the following that Al scales with Si in the two cases listed above. }

An array of simulations as a function of ionization parameter and density was computed assuming the conditions (1) and (2) listed in the previous paragraph.  As expected, if the solar metallicity is simply scaled by a factor (5Z) we find that the ratio \aliii/\siiii\ is not strongly dependent on $Z$: the ratio increases by about 40\%\ passing from $Z = 1 Z_{\odot}$\ to $Z = 5 Z_{\odot}$, for $\log $\nh $\approx 12$ and  $\log U \approx$--2. The same is true for the \siii/\siiii\  and \siiv/\siiii\ ratios. Since the first ratio sets \nh, and the last  two $U$, the ratios mentioned in this paragraph should be preferred because they provide \nh\ and $U$\ values that are weakly affected by a factor 5 change in metallicity.  A posteriori we confirm that the effect  on the product $U$\nh\ derived also with ratios involving \civ\ is negligible in case 5Z (and it should be even more so if a metallicity increase is $Z_{\odot}\la Z \la 5 Z_{\odot}$). 

The two extreme cases seem to be revealing also as far as metallicity is concerned. Ratios converge to a fairly well defined point in the case of 3C 390.3 (see Fig. \ref{fig:neu_extreme}, left panel, $Z = 1 Z_{\odot}$\ assumed). In this case there is no major evidence of supersolar metallicity. The converse is true in the case of SDSS J120144.36+011611.6. The ratios involving \civ\ indicate a lower ionization level, with $ \log U \sim -3$.  {\bf This is because C$^{3+}$ changes ionization state to C$^{2+}$ for smaller ionization parameters, and so \civ\ rapidly disappears.} The  \civ\ intensity depends weakly on $Z$\ while the \aliii\ and \siiii\ lines are more sensitive. The 5Z case yields values closer to the ones obtained from the Si and Al line ratios, but not yet in concordance.  If we pass to case 5ZSA with a factor 3 Si overabundance, then the concordance of the line ratios is good, especially if no correction because of low density emission is applied. As a further confirmation we checked that the metallicity-dependent \civ/\siiv\ ratio is in very good agreement with the crossing point of the other lines. Therefore, in this case, we have independent measures of metallicity, \nh, and $U$.   The case  of even higher metallicity, say $Z \sim 10 Z_{\odot}$, remains to be explored
but may not be appropriate considering the good agreement with 5ZSA. 

In principle,  the discrepancy in the intersection point, with \siiii/\civ\ and \aliii/\civ\ yielding lower $U$\ than the Al-Si ratios, should signal a significant enrichment in Si and Al  of the BLR gas. In this case the \siiv/\civ\ ratio should be helpful, as it can be assumed to be dependent mainly on the Si abundance relative to C. Here, more than precisely  determining the exact abundance value we are interested in analyzing the effect of large metallicity changes on $U$, \nh, and their product.  Appreciable discrepancy is visible in the contour plots of the BAL quasar J01225+1339, J03036-0023, J20497-0554 and J23509-0052 if $Z = Z_{\odot}$ is assumed. 

In the same plots made for  $Z = 5 Z_{\odot}$SA, the agreement becomes  better (see Fig. \ref{fig:z5}). High metallicity yields higher $U$\ and smaller \nh\ if emission line ratios involving \civ\ are considered. This reflect the increase in abudance of Si and Al with respect to C, and the fact  the \siii, \siiii, \aliii\  lines are emitted at  lower ionization than \civ.  In the case of sources like J02390-0038 and J20497-0554, the discrepancy of the crossing lines might indicate $Z_{\odot}\la Z \la 5 Z_{\odot}$, more than the extreme enrichment like the one assumed in 5ZSA, as also suggested by the \civ/\siiv\ ratio.   

 In Fig. \ref{fig:z5} and \ref{fig:z5corr} the contour plots are shown for the case 5ZSA. We consider uncorrected and corrected line ratios as two independent cases.   The halftone bands also show the importance of accurate \siii\ measurements to infer unambiguous constrain on metallicity (and, by extension, to estimate \nh\ and $U$\ independently).   The product \nh$U$\ is however much less affected than \nh\ and $U$\ individually (Tab. \ref{tab:neu}). 
 
A first \rb\ and \mbh\ estimate can be  obtained considering only the \siii/\siiii\ and \aliii/\siiii\ ratios. We conclude that the effect of scaling the metallicity up to $Z=5 Z_{\odot}$\ is within the uncertainty of the method, and particularly small if the ratios \siii/\siiii, \aliii/\siiii, and \siiv/\siiii\ are considered to compute \nh\ and $U$. It is significant if strong enrichment of Al and Si over C occurs. A more refined approach   could exploit the dependence of   \aliii/\civ\ (and \siiii/\civ\ and \siiv/\civ) on $Z$\ to build a 3D diagram where \nh, $U$,  $Z$\ along with Si-Al enrichment can be determined independently. 

\subsection{Determining the best estimate of \nh, $U$}

{\bf There are two sets of line ratios for each object: one comes from the {\sc specfit} results, and the other is the one computed after correcting the {\sc specfit} results because of low density contributions. Solutions with corrected value usually point toward very high density and low ionization, predicting \siii\ emission even twice as strong as \siiii. However, increasing the metallicity for corrected ratios leads to better agreement among lines and more reasonable \nh, $U$\ values, while leaving the product fairly unaffected. Especially the assumption of Silicon - Aluminium enrichment improves the concordance in the intriguing cases of SDSS J12014+0116, J01225+1339, J02287+0002 (using $z_{oi}$), the extreme objects and the two BAL QSOs in our sample. However, apart from the case of SDSS J12014+0116, the enrichment is probably excessive.  From the discussion above the independent determination of  \nh\ and  $U$\ seems possible only if metallicity is at least roughly known.  We exclude metallicity cases where we find a sizeable disagreement in the crossing points (with the exceptions of ratios involving \siii: the ionization level can be estimated in a $Z$-independent way using the \siiv/\siii\ and the \siiv).  We consider each individual source with line intensity before and after correction to obtain two independent sets of product \nh $U$\ values. As mentioned, changing metallicity is not affecting  the product \nh $U$\ as much as \nh\ and $U$\ individually. In case of concordance of the crossing points  and of high accuracy in the \siii\ ratio, \nh\ $U$, and $Z$ can be considered independently determined. In Table \ref{tab:neu} we indicate the values that are deemed most appropriate. }

\section{A  Photoionization Method to Compute the Broad Line Region  Distance and the Black Hole Mass.}
\label{rblr}

The distance of the broad line region (\rb) from the central continuum source and the black hole mass (\mbh) are key parameters that let us understand the dynamics of the gas in the emitting region and the quasar behavior and evolution. In this work we will use a  method based on the determination of \nh\ and $U$\  to compute \rb.  Eq. \ref{eq:u} can be rewritten as
\begin{equation}
r_{BLR} = \left[ \frac {\int_{\nu_0}^{+\infty}  \frac{L_\nu} {h\nu} d\nu} {4\pi Un_\mathrm{H} c} \right]^{1/2}
\end{equation}
and also as
\begin{equation}
r_{BLR} = \frac 1{h^{1/2} c} (U n_\mathrm{H})^{-1/2} \left( \int_{0}^{\lambda_{Ly}} f_\lambda \lambda d\lambda \right) ^{1/2} d_p \label{eq:rblr1}
\end{equation}
where $h$ is the Plank constant, $c$ is the light speed, $d_\mathrm{p}$ is the proper distance. The integral is carried out from the Lyman limit to the shortest wavelengths on the {\em rest frame} specific flux $f_{\lambda}$. For the integral we will use two Spectral Energy Distributions (SEDs): one described by Mathews \& Ferland (1987) and one by Laor et al. (1997a), also reproduced in Fig. \ref{fig:Laor_Mathews}. 

{\bf Eq. \ref{eq:rblr1} becomes:

\begin{equation}
r_{\rm BLR} \approx 93 \cdot  (U n_{\mathrm e})_{10}^{-\frac{1}{2}} \cdot  f_{\lambda_0,-15}^\frac{1}{2} \tilde{Q}_{H,0.1}^\frac{1}{2} \zeta(z, 0.3, 0.7) ~~\mathrm{ld} 
\end{equation}

where $\tilde{Q}_{H} = \int_{0}^{\lambda_\mathrm{Ly}} \tilde{n}_\lambda \lambda d\lambda$, and $\zeta(z, 0.3, 0.7)$\ is  an interpolation function for $d_{\mathrm p}$\ as a function of redshift.

$\tilde{Q}_{H}$\ is { 0.00963} cm\AA\  in the case the continuum of {Laor et al. (1997) is considered; $\tilde{Q}_{H} \approx$0.02181} cm\AA\ for Mathews~ \& Ferland (1987). We use their average value, since the derived $U$\ and \nh\ are not sensitive to the two different shapes to a first approximation.\footnote{Since the Laor et al. (1997) continuum produces a fewer ionizing photons, the same value of $U$ is obtained at a smaller distance. } 
}

Knowing \rb\ we can calculate the $M_\mathrm{BH}$ assuming virial motions of the gas
\begin{equation}
\label{eq:vir}
M_\mathrm{BH} = f \frac{\Delta v^2 r_\mathrm{BLR}}G.
\end{equation}
or,
\begin{equation}
M_\mathrm{BH} = \frac 3{4G} f_{0.75} (FWHM)^2 r_{BLR} 
\end{equation}
{\bf with the geometry term $f \approx 0.75$, corresponding to $f_{0.75} \approx 1.0$ \  (Graham et al. 2011, see also Onken et al. 2004 and Woo et al. 2010)}. Collin et al. (2006)  suggest that $f$\ is significantly different for Pop.A and B sources; we do not consider here their important result for the sake of comparison with previous work (\S \ref{vp06}). The resulting \rb\ and \mbh\ are reported in Columns 7 to 14 of Table \ref{tab:r_m}. Errors are at 2$\sigma$\ confidence level and have been computed propagating quadratically the major sources of uncertainty. More precisely, in addition to the error on $\log$(\nh $U$), the \rb\ determination is affected by the uncertainty in the spectrophotometry (specific fluxes of Col. 3 of Tab. \ref{tab:r_m}), and errors on the shape of the ionizing continuum. The two SEDs that we assumed as extreme yield a difference in ionizing photons of a factor 2.2. At a 2$\sigma$\ confidence level this corresponds to an uncertainty of $\pm$0.065 in logarithm. An additional source of uncertainty affects \mbh\ due to the FWHM determination. Errors on FWHM are  quadratically added to the uncertainty on \rb\ in the values reported as $\log$\mbh\ errors.

\section{Discussion} 
\label{sec:discussion}

\subsection{Previous work}
\label{sec:previous_work}

There have been several studies aimed at computing \rb\ and  \mbh. 
A direct measure of \rb\ through reverberation mapping  requires an enormous amount of observational effort and has only been applied to a relatively small number of quasars:  slightly less than 50  objects with $z \la 0.4$\ (Kaspi et al. 2000, 2005, Peterson et al. 2004; Bentz et al. 2010).  A second way to measure \rb\ uses a less direct method. Kaspi et al. (2000, 2005) and Bentz et al. (2009) used reverberation mapping results to find, in an empirical way, a relationship between \rb\ and the optical continuum luminosity at 5100\AA, 
\begin{equation}
\label{eq:r_l}
r_{\mathrm{BLR}} \propto L^\alpha
\end{equation}
with $\alpha \approx 0.52$.  Vestergaard \& Peterson (2006) obtained a similar result for the optical continuum luminosity with an $\alpha \approx 0.50$ and for the UV continuum at 1350\AA, $\alpha \approx 0.53$. These relationships have been used to compute the \rb\ not only for nearby objects, but also for  high redshift, high luminosity objects. There are other works that use single epoch spectra and the continuum at 3000\AA, obtaining an $\alpha \approx 0.47$ (McLure \& Jarvis 2002).

We can rewrite Eq. \ref{eq:vir} as 
\begin{equation}
M_{\mathrm{BH}} \propto f \frac{\mathrm{FWHM}^2 L^\alpha}{G}.
\end{equation}

\hb\ is a low ionization strong line  whose FWHM has been widely used to determine the \mbh\ for objects mainly up to $z\la 0.9$; above this limit IR spectrometers and large telescopes are needed to cover the redshifted line. For distant objects ($z\sim2$), an alternative is to use  \civ, a high ionization line emitted in the UV. However, this line should be used with caution  because  the line is often blueshifted. This  means that at least part of this line is likely emitted in an outflow (Sulentic et al. 2007; Richards et al. 2010). Thus the estimation of  \mbh\ using FWHM(\civ) tend to be systematically higher than those using FWHM(\hb), especially for objects of Population A.

\subsection{Comparison with Vestergaard and Peterson (2006)}
\label{vp06}

Vestergaard \& Peterson (2006) used the relationship $r_{\mathrm{BLR}} \propto L^{0.53}$ \ to obtain the following formula that relates \mbh\ to the FWHM(\civ$_{}$) and the continuum luminosity at 1350\AA:
\begin{equation}
\label{eq:vestergaard}
\log M_\mathrm{BH}(\mathrm{C\sc{\i v}}) = \log \left\{ \left[ \frac {\mathrm{FWHM}(\mathrm{C \sc{\i v}})} {1000 \> \mathrm{km} \> \mathrm{s}^{-1}} \right]^2 \left[ \frac{\lambda L_\lambda(1350 \mathrm{\AA})} {10^{44} \> \mathrm{ergs} \> \mathrm{s}^{-1}} \right]^{0.53} \right\} + (6.66 \pm 0.01) - s_\mathrm{f}.
\end{equation}

{\bf The scale factor $s_\mathrm{f} \approx -0.27$\ sets the masses to the $f$\ value obtained by Graham et al. (2011)}.  In Cols. 11 and 15 of Table \ref{tab:r_m} and in Fig. \ref{fig:MBHcomparison} we compare our \mbh\ results with those using Eq. \ref{eq:vestergaard}. We do not apply corrections for radiation-pressure effects that are likely relevant especially for objects radiating at large Eddington ratio (Netzer 2009; Netzer \& Marziani 2010). The difference between this computation and the one reported in Sulentic et al. (2007) is that in the latter work the blueshifted component was not separated from the broad component of  \civ\ just to show how larger values of FWHM(\civ) yielded \mbh\ much larger than the ones derived from FWHM(\hb) in Pop. A objects. 

We compare the masses obtained using our photoionization method with those of Vestergaard \& Peterson (2006) in Fig. \ref{fig:MBHcomparison}. We use the FWHM of the BC as an estimator of the virial line broadening. {\bf  From our results we can see that the masses agree within less than 1$\sigma$\ uncertainty in the luminosity correlation (0.33). There is a systematic offset of 0.17$\pm$0.10 if uncorrected ratios are used.  The mass values obtained after correction for low density gas  are systematically lower. This happens because the correction increases the product $U$\nh, lowering \rb\ and hence \mbh. The \mbh\ obtained after corrections are within the error bars. The systematic offset is then $0.13 \pm 0.12$.  It is important to consider that the computed correction is in many ways a maximum correction. \ciii\ emission is assumed to have the same FWHM of \siiii\ while it could be significantly narrower; in addition part of the \ciii\ emission could be due to \feiii\ $\lambda$ 1914. In many sources of Pop. A the correction could be ignored altogether.  
} The present results indicate that the Kaspi et al. (2000) relationships can be extended to be used in high redshift objects (or at least until z$\sim$3) if the FWHM of the core broad line region can be well determined and measured. In order to do this, we need:

\begin{itemize}
\item [$\bullet$] Spectra with S/N high enough to see the profile shape that allows decomposition of the \civ\ line, especially to separate the blue component from the broad core;

\item [$\bullet$] to follow the methodological considerations explained in \S \ref{sec:data_analysis} .
\end{itemize}

Fig. \ref{fig:MBHcomparison} should be looked at with two cautions. First, the correlation is dominated by the luminosity dependence of \rb, used to compute \mbh\ in both cases. Second, the spread of \mbh\ values is small, less than one order of magnitude (and most objects have statistically indistinguishable masses).  Our estimated error bars are however smaller compared to the spread expected on the basis of the \rb-$L$\ correlation which is, according to  Vestergaard \& Peterson (2006), $\pm$0.66 at a 2$\sigma$ confidence level. The two shaded bands of Fig. \ref{fig:MBHcomparison} limit the region where we can expect to find data points on the basis of the  \rb-$L$\  correlation. Clearly, a proper interpretation of the \civ\ profile may help to reduce the scatter. In any case, our method should provide  \mbh\  estimates with somewhat lower uncertainty. It is interesting to note that SDSS J12014+0116 appear at the largest \mbh, and the agreement with the \rb-$L$\ is very good.  The \rb\ value of 3C 390.3  we obtain is fairly uncertain due to the low S/N. {\bf The size computed in the present paper is  larger than the reverberation mapping derived \rb\ for \hb\ and \civ, although consistent with the value derived from \lya. 3C 390.3 has the unusual property of having a response time longer in \civ\ than in \hb, although the large error bars do not exclude that the two lines respond with similar time. This behavior might be related to different physical conditions found for this object.}

\subsection{The LIL-BLR}

While the existence of high density and low ionization has been invoked since long to explain \feii\  emission (especially by S. Collin and collaborators, as mentioned in the introduction), we have  provided additional evidence that high density and low ionization are indeed diagnosed from emission lines other than \feii\  blends, and that conditions are the one producing most line emission in extreme Pop. A quasars. Last but not least, both $U$\ and \nh\ can be observationally determined with reasonable accuracy from the diagnostic ratios \siii/\siiii, \siiv/\siiii, \aliii/\siiii, and  \civ/(\aliii\ or \siiii). Metallicity can be also constrained from the previous ratios as well as from \civ/\siiv. These ratios are pretty well defined, while estimating  \ciii\  and \feiii\l1914 relative contribution is not relevant to our method (with the exceptions of   sources like 3C 390.3). This low ionization BLR (or LIL BLR) has very similar properties to the O{\sc i}  and Ca{\sc ii} emitting region identified by Matsuoka et al. (2008). The LIL-BLR seems to be present in the vast majority of quasars, probably all the ones with significant \feii\ emission (Marziani et al. 2010). The low values of $U$\ could be a consequence of the  high density rather than of a far away location of the emitting region. The assumption of a single well defined value of $U$\ and \nh\ is probably an idealization even for the LIL-BLR taken alone; however the convergence of emission line ratios toward a well defined point, along with the ability to qualitatively explain \feii\ and most \hb\ emission, indicate that the LIL-BLR might be a region with a small range  of low $U$\ and high \nh.

Do density and ionization parameter in the LIL BLR converge to a  single well defined value with a small dispersion? A tentative answer comes from the values reported in Tab. \ref{tab:neu}. Excluding 3C390.3, the range $U$ and \nh\ span is not very large (even considering changes in metallicity): less than one order of magnitude around $\log $ \nh $\sim$ 12.5, and ionization parameter $\sim - 2.75$, and the spread is not much larger than the uncertainties in the individual measurements. The product $U\cdot$\nh\ seems to be fairly  stable. We have $<\log (U \cdot$\nh)$> \approx 9.5$, with a sample dispersion of 0.15 (excluding 3C 390.3). We applied the same method to 14 low-$z$\ quasars (Negrete et al. 2010), and we obtain $<\log (U \cdot$\nh)$> \approx 9.7$, with a dispersion of 0.3.  Assuming $\log (U \cdot$\nh) $\approx 9.6$  could be a good approach to estimate \mbh\ if  elaborate measurements on \civ\ and \ciii\ are not possible, and only a rough estimate of the \siiii\ BC FWHM is available. On the other hand, there is still a major effect of measurement errors on the uncertainty derived for $\log (U \cdot$\nh); instrumental improvements may lead to a significant appreciation of object-by-object diversity. 

The $U\cdot$\nh\ values reported in Table \ref{tab:neu} are not very far from the average value obtained by Padovani and Rafanelli (1988). This is not surprising since the spectra at low and high $z$\ seems to show the same diversity, classified through the E1 sequence and the Pop. A/Pop. B distinction. In other words, NLSy1-like sources whose spectrum is similar to I Zw1 appear to be present at  high redshift, meaning that the LIL-BLR  remains strong and prominent over a wide range of redshifts. A second reason is that Padovani and Rafanelli (1988)  considered \hb. Emission of \hb\ can be significant under a much wider range of $U$\ and \nh; however, the \aliii, \mgii, \feii\ emitting region should be also a strong producer of \hb. This region with well-defined physical conditions is expected to be the emitter of the core of \hb, i.e., the part of the line responding more strongly to continuum changes.

\subsubsection{Verification on EW and Line Luminosity} 

Considering that we have  very low ionization parameter values, a legitimate question is whether we have a sufficient number of photons to explain the EW and luminosity of the emission lines. We  made a preliminary check of consistency for the EW from {\sc CLOUDY} simulations. A second, a-posteriori test was to consider the predicted line luminosity assuming the actual luminosity of the quasar, the density and the distance \rb\ derived from our method, and  spherical geometry. 

A remarkable property common to all spectra is the  low EW of the emission lines. The extreme source  SDSS J120144.36+011611.6 has total $W$(\civ) $\approx$ 19 \AA, which becomes $\approx$  7 \AA\   if only the LIL component is considered. The whole 1900 \AA\ blend has  $ W \approx$ 20 \AA; the individual  EWs of \siiii\ and \aliii\ are just a few \AA. Similar considerations apply to the other high-$z$\ quasars,  especially for Pop. A sources. In this case we have $W$(\civ)$\approx 20 - 30$ \AA, the equivalent width of the whole 1900 \AA\  $30 - 40$\AA\ including the uncertain contribution of \feiii\ and \ciii.  This has the important implication that the ionization parameter cannot be very large. From our simulations, we deduce that the LIL-BLR $\log U $\ is always $ \la -2.0$. On the other hand, toward the low $U$, high density limit emission lines tend to disappear altogether. The predicted EWs become too low to account for the observed EW with a covering factor $f_{\mathrm c}\la 0.5$\ if $\log U \la -3$. The  values obtained after correction $\log U \approx -3.25$\ are still possible within the condition of $f_{\mathrm c}\la 0.5$\ since the \civ\ EW is greatly diminished.  Cases of very low $U$\ are rather indicative of strong metal enhancement than of extremely low ionization level (see \S \ref{metals}).  If we take  $\log$ \nh $\approx$12.5, $\log U   = -2.75$, the predicted equivalent width of is $W$(\civ)$\approx$37 \AA\ for  a covering factor 0.5, close to the largest values observed in our quasars. However, The \civ\ EW of   SDSS J12014+0116 can be accounted  for by the values of $\log  U \approx -3.00$ and  $\log$ \nh $\sim$ 12.75 -- 12.50. 

If we model the BLR with a spherical geometry where the emitting gas covers a fraction $f_{\mathrm c}$ of the continuum, {\sc CLOUDY} computations confirm that the line luminosity can be accounted for. The luminosity at 1700 \AA\ of J00103-0037 is  $\log \lambda L_{\lambda} \approx$ 46.6; the predicted luminosity of \civ\ is $\log L$(\civ) $\approx$ 44.5, under the assumption of $Z = Z_{\odot}$, $f_{\mathrm c} = 0.1$, and Mathews \& Ferland continuum shape.  The observed \civ\ line luminosity of  J00103-0037, $\approx 9\cdot 10^{44}$ ~\ergss, is obtained with $f_{\mathrm c} \approx 0.3$.

\subsubsection{Analogy with $\eta$ Carin\ae} 

The physical conditions we envisage for the BC of quasars find a correspondence in the so-called ÒWeigelt blobsÓ of $\eta$ Carin\ae, located in the equatorial plane of the system, perpendicular to the symmetry axis of the bipolar lobes forming the ÒhomunculusÓ nebula (cf. Marziani et al. 2010). Unlike the gas of the bipolar lobes, predominantly shock heated, the Weigelt blobs are believed to be dense gas photoionized by the radiation associated to the central, massive star and to a possible companion (e.g., Johansson et al. 2000; Davidson 2005). The spectrum of the Weigelt blobs shows very weak  \ciii\  along with a prominent line at \l1914, ascribed to the $z^7 P_3^0 \rightarrow a^7 S_3$ \feiii\  transition. The line appears very strong because the upper level is populated by \lya\ fluorescence. This very same process is expected to be present also in quasars. Indeed, in I Zw 1, where lines are narrow, and in SDSS J120144.36+011611.6  the peak emission at around 1914 \AA\ is actually visible. The amount of \lya\ pumping to the upper level (z$^{7}P^{\mathrm o}_{3}$) of the UV 34  cannot be estimated through the standard edition of {\sc CLOUDY} (the relevant levels of the UV 34 multiplets of Fe$^{+2}$ ion are not included). Additional photoionization computations including a suitable Fe$^{+2}$ model and line transfer should be considered. This is beyond the aim of the present study; we can conclude in a qualitative fashion that the spectrum of the $\eta$ Carin\ae\ blobs supports a view of the low-ionization part of the BLR that is not conventional: very high density gas, at very low ionization.

\section{Conclusions} 
\label{sec:conclusions}

In this paper we presented new observations of eight high redshift quasars. The spectra were meant to provide high S/N, moderate resolution data on which  the \civ, \siiii, \aliii, and \siii\ emission line profiles could be accurately analyzed. Line profile fits allowed us to isolate a specific component whose  intensity ratios were used to derive consistent values for electron density and ionization parameter. This line component (LIL BC) seems to be emitted predominantly by low ionization, high density gas {\bf in the majority of quasars studied thus far by us}. 

These results permitted us to compute the  product \nh $\cdot U$\ and hence the size of the Broad Line Region and the central black hole mass. The method 
described in this paper rests on the assumption of photoionization as the mechanism of gas heating; on the assumption of isotropic luminosity, and on line ratios predicted by {\sc cloudy} simulations.  The photoionization method explored in this paper offers an estimate of \rb\  for each quasar, with some advantages on the \rb\ valued derived from the luminosity-size correlation. The luminosity correlation suffers from large scatter and is simply extrapolated to very high luminosity without any support since there are, unfortunately, no conclusive results on reverberation of high luminosity quasars even if heroic efforts are underway (e.g., Trevese et al. 2007, Botti et al. 2010). We found that the black hole masses derived from the computed \rb\ and from the virial assumption are in good agreement with the ones derived from the luminosity-size relationship. Actually, Fig. \ref{fig:MBHcomparison} suggests  that we might have reduced the errors of the \mbh\ by a factor of two with respect to the expectation from the  Vestergaard \& Peterson (2006) relationship.

We repeat that our \mbh\ and \rb\ results are based on the product \nh$ \cdot U$\ and not on  values of \nh\ and of $U$\ taken separately. It seems that this product converges to two typical ranges of values, one of them associated to low-ionization, high density gas (the LIL-BLR).  For our \nh\ and $U$\ determinations we do not use ratio \ciii/\siiii\ except for 3C 390.3. As we discussed in \S \ref{sec:3c390}, this ratio  should not be considered at high density because  \ciii\  is collisionally quenced if \nh\ $ \ga 10^{10}$ \cm3. \ciii\  is produced in conditions that are very different from the ones  we found for the LIL-BLR. While the method can be applied to most quasars, the application seems to be especially straightforward to quasars whose spectrum is like SDSS J120144.36+011611.6 (if high metallicity is properly taken into account) or, at the other end, 3C390.3. In the first case we have dominance by the LIL-BLR, in the second case the LIL-BLR seems to be completely absent and physical conditions look radically different: high ionization and moderate density. An inspection of SDSS spectra covering both the 1900 \AA\ blend and \civ\ (up to $z\approx 3.5$) shows that SDSS J120144.36+011611.6 has many replicas at high redshift, accounting for at least a few percent of all quasars. These high-metallicity objects should be the first candidates to expand black hole mass computations to high redshift without relying on the \rb\ - $L$\ correlation. 

To apply the photoionization method in the most effective way, determining \nh\ and $U$\ with the lowest uncertainty, spectral data should be of moderate resolution ($\lambda/\Delta \lambda \sim 1000$) as well as of high S/N. If the \siii\ line can be measured in an accurate way, it would be possible to derive independent estimates of $U$, \nh, and $Z/Z_\odot$\  in most quasars.  

{\bf Especially the most extreme (in terms of \aliii\ strength) objects in bin A3 and A2 hold the promise to make possible an independent estimate of \nh, $U$, and metallicity.   Clearly, objects in bin A1 resembling their median spectrum are not well suited for an application of the method. Also, any source with \ciii/\siiii $>$ 1 is subject to a large correction. In light of the many uncertainty, an average value of the product $U n$\  (obtained from the objects of the other spectral types)  could be considered. 
}
{\bf Pop. B objects should not avoided entirely, especially whenever \siiii\ $\simgt$ \ciii\ after VBC removal.  }

{\bf The present exploratory analysis emphasized several sources of uncertainty. However, the parameter needed for \rb\ and \mbh\ computation, the product $U $\nh, seems to be fairly stable and well-defined. Even with an error of a 0.3 in logarithm, the square root will be subject to a 0.15 uncertainty in logarithm, much lower than the uncertainty associated with the \rb\-luminosity correlation. The large intrinsic spread of the correlation at low luminosity, its uncertain extrapolation at very high luminosity   make preferable a one-by-one determination based on  physical properties of an emitting region that remains similar to itself.  
}

\acknowledgments
A. Negrete and D.Dultzin acknowledge support form grant IN111610-3 PAPIIT, DGAPA UNAM.      Funding for the SDSS and SDSS-II has been provided by the Alfred P. Sloan Foundation, the Participating Institutions, the National Science Foundation, the U.S. Department of Energy, the National Aeronautics and Space Administration, the Japanese Monbukagakusho, the Max Planck Society, and the Higher Education Funding Council for England. The SDSS Web Site is http://www.sdss.org. The SDSS is managed by the Astrophysical Research Consortium for the Participating Institutions listed at the \href{http://www.sdss.org}{SDSS Web Site}.

\appendix

\section*{References}
\def\REF{\par\noindent\hangindent 20pt} 

\REF Aoki, K., Yoshida, M. 1999, ASPC, 162, 385
\REF Appenzeller, I., et al. 1998, The Messenger, 94, 1
\REF Bachev et al. 2004, ApJ, 617, 171
\REF Baldwin, Ferland, Korista and Verner, 1995, ApJ, 455, L119
\REF Baldwin J. A. et al. 1996, ApJ, 461, 682
\REF Baskin, A. \& Laor, A. 2005, MNRAS, 356, 1029
\REF Bentz, M. C., et al. 2010, ApJ, 716, 993
\REF Bevington, P.~R.\ 1969, Data reduction and error analysis for the physical sciences, New  York: McGraw-Hill, 1969
\REF Boroson, T.A. \& Green, R.F., 1992, ApJS, 80, 109
\REF Botti, I., Lira, P., Netzer H., Kaspi, S. 2010, IAU Symposium, 267, 198
\REF Bottorff M.C. \& Gary J.F. 2000, MNRAS, 316, 103
\REF Bruhweiler, F. \& Verner, E., 2008, ApJ, 675, 83
\REF Clavel, J., et al. 1991, ApJ, 366, 64
\REF Clayton, D.D., 1983, Principles of  Stellar Evolution \& Nucleosynthesis, Chicago:University of Chicago Press,  Ch. 7
\REF Collin-Souffrin S. et al. 1988, MNRAS, 232, 539
\REF Collin, S., et al. 2006, A\&A, 456, 75
\REF Constantin A. et al. 2002, ApJ, 565, 50
\REF Davidson, K. \& Netzer, H., 1979, Rev. Mod. Phys. 51, 715
\REF Davidson, K. 2005, ASPC, 332, 101
\REF Dumont, A. M. \& Mathez, G., 1981, A\&A, 102, 1
\REF Dumont, A. M, \& Collin-Souffrin, S., 1990 A\&A 229, 292
\REF Edl\'en \& Swings, 1942, ApJ, 95, 532
\REF Ekberg J.O. 1993, A\&AS, 101, 1
\REF Espey, B. R., Carswell, R. F., Bailey, J. A., Smith, M. G., \& Ward, M. J. 1989, ApJ, 342, 666
\REF Evans \& Koratkar 2004, ApJS, 150, 73
\REF Francis P.J. et al. 1991, ApJ, 373, 465
\REF Feibelman, W.A., and Aller, L.H., 1987, ApJ, 319, 407
\REF Feldman U. et al. 1992, ApJS, 81, 387 
\REF Ferland G. J. et al. 1998, PASP, 110, 761
\REF Gaskell, C. M. 1982, ApJ, 263, 79
\REF Gaskell, M. et al. 1999, ASPC, 175, 423
\REF Graham, A. W. et al. 2011, MNRAS, 48
\REF Hartig, G. F. \& Baldwin, J. A. 1986, ApJ, 302, 64
\REF Hu, C., et al. 2008, ApJ, 687, 78
\REF Johansson S. et al. 2000, A\&A, 361, 977
\REF Joly, M., 1987, A\&A, 184,33
\REF Kaspi et al. 2000, ApJ, 533, 631
\REF Kaspi et al. 2005, ApJ, 629, 61
\REF Kriss G.A., 1994, ASPC, 61, 437
\REF Korista K. et al. 1997, ApJS, 108, 401
\REF Kuraszkiewicz, J. K. et al. 2002 ApJS, 143,257 
\REF Laor A. et al. 1994, ApJ, 420, 110
\REF Laor A. et al. 1997a, ApJ, 477, 93
\REF Laor A. et al. 1997b, ApJ, 489, 656
\REF Leitherer, C., et al. 1999, ApJS, 123, 2
\REF McLure \& Jarvis 2002, MNRAS, 337, 109
\REF Marziani, P., Sulentic, J. W., Dultzin-Hacyan, D., Calvani, M., \& Moles, M. 1996, ApJS, 104, 37
\REF Marziani P. et al. 2003, ApJS, 145, 199
\REF Marziani P. et al. 2003b, MNRAS, 345, 1133
\REF Marziani P. et al. 2008, RMAA serie de conferencias, 32, 69
\REF Marziani P. et al. 2009, A\&A, 495, 83
\REF Marziani P. et al. 2010, MNRAS, arXiv:1007.3187v1
\REF Matsuoka, Y., Kawara, K., \& Oyabu, S.\ 2008, \apj, 673, 62
\REF Mathews \& Ferland 1987, ApJ, 323, 456
\REF Metzroth, K. G., Onken, C. A., Peterson B. M. 2006, ApJ, 647, 901
\REFÊNegrete, C. A., Dultzin, D., Marziani, P., Sulentic J. W. 2010, ApJL, submitted
\REF Netzer, H., \& Trahktenbrot, B. 2007, ApJ, 654, 754
\REF Netzer, H. 2009, ApJ, 695, 793
\REF Netzer, H. \& Marziani, P. 2010, ApJ, in press, arXiv:1006.3553v2  
\REF Onken, C. A., Peterson, B. M. 2002, ApJ, 572, 746
\REFÊ Onken, C. A., Ferrarese, L., Merritt, D., Peterson, B. M., Pogge, R. W., Vestergaard, M., \& Wandel, A. 2004, ApJ, 615, 645
\REF  Osterbrock, D.~E., \& Ferland, G.~J.\ 2006, Astrophysics of gaseous nebulae and active galactic nuclei, 2nd.~ed.~by D.E.~Osterbrock and G.J.~Ferland.~Sausalito, CA: University Science Books, 2006,  
\REF Padovani, P., 1988, A\&A, 192, 9
\REF Padovani, P. \& Rafanelli, P. 1988, A\&A, 205, 53
\REF Padovani, P., Burg, R.I., Edelson, R.A., 1990, ApJ, 353, 438
\REF Peterson B. M. et al. 2004, ApJ, 613, 682. 
\REF Punsly, B. 2010, ApJ, 713, 232
\REF Rees M.J. 1987, MNRAS, 228, 47
\REF Richards, G. T., et al. 2010, arXiv preprint 2010arXiv1011.2282R
\REF Richards, G. T., Vanden Berk, D. E., Reichard, T. A., Hall, P. B., Schneider, D. P., SubbaRao, M., Thakar, A. R., \& York, D. G. 2002, AJ, 124, 1
\REF Sigut T.A. et al. 2004, ApJ, 611, 81
\REF Sulentic J. W. et al. 2000, ApJ, 536, L5
\REF Sulentic J. W. et al. 2001, AIP Conference Proceedings, 599,  963
\REF Sulentic J. W. et al. 2002, ApJ, 566, 71
\REF Sulentic J. W. et al. 2006, RMxAA, 42, 23
\REF Sulentic J. W. et al. 2006b, A\&A, 456, 929
\REF Sulentic J. W. et al. 2007, ApJ, 666, 757.
\REF Trevese, D., Paris, D., Stirpe, G. M., Vagnetti, F., and  Zitelli, V. 2007, A\&A, 470, 491
\REF Tytler, D., \& Fan, X.-M. 1992, ApJS, 79, 1
\REF Vanden Berk et al. 2001, AJ, 122, 549
\REF Verner E. et al. 1999, ApJS, 120, 101
\REF Verner E. et al. 2003, ApJ, 592, 59
\REF Verner E. et al. 2004, ApJ, 611,780
\REF Vestergaard M. and Peterson B.M. 2006, ApJ, 641, 689.
\REF Vestergaard \& Wilkes, 2001, ApJS, 134, 1
\REF Wandel, A., Peterson, B. M., Malkan, M. 1999, ApJ, 526, 579
\REF Wills, B. J., et al. 1999, ApJ, 515, L53
\REF Woo, J.-H., et al. 2010, ApJ, 716, 269
\REF Woosley,  S. E., Weaver, T. A. 1995, ApJS, 101, 181
\REF Zamfir, S., Sulentic, J. W., Marziani, P., \& Dultzin, D. 2010, MNRAS, 403, 1759

\clearpage

\begin{deluxetable}{cccccccl}
\tabletypesize{\scriptsize}
\tablecaption{Line Components in the $\lambda$1900 blend. \label{tab:lines}}
\tablewidth{0pt}
\tablehead{
\colhead{Ion} & \colhead{$\lambda$} & \colhead{$X$} &\colhead{$E_l - E_u$} & 
\colhead{Transition} & \colhead{$A_{ki}$} & \colhead{$n_\mathrm{c}$} &\colhead{Note}\\
\colhead{} & \colhead{\AA} & \colhead{eV} &\colhead{eV} & 
\colhead{} & \colhead{s$^{-1}$} & \colhead{cm$^{-3}$} &\colhead{}
}
\startdata
Si II	&	1808.00	&	8.15		&	0.000	-	6.857	&	${}^2 D^o_{3/2} \rightarrow {}^2 P_{1/2}$	&	$2.54 \cdot 10^6$	&	\nodata	& 1 	\\
Si II	&	1816.92	&	8.15		&	0.036	-	6.859	&	${}^2 D^o_{5/2} \rightarrow {}^2 P_{3/2}$	&	$2.65 \cdot 10^6$	&	\nodata	& 1		\\
Al III	&	1854.716	&	18.83	&	0.000	-	6.685	&	${}^2 P^o_{3/2} \rightarrow {}^2 S_{1/2}$	&	$5.40 \cdot 10^8$	&	\nodata	& 1		\\
Al III	&	1862.790	&	18.83	&	0.000	-	6.656	&	${}^2 P^o_{1/2} \rightarrow {}^2 S_{1/2}$	&	$5.33 \cdot 10^8$	&	\nodata	& 1		\\

\[[Si III]&	1882.7	&	16.34	&	0.000	-	6.585	&	${}^3 P^o_2 \rightarrow {}^1 S_0$		&	0.012			&	$6.4 \cdot 10^{4}$	& 1,2,3	\\
Si III]	&	1892.03	&	16.34	&	0.000	-	6.553	&	${}^3 P^o_1 \rightarrow {}^1 S_0$		&	16700			&	$2.1 \cdot 10^{11}$	& 1,4,5	\\

\[[C III]&	1906.7	&	24.38	&	0.000	-	6.502	&	${}^3 P^o_2 \rightarrow {}^1 S_0$		&	0.0052			&	$7.7 \cdot 10^{4}$	&  1,2,6 	\\
C III]	&	1908.734	&	24.38	&	0.000	-	6.495	&	${}^3 P^o_1 \rightarrow {}^1 S_0$		&	114				&	$1.4 \cdot 10^{10}$	&  1,2,4,5	\\

Fe III	&	1914.066	&	16.18	&	3.727	-	10.200	&	$z^7 P^o_3 \rightarrow a^7 S_3$		&	$6.6 \cdot 10^8$	&	   \nodata	&	7	\\
\enddata
\tablecomments{All wavelengths are in vacuum. (1)  Ralchenko, Yu., Kramida, A.E., Reader, J.,  and NIST ASD Team (2008). NIST Atomic Spectra Database (version 3.1.5). Available at: http://physics.nist.gov/asd3.  2: Feibelman \& Aller (1987). 3: $n_{\mathrm{c}}$\ computed following Shaw \& Dufour (1995).  4:  Morton (1991).  5: Feldman (1992).  6:  Zheng (1988).  7: Wavelength and $A_{ki}$\ from Ekberg (1993), energy levels from Edl\'en and Swings (1942). }
\end{deluxetable}

\begin{deluxetable}{ccccccccccc}
\tabletypesize{\scriptsize}
\tablecaption{Basic Properties of Sources and Log of Observations.\label{tab:obs}}
\tablewidth{0pt}
\tablehead{
\colhead{Object name} &\colhead{$\rm m_B$} & \colhead{$z$} & \colhead{Line} & \colhead{$M_{\mathrm B}$} &
\colhead{Flux 6cm (mJy)} & \colhead{Date} & \colhead{DIT} & \colhead{N$_{\rm exp}$} & \colhead{Airmass} & \colhead{S/N} \\
\colhead{(1)}&\colhead{(2)}&\colhead{(3)}&\colhead{(4)}&\colhead{(5)}&\colhead{(6)}&
\colhead{(7)}& \colhead{(8)}&\colhead{(9)}&\colhead{(10)}&\colhead{(11)}
}
\startdata
J00103-0037	&	18.39	&	3.1546	&	1	&	-25.68	&	0.40	&	2006-11-08	&	1139	&	3	&	1.16, 1.13, 1.11	&	60	\\
J00521-1108	&	18.70	&	3.2364	&	2	&	-25.39	&	0.43 	&	2007-01-01	&	1199	&	3	&	1.15, 1.21, 1.29	&	41	\\
J01225+1339	&	18.24	&	3.0511	&	1	&	-25.80	&	*	&	2006-11-08	&	1259	&	2	&	1.36, 1.32	&	92	\\
J02287+0002	&	18.20	&	2.7282	&	1	&	-25.72	&	0.35 	&	2006-12-16	&	1259	&	2	&	1.10, 1.12	&	67	\\
J02390-0038	&	18.68	&	3.0675	&	1	&	-25.36	&	0.43	&	2006-11-07	&	1199	&	3	&	1.35, 1.46, 1.60	&	57	\\
J03036-0023	&	17.65	&	3.2319	&	1	&	-26.44	&	0.34	&	2006-12-16	&	1259	&	2	&	1.11, 1.14	&	88	\\
J20497-0554	&	18.29	&	3.1979	&	1	&	-25.79	&	*	&	2006-11-04	&	1259	&	2	&	1.52, 1.70	&	54	\\
J23509-0052	&	18.67	&	3.0305	&	1	&	-25.36	&	0.41	&	2006-11-07	&	1199	&	3	&	1.10, 1.11, 1.14	&	62	\\
\enddata
\tablenotetext{*}{Not in FIRST}
\end{deluxetable}

\begin{deluxetable}{lcccccccccccc}
\setlength{\tabcolsep}{1.5pt}
\tablecaption{Line Fluxes $^{a}$ \label{tab:fluxes}}
\tabletypesize{\tiny}
\rotate
\tablewidth{0pt}
\tablehead{
 & \multicolumn{2}{c}{\ciii}  & &&  &   \multicolumn{3}{c}{\civ}  & &  \multicolumn{3}{c}{\siiv}  \\
\cline{2-3}
\cline{7-9}
\cline{11-13} 
\colhead{Object} & \colhead{$_{BC}$} & \colhead{$_{VBC}$} & \colhead{\siiii} & \colhead{\aliii}& \colhead{\siii} &  \colhead{$_{BC}$} & \colhead{$_{blue}$} & \colhead{$_{VBC}$} & \colhead{} & \colhead{$_{BC}$} & \colhead{$_{blue}$} & \colhead{$_{VBC}$} 
}
\startdata
J00103-0037	&	4.99	$\pm$	2.11	&	1.18	$\pm$	1.38	* &	2.91	$\pm$	1.03	&	1.99	$\pm$	0.81	&	1.13	$\pm$	0.82	:	&	14.25	$\pm$	8.34	&	6.32	$\pm$	1.80	&	8.71	$\pm$	8.24	&&	4.27	$\pm$	2.38	&	0.78	$\pm$	1.52	&	0.54	$\pm$	0.89	\\
J00521-1108	&	3.08	$\pm$	0.36	&	0.01	$\pm$	0.12	*&	2.77	$\pm$	0.66	&	1.55	$\pm$	0.85	&	1.06	$\pm$	0.90		&	10.89	$\pm$	2.37	&	1.40	$\pm$	2.23	&	8.27	$\pm$	2.63	&&		\ldots		&		\ldots		&		\ldots		\\
J01225+1339	&	10.71	$\pm$	1.35	&		\ldots		&	8.26	$\pm$	1.33	&	4.35	$\pm$	2.09	&	1.01	$\pm$	0.98		&	22.73	$\pm$	6.04	&	14.21	$\pm$	1.52	&		\ldots		&&	11.21	$\pm$	2.93	&	6.66	$\pm$	1.93	&		\ldots		\\
J02287+0002 $^{(1)}$	&	5.28	$\pm$	1.56	&	0.01	$\pm$	0.02	* &	4.70	$\pm$	2.01	&	2.29	$\pm$	1.13	&	0.98	$\pm$	0.79		&	7.64	$\pm$	5.00	&	7.66	$\pm$	2.32	&	0.77	$\pm$	1.63	* &&	4.96	$\pm$	1.04	&	1.35	$\pm$	0.80	&	0.43	$\pm$	0.56	\\
J02287+0002 $^{(2)}$	&	6.86	$\pm$	1.56	&	0.02	$\pm$	0.02	* &	2.78	$\pm$	2.01	&	1.77	$\pm$	1.13	&	0.94	$\pm$	0.79		&	11.48	$\pm$	5.00	&	2.99	$\pm$	2.32	&	0.64	$\pm$	1.63	* &&	4.88	$\pm$	1.04	&	1.24	$\pm$	0.80	&	0.57	$\pm$	0.56	\\
J02390-0038	&	3.57	$\pm$	0.78	&	1.26	$\pm$	0.88	&	3.50	$\pm$	0.40	&	2.41	$\pm$	0.64	&	0.90	$\pm$	0.83		&	7.51	$\pm$	1.25	&	7.92	$\pm$	1.37	&	2.04	$\pm$	1.52	&&	2.64	$\pm$	0.53	&	1.49	$\pm$	0.81	&	0.23	$\pm$	0.39	\\
J03036-0023	&	13.24	$\pm$	1.09	&		\ldots		&	11.82	$\pm$	1.21	&	5.17	$\pm$	1.48	&	1.53	$\pm$	1.16	:	&	29.46	$\pm$	3.54	&	20.60	$\pm$	3.34	&		\ldots		&&	11.13	$\pm$	1.86	&	7.37	$\pm$	4.12	&		\ldots		\\
J20497-0554	&	8.04	$\pm$	1.09	&		\ldots		&	7.43	$\pm$	0.56	&	3.01	$\pm$	1.25	&	1.52	$\pm$	1.51	:	&	18.39	$\pm$	2.17	&	9.23	$\pm$	2.24	&		\ldots		&&	6.65	$\pm$	2.51	&	1.95	$\pm$	1.06	&		\ldots		\\
J23509-0052	&	5.24	$\pm$	1.21	&		\ldots		&	4.61	$\pm$	1.62	&	1.50	$\pm$	0.50	&	0.40	$\pm$	0.36		&	9.24	$\pm$	1.17	&	7.80	$\pm$	2.36	&		\ldots		&&	3.61	$\pm$	1.49	&	2.93	$\pm$	1.34	&		\ldots		\\
\cline{1-13}  																																														
Extreme Objects	&				&				&				&				&					&				&				&				&&				&				&				\\
J12014+01161	&	5.52	$\pm$	1.07	&		\ldots		&	16.16	$\pm$	5.90	&	14.03	$\pm$	2.54	&	4.90	$\pm$	2.95		&	21.18	$\pm$	12.48	&	29.37	$\pm$	9.98	&		\ldots		&&	18.45	$\pm$	6.55	&	11.81	$\pm$	9.08	&		\ldots		\\
3C 390.3	&	3.74	$\pm$	1.15	&	4.30	$\pm$	0.68	&	2.32	$\pm$	0.43	&	0.23	$\pm$	0.32	* &	0.40	$\pm$	0.48		* &	30.04	$\pm$	4.99	&		\ldots		&	56.44	$\pm$	10.66	&&		\ldots		&		\ldots		&		\ldots		\\
\enddata
\tablecomments{(a) Units are $10^{-14}$ ergs s$^{-1}$ cm$^{-2}$ \AA$^{-1}$.  (1) Considering $z_{OI\lambda1304}$. (2) Considering $z_{CIII]\lambda1909}$. (:) \siii\ approximated values due the line is affected by telluric absorptions (see Fig. \ref{fig:sample_ab}). We do not measure \siiv\ for J00521-1108 and 3c390.3 because they have low S/N. (*) Consistent with no emission.}
\end{deluxetable}

\begin{deluxetable}{lcccc}
\tablecaption{Weak lines around \civ. $^{a}$ \label{tab:weak}}
\tabletypesize{\scriptsize}
\tablewidth{0pt}
\tablehead{
\colhead{Object} & \colhead{\niv} & \colhead{\siiiuv} &  \multicolumn{2}{c} {\heiiuv}\\
\cline{4-5}
&  &  &  \colhead{$_{BC}$} & \colhead{$_{blue}$} 
}
\startdata
J00103-0037	&	2.4	$\pm$	1.9	&	1.1	$\pm$	0.8	&	1.3	$\pm$	0.4	&	2.0	$\pm$	1.8	\\
J00521-1108	&	0.1	$\pm$	0.2	&	1.1	$\pm$	1.1	&	1.6	$\pm$	1.2	&	0.1	$\pm$	0.2	\\
J01225+1339	&		\ldots		&	1.0	$\pm$	1.8	&	3.2	$\pm$	3.4	&	4.3	$\pm$	1.8	\\
J02287+0002 $^{(1)}$	&		\ldots		&	1.0	$\pm$	0.8	&	0.3	$\pm$	0.1	&	0.2	$\pm$	0.2	\\
J02287+0002 $^{(2)}$	&		\ldots		&	0.9	$\pm$	0.8	&	0.2	$\pm$	0.1	&	0.0	$\pm$	0.2	\\
J02390-0038	&	0.5	$\pm$	0.9	&	0.9	$\pm$	0.9	&	0.2	$\pm$	0.4	&	1.6	$\pm$	1.1	\\
J03036-0023	&		\ldots		&	1.5	$\pm$	1.2	&	0.5	$\pm$	0.5	&	9.2	$\pm$	3.4	\\
J20497-0554	&	0.3	$\pm$	0.6	&	1.5	$\pm$	1.5	&	1.8	$\pm$	0.8	&	4.6	$\pm$	2.7	\\
J23509-0052	&		\ldots		&	0.4	$\pm$	0.4	&	0.2	$\pm$	0.9	&	2.5	$\pm$	1.6	\\
\cline{1-5}  																	
Extreme Objects	&				&				&				&				\\
J12014+01161	&	3.0	$\pm$	3.5	&	4.9	$\pm$	5.8	&	1.5	$\pm$	1.7	&	10.3	$\pm$	9.2	\\
3C 390.3	&	5.6	$\pm$	2.7	&	1.6	$\pm$	1.5	&	2.8	$\pm$	1.5	&		\ldots		\\
\enddata
\tablecomments{(a) Units are $10^{-14}$ ergs s$^{-1}$ cm$^{-2}$ \AA$^{-1}$. (1) Considering $z_{OI\lambda1304}$. (2) Considering $z_{CIII]\lambda1909}$. We do not show \heiiuv$_{VBC}$ because is very weak, when is considered.}
\end{deluxetable}

\begin{deluxetable}{lccccccccc}
\rotate
\tablecaption{Equivalent Widths.\label{tab:ew} }
\tabletypesize{\scriptsize}
\tablewidth{0pt}
\tablehead{
\colhead{Object} & \colhead{\ciii$_{BC}$} & \colhead{\ciii$_{Tot}$} & \colhead{\siiii} & \colhead{\aliii}& \colhead{\siii} &  \colhead{\civ$_{BC}$} &\colhead{\civ$_{Tot}$} &  \colhead{\siiv$_{BC}$} &\colhead{\siiv$_{Tot}$} 
}
\startdata
J00103-0037	&	13.7	$\pm$	5.8	&	17.1	$\pm$	7.2	&	7.9	$\pm$	3.0	&	5.3	$\pm$	2.8	&	2.9	$\pm$	2.5	:	&	29.0	$\pm$	16.1	&	59.6	$\pm$	24.4	&	7.4	$\pm$	4.32	&	9.68	$\pm$	6.74	\\
J00521-1108	&	8.2	$\pm$	1.5	&	8.2	$\pm$	2.4	&	7.3	$\pm$	2.5	&	4.0	$\pm$	2.6	&	2.6	$\pm$	3.1		&	21.0	$\pm$	18.9	&	40.1	$\pm$	26.8	&		\ldots		&		\ldots		\\
J01225+1339	&	15.3	$\pm$	3.5	&		\ldots		&	11.6	$\pm$	3.1	&	6.0	$\pm$	3.6	&	1.4	$\pm$	2.3		&	25.5	$\pm$	9.1	&	41.6	$\pm$	9.7	&	11.1	$\pm$	3.95	&	17.74	$\pm$	4.71	\\
J02287+0002 $^{(1)}$	&	15.9	$\pm$	5.3	&	15.9	$\pm$	5.3	&	14.1	$\pm$	6.6	&	6.8	$\pm$	4.0	&	2.9	$\pm$	2.6		&	20.4	$\pm$	10.5	&	42.9	$\pm$	16.8	&	12.7	$\pm$	3.04	&	17.30	$\pm$	4.03	\\
J02287+0002 $^{(2)}$	&	20.5	$\pm$	5.3	&	20.5	$\pm$	5.3	&	8.3	$\pm$	6.6	&	5.2	$\pm$	4.0	&	2.7	$\pm$	2.6		&	30.8	$\pm$	10.5	&	40.6	$\pm$	16.8	&	12.7	$\pm$	3.04	&	17.85	$\pm$	4.03	\\
J02390-0038	&	10.8	$\pm$	3.3	&	14.7	$\pm$	4.5	&	10.4	$\pm$	2.0	&	7.0	$\pm$	2.4	&	2.5	$\pm$	2.5		&	16.3	$\pm$	4.1	&	37.7	$\pm$	7.0	&	4.9	$\pm$	1.32	&	11.72	$\pm$	4.15	\\
J03036-0023	&	12.4	$\pm$	1.8	&		\ldots		&	10.9	$\pm$	1.8	&	4.6	$\pm$	1.6	&	1.3	$\pm$	1.1	:	&	19.7	$\pm$	3.7	&	33.3	$\pm$	4.8	&	6.4	$\pm$	1.46	&	10.71	$\pm$	2.94	\\
J20497-0554	&	15.5	$\pm$	3.5	&		\ldots		&	14.1	$\pm$	2.3	&	5.6	$\pm$	2.9	&	2.7	$\pm$	2.9	:	&	25.4	$\pm$	5.6	&	38.1	$\pm$	7.1	&	8.0	$\pm$	3.75	&	10.34	$\pm$	4.04	\\
J23509-0052	&	15.8	$\pm$	4.0	&		\ldots		&	13.7	$\pm$	5.3	&	4.4	$\pm$	1.8	&	1.1	$\pm$	0.9		&	22.0	$\pm$	4.2	&	40.5	$\pm$	10.6	&	7.7	$\pm$	3.73	&	14.02	$\pm$	5.10	\\
\cline{1-10}  																																						
Extreme Objects	&				&				&				&				&					&				&				&				&				\\
J12014+0116	&	2.9	$\pm$	0.9	&		\ldots		&	8.4	$\pm$	4.1	&	7.1	$\pm$	2.0	&	2.4	$\pm$	1.9		&	8.2	$\pm$	5.6	&	19.4	$\pm$	7.4	&	6.03	$\pm$	2.76	&	9.85	$\pm$	4.37	\\
3C 390.3	&	13.1	$\pm$	4.3	&	28.4	$\pm$	6.2	&	7.9	$\pm$	3.3	&	0.7	$\pm$	0.8	&	1.2	$\pm$	1.3		&	49.1	$\pm$	8.1	&	147.2	$\pm$	19.4	&		\ldots		&		\ldots		\\
\enddata
\tablecomments{(1) Considering $z_{OI\lambda1304}$. (2) Considering $z_{CIII]\lambda1909}$. (:) \siii\ approximated values due the line is affected by telluric absorptions.}
\end{deluxetable}

\clearpage

\begin{deluxetable}{lcccccccccccccc}
\rotate
\setlength{\tabcolsep}{1.5pt}
\tablecaption{Hydrogen Density and Ionization Parameter.\label{tab:neu} }
\tabletypesize{\tiny}
\tablewidth{0pt}
\tablehead{
&\multicolumn{4}{c} {Log$n_H$}&&\multicolumn{4}{c} {Log$U$}&&\multicolumn{4}{c} {Log$n_H\cdot U$}\\
\cline{2-5}
\cline{7-10}
\cline{12-15}
\colhead{Object} &
\colhead{$1Z_\odot$} & \colhead{$1Z_\odot$ low dens} & \colhead{$5Z_\odot$} & \colhead{$5Z_\odot$ low dens} && \colhead{$1Z_\odot$} & \colhead{$1Z_\odot$ low dens} & \colhead{$5Z_\odot$} & \colhead{$5Z_\odot$ low dens} && \colhead{$1Z_\odot$}& \colhead{$1Z_\odot$ low dens} & \colhead{$5Z_\odot$} & \colhead{$5Z_\odot$ low dens} 
}
\startdata
J00103-0037	*&	{\bf 	12.50	$\pm$	0.17	}	&		\ldots		&			\ldots			&		\ldots		&&	-2.79	$\pm$	0.19	&		\ldots		&		\ldots		&		\ldots		&&		9.71	$\pm$	0.22		&			\ldots	\\
J00521-1108	&	{\bf 	12.43	$\pm$	0.26	}	&	12.84	$\pm$	0.23	&			\ldots			&		\ldots		&&	-2.86	$\pm$	0.15	&	-3.00	$\pm$	0.11	&		\ldots		&		\ldots		&&		9.58	$\pm$	0.26		&		9.85	$\pm$	0.22\\
J00103-0037	*&	{\bf 	12.50	$\pm$	0.17	}	&		\ldots		&			\ldots			&		\ldots		&&	-2.79	$\pm$	0.19	&		\ldots		&		\ldots		&		\ldots		&&		9.71	$\pm$	0.22		&			\ldots			&			\ldots			&		\ldots		\\
J00521-1108	&	{\bf 	12.43	$\pm$	0.26	}	&	12.84	$\pm$	0.23	&			\ldots			&		\ldots		&&	-2.86	$\pm$	0.15	&	-3.00	$\pm$	0.11	&		\ldots		&		\ldots		&&		9.58	$\pm$	0.26		&		9.85	$\pm$	0.22		&			\ldots			&		\ldots		\\
J01225+1339	&		12.45	$\pm$	0.22		&	13.24	$\pm$	0.20	&	{\bf 	11.47	$\pm$	0.28	}	&	12.62	$\pm$	0.25	&&	-2.96	$\pm$	0.09	&	-3.42	$\pm$	0.18	&	-1.83	$\pm$	0.11	&	-2.79	$\pm$	0.68	&&		9.49	$\pm$	0.21		&		9.82	$\pm$	0.23		&		9.64	$\pm$	0.27		&	9.83	$\pm$	0.66	\\
J02287+0002 $^{(1)}$	&	{\bf 	12.35	$\pm$	0.16	}	&		\ldots		&		11.64	$\pm$	0.33		&	12.28	$\pm$	0.28	&&	-2.81	$\pm$	0.28	&		\ldots		&	-2.33	$\pm$	0.25	&	-2.58	$\pm$	0.98	&&		9.55	$\pm$	0.28		&			\ldots			&		9.31	$\pm$	0.36		&	9.70	$\pm$	0.94	\\
J02287+0002 $^{(2)}$	&		12.47	$\pm$	0.16		&		\ldots		&			\ldots			&		\ldots		&&	-2.81	$\pm$	0.28	&		\ldots		&		\ldots		&		\ldots		&&		9.67	$\pm$	0.28		&			\ldots			&			\ldots			&		\ldots		\\
J02390-0038	&		12.75	$\pm$	0.12		&	13.42	$\pm$	0.21	&		11.97	$\pm$	0.15		&	13.20	$\pm$	0.13	&&	-3.18	$\pm$	0.05	&	-3.60	$\pm$	0.07	&	-2.19	$\pm$	0.08	&	-3.29	$\pm$	0.13	&&	{\bf 	9.57	$\pm$	0.11	}	&		9.82	$\pm$	0.20		&	{\bf 	9.78	$\pm$	0.15	}	&	9.91	$\pm$	0.15	\\
J03036-0023	&		12.32	$\pm$	0.14		&	12.92	$\pm$	0.15	&			\ldots			&		\ldots		&&	-2.92	$\pm$	0.06	&	-3.34	$\pm$	0.05	&		\ldots		&		\ldots		&&	{\bf 	9.40	$\pm$	0.14	}	&	{\bf 	9.58	$\pm$	0.15	}	&			\ldots			&		\ldots		\\
J20497-0554	&	{\bf 	12.26	$\pm$	0.25	}	&	12.82	$\pm$	0.18	&			\ldots			&		\ldots		&&	-2.89	$\pm$	0.12	&	-3.29	$\pm$	0.13	&		\ldots		&		\ldots		&&		9.37	$\pm$	0.25		&		9.53	$\pm$	0.19		&			\ldots			&		\ldots		\\
J23509-0052	&		12.13	$\pm$	0.24		&	12.90	$\pm$	0.17	&		10.86	$\pm$	0.23		&	12.12	$\pm$	0.58	&&	-2.89	$\pm$	0.09	&	-3.51	$\pm$	0.21	&	-1.93	$\pm$	0.13	&	-2.99	$\pm$	0.82	&&		9.24	$\pm$	0.24		&		9.39	$\pm$	0.23		&		8.93	$\pm$	0.23		&	9.13	$\pm$	0.86	\\
\cline{1-15}  																																																											
Extreme Objects	&						&				&						&				&&				&				&				&				&&						&						&						&				\\
J12014+01161	&		12.81	$\pm$	0.29		&	12.95	$\pm$	0.16	&	{\bf 	12.34	$\pm$	0.11	}	&	12.90	$\pm$	0.42	&&	-2.85	$\pm$	0.18	&	-2.87	$\pm$	0.08	&	-2.41	$\pm$	0.25	&	-3.20	$\pm$	0.58	&&		9.97	$\pm$	0.30		&		10.09	$\pm$	0.16		&		9.93	$\pm$	0.25		&	9.70	$\pm$	0.62	\\
3C 390.3	&	{\bf 	10.05	$\pm$	0.34	}	&		\ldots		&			\ldots			&		\ldots		&&	-1.48	$\pm$	0.34	&		\ldots		&		\ldots		&		\ldots		&&		8.57	$\pm$	0.41		&			\ldots			&			\ldots			&		\ldots		\\
\enddata
\tablecomments{We also show the values considering the correction by the contribution of low density regions (\S \ref{sec:ciii_contr}), $Z=5Z_\odot$ (\S \ref{metals}), and $Z=5Z_\odot$ with the correction by the contribution of low density regions. (1) Considering $z_{OI\lambda1304}$. (2) Considering $z_{CIII]\lambda1909}$. (*) For J00103-0037 the correction is too large to be reliable. We show in bold numbers the ones that we consider the best.}
\end{deluxetable}

\begin{deluxetable}{lccccccccccccccc}
\tabletypesize{\tiny}
\rotate
\setlength{\tabcolsep}{1.3pt}
\tablecaption{The Size of the Broad Line Region and the Black Hole Masses.\label{tab:r_m}}
\tablewidth{0pt}
\tablehead{
\colhead{Object} & \colhead{$d_p [Mpc]$} & \colhead{f(1700\AA)$^{\mathrm{a}}$}& \colhead{f(1350\AA)$^{\mathrm{a}}$} & \colhead{FWHM$_{BC}$} &\colhead{Pop.} &  \multicolumn{4}{c} {Log($r_{BLR}$) [cm]$^{\mathrm{b}}$} & & \multicolumn{5}{c}{Log($M_{BH}$) [$M_{\odot}$]$^{\mathrm{b}}$}\\
\cline{7-10}
\cline{12-16}
\colhead{}  & 
\colhead{$X10^{15}$} & 
\colhead{$X10^{-15}$} & 
\colhead{$X10^{-15}$} & 
\colhead{[km s$^{-1}$]}&
\colhead{} & 
\colhead{original} &
\colhead{low dens} &
\colhead{$5Z_\odot$} &
\colhead{$5Z_\odot$low dens} &
\colhead{} &
\colhead{original} &
\colhead{low dens} &
\colhead{$5Z_\odot$} &
\colhead{$5Z_\odot$low dens} &
\colhead{V\&P (2006)$^{\mathrm{c}}$}\\
\colhead{(1)}&\colhead{(2)}&\colhead{(3)}&\colhead{(4)}&\colhead{(5)}&\colhead{(6)}&\colhead{(7)}&\colhead{(8)}&\colhead{(9)}&\colhead{(10)}&&\colhead{(11)}&\colhead{(12)}&\colhead{(13)}&\colhead{(14)}&\colhead{(15)}
}
\startdata
J00103-0037	&		6.51	&	4.8		$\pm$	1.0	&	6.8	$\pm$	1.4	&	4500	$\pm$	800	&	B	&	18.10	$\pm$	0.12	&		\ldots		&		\ldots		&		\ldots		&&	9.16	$\pm$	0.20	&		\ldots		&		\ldots		&		\ldots		&	9.11	\\
J00521-1108	&		6.59	&	6.1		$\pm$	1.5	&	8.8	$\pm$	2.1	&	5300	$\pm$	1600	&	B	&	18.23	$\pm$	0.14	&	18.09	$\pm$	0.13	&		\ldots		&		\ldots		&&	9.43	$\pm$	0.30	&	9.29	$\pm$	0.29	&		\ldots		&		\ldots		&	9.32	\\
J01225+1339	&		6.42	&	8.1		$\pm$	1.6	&	10.9	$\pm$	2.2	&	4400	$\pm$	1000	&	A$^{\dagger}$	&	18.32	$\pm$	0.12	&	18.16	$\pm$	0.13	&	18.25	$\pm$	0.15	&	18.15	$\pm$	0.33	&&	9.36	$\pm$	0.23	&	9.20	$\pm$	0.23	&	9.29	$\pm$	0.25	&	9.19	$\pm$	0.39	&	9.19	\\
J02287+0002 $^{(1)}$	&		6.09	&	7.4		$\pm$	2.7	&	8.4	$\pm$	3.0	&	4700	$\pm$	1000	&	A$^{\dagger}$	&	18.25	$\pm$	0.16	&		\ldots		&	18.37	$\pm$	0.20	&	18.17	$\pm$	0.48	&&	9.35	$\pm$	0.25	&		\ldots		&	9.47	$\pm$	0.27	&	9.27	$\pm$	0.51	&	9.17	\\
J02287+0002 $^{(2)}$	&			&					&				&				&		&	18.19	$\pm$	0.16	&		\ldots		&		\ldots		&		\ldots		&&	9.29	$\pm$	0.25	&		\ldots		&		\ldots		&		\ldots		&		\\
J02390-0038	&		6.45	&	6.9		$\pm$	2.1	&	9.9	$\pm$	3.0	&	5400	$\pm$	1000	&	B	&	18.25	$\pm$	0.09	&	18.12	$\pm$	0.12	&	18.14	$\pm$	0.10	&	18.08	$\pm$	0.10	&&	9.46	$\pm$	0.18	&	9.34	$\pm$	0.20	&	9.36	$\pm$	0.19	&	9.30	$\pm$	0.19	&	9.35	\\
J03036-0023	&		6.58	&	20.5		$\pm$	5.7	&	30.0	$\pm$	8.4	&	3700	$\pm$	600	&	A	&	18.58	$\pm$	0.10	&	18.49	$\pm$	0.10	&		\ldots		&		\ldots		&&	9.47	$\pm$	0.17	&	9.38	$\pm$	0.17	&		\ldots		&		\ldots		&	9.30	\\
J20497-0554	&		6.55	&	6.6		$\pm$	1.3	&	9.5	$\pm$	1.9	&	3800	$\pm$	600	&	A	&	18.34	$\pm$	0.13	&	18.26	$\pm$	0.11	&		\ldots		&		\ldots		&&	9.25	$\pm$	0.19	&	9.17	$\pm$	0.18	&		\ldots		&		\ldots		&	9.04	\\
J23509-0052	&		6.40	&	4.8		$\pm$	1.0	&	6.1	$\pm$	1.2	&	3600	$\pm$	800	&	A	&	18.33	$\pm$	0.13	&	18.25	$\pm$	0.12	&	18.49	$\pm$	0.13	&	18.38	$\pm$	0.43	&&	9.19	$\pm$	0.23	&	9.12	$\pm$	0.23	&	9.35	$\pm$	0.23	&	9.25	$\pm$	0.47	&	8.88	\\
\cline{1-16}																																																					
Extreme Objects	&			&					&				&				&		&				&				&				&				&&				&				&				&				&		\\
J12014+01161	&		6.58	&	21.8		$\pm$	2.3	&	31.9	$\pm$	3.2	&	4000	$\pm$	800	&	A	&	18.31	$\pm$	0.15	&	18.25	$\pm$	0.09	&	18.33	$\pm$	0.13	&	18.44	$\pm$	0.31	&&	9.26	$\pm$	0.23	&	9.20	$\pm$	0.19	&	9.28	$\pm$	0.22	&	9.40	$\pm$	0.43	&	9.37	\\
3C 390.3	&		0.24	&	4.2		$\pm$	0.7	&	9.9	$\pm$	1.6	&	6400	$\pm$	2200	&	B	&	17.23	$\pm$	0.21	&		\ldots		&		\ldots		&		\ldots		&&	8.60	$\pm$	0.37	&		\ldots		&		\ldots		&		\ldots		&	7.99	\\
\enddata
\tablecomments{(a) Units of the flux at 1350 and 1700\AA\, are in ergs s$^{-1}$ cm$^{-2}$ \AA$^{-1}$. (b) The showed values are the average $\pm$ 0.17 dex of the computation using both SED of Laor (1997) and Mathews \& Ferland (1987) (see Fig. \ref{fig:Laor_Mathews}. (c) We show the comparison between our computations and those using the Vestergaard \& Peterson (2006) mehtod. They report an uncertainty of 0.66 dex. ($^\dagger$) Acording to the FWHM it is clasified as pop B, but has other spectral caracteristics of pop. A objects. See \S \ref{sec:component_analysis}. (1) Considering $z_{OI\lambda1304}$. (2) Considering $z_{CIII]\lambda1909}$}
\end{deluxetable}

\clearpage

\begin{figure}[h]
 \begin{center}
\includegraphics[scale=0.85, angle=0]{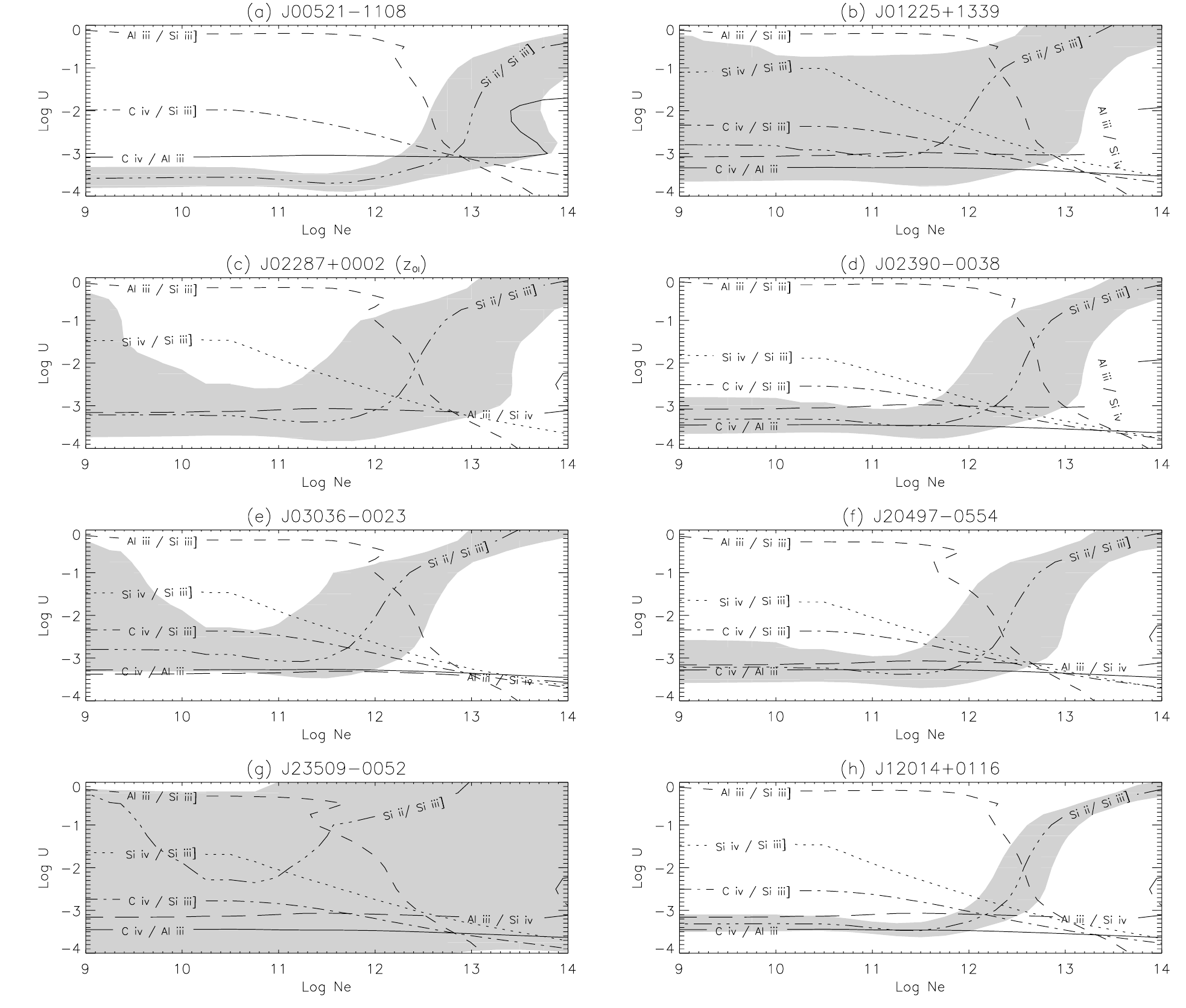}
\caption{Contour plots for corrected values considering the low density emission contribution of \ciii. Contour plot for J00103-0037 is not shown because the correction is so large to be reliable. For the objects in panels (e) and (f) the \siii\ line is also affected by the telluric absorption. Abscissa, ordinate and symbols are the same as Fig. \ref{fig:neu_extreme}.\label{fig:neucorrected}}
\end{center}
\end{figure}

\begin{figure}
\epsscale{1.}
\plotone{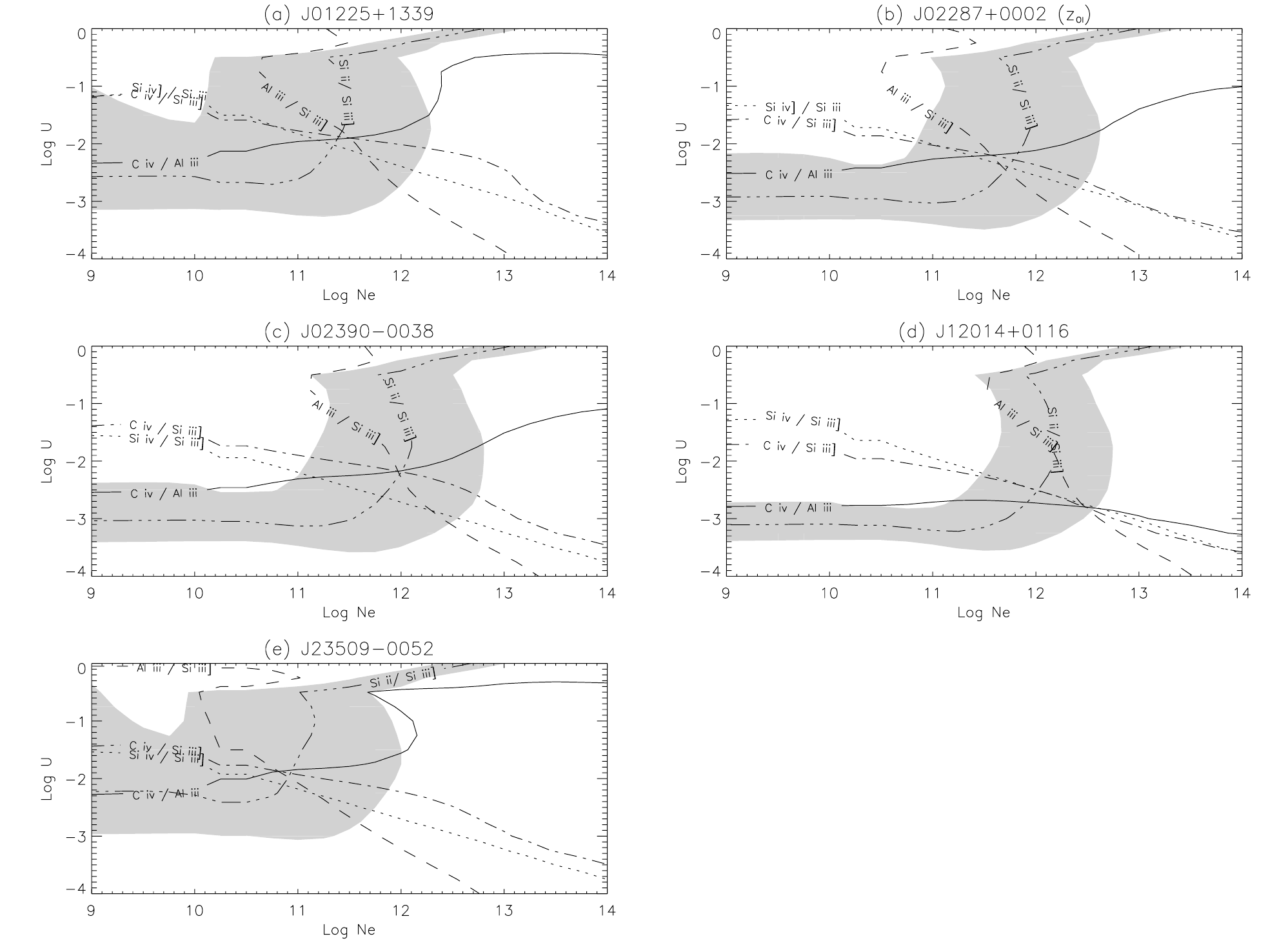}
\caption{Contour plots for (a) J01225+1339, (b) J02287+0002 using $z_{CIII]\lambda1909}$, (c) J02390-0038, the extreme object (d) J12014+0116, and (e) J23509-0052, from the array of simulations computed for $Z = 5Z_{\odot}$.  Coordinates and symbols are as for Fig. \ref{fig:neu_extreme}. The  intersection point improves in certain cases, but in others is the same.\label{fig:z5}}
\end{figure}

\begin{figure}
\epsscale{1.}
\plotone{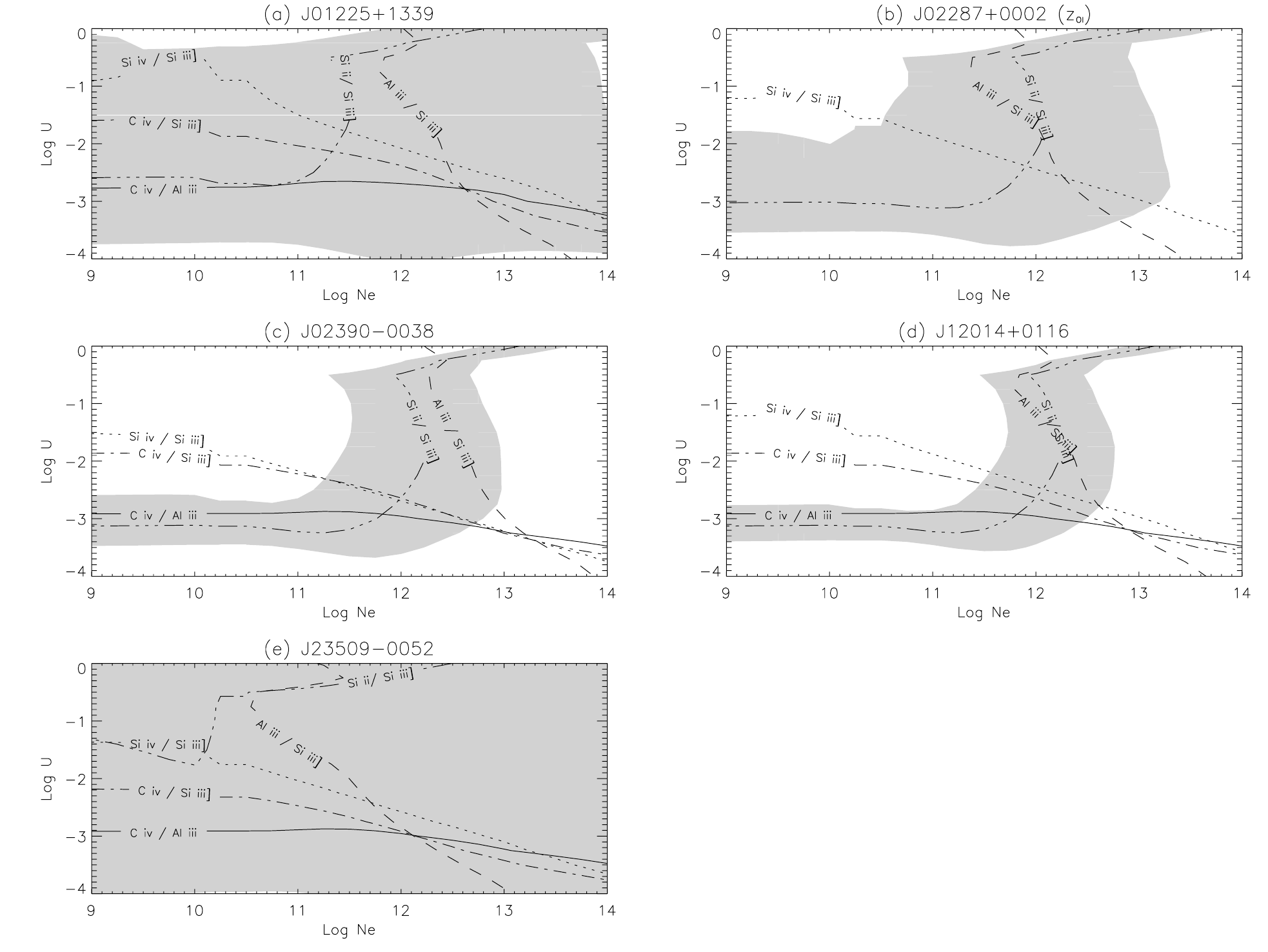}
\caption{Contour plots for the same objects of the previous figure from the array of simulations computed for $Z = 5Z_{\odot}$\ with ratios corrected because of low-density emission.  Coordinates and symbols are as for Fig. \ref{fig:neu_extreme}. \label{fig:z5corr}}
\end{figure}

\begin{figure}
\epsscale{0.8}
\plotone{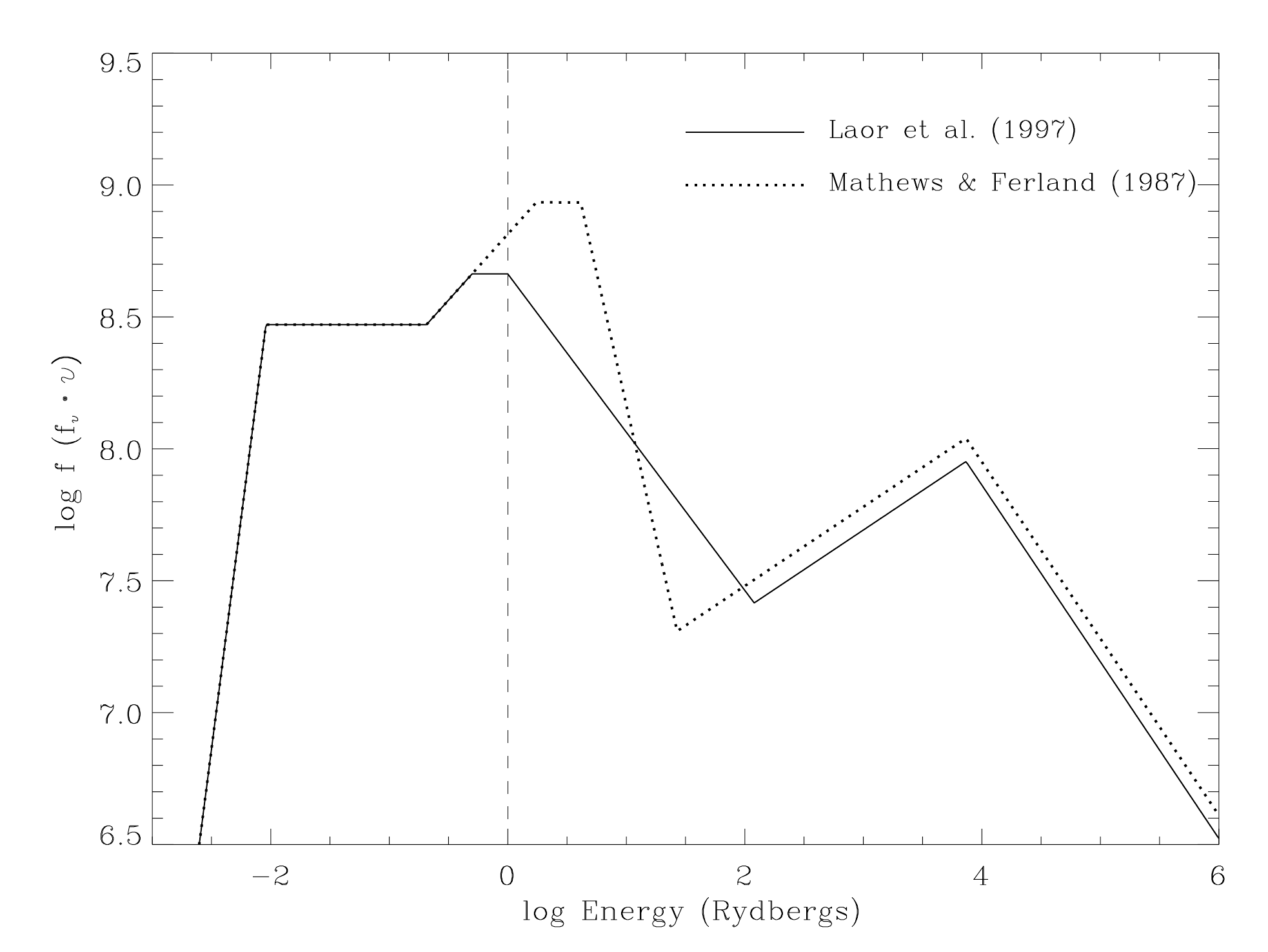}
\caption{Spectral energy distribution used to compute the number of ionizing photons for Laor et al. (1997) in solid line and Mathews \& Ferland (1987) in dotted line. Dashed line shows the Lyman limit. \label{fig:Laor_Mathews}}
\end{figure}

\begin{figure}
\epsscale{0.75}
\plotone{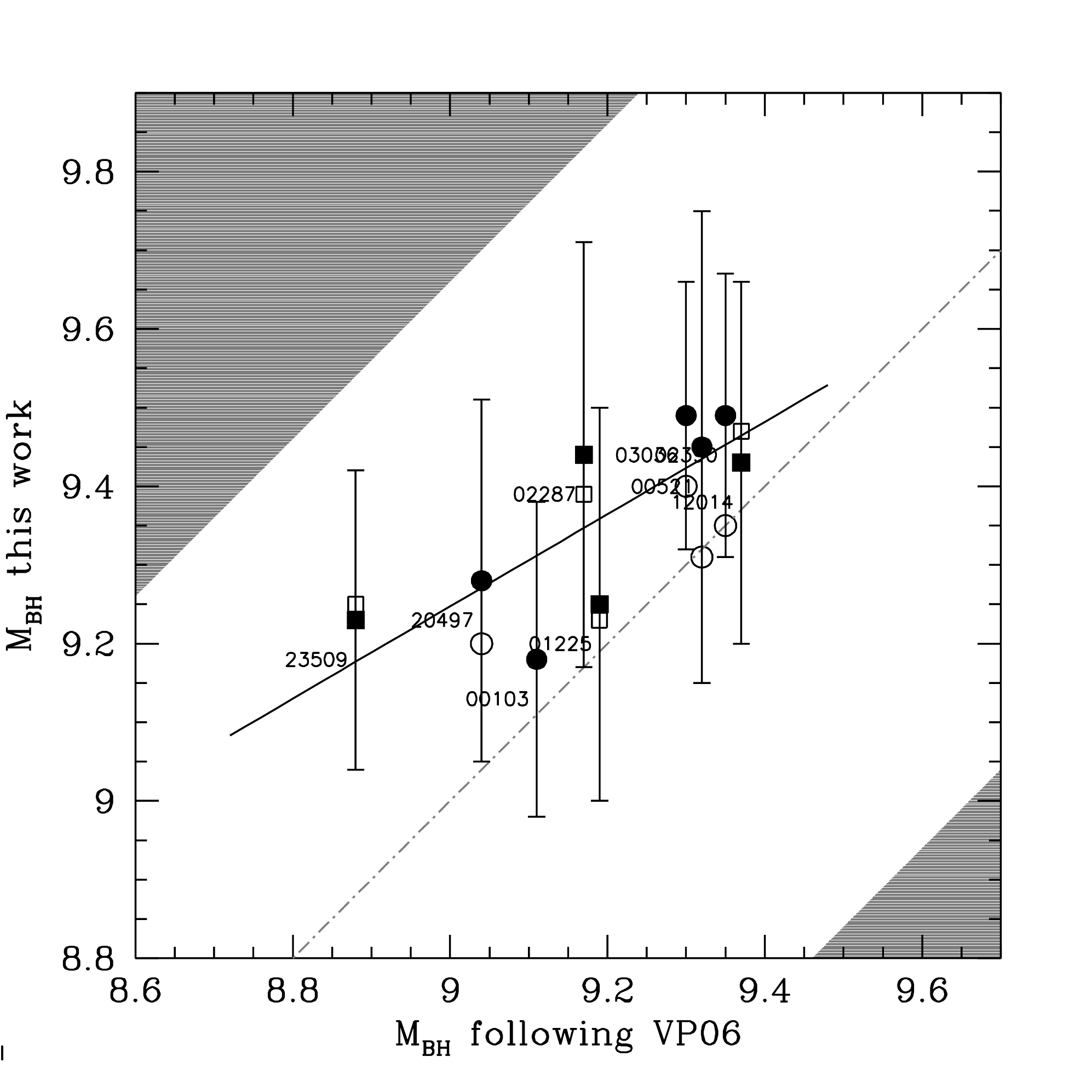}
\caption{\mbh\ comparison for the high-$z$ sample. Filled symbols refer to uncorrected intensity ratios; open symbols are for intensity ratios corrected because of low-density emission. Circles refer to solar metallicity; squares to 5 times solar metallicity and Si-Al enrichment.  Abscissa and ordinate are   logarithm of   \mbh\ in solar masses. Each point is labeled with the object name in short format. In ordinate we report \mbh\ values obtained with the method of this paper; in abscissa those obtained employing the  Vestergaard \& Peterson relationship described in the text.  The shaded bands  limit the $2\sigma$ confidence level spread expected on the basis of the Vestergaard \& Peterson relationship.\label{fig:MBHcomparison}}
\end{figure}

\end{document}